\documentclass[12pt]{article}
\usepackage{amsmath,amssymb}
\usepackage{bm}
\usepackage{color}
\usepackage{graphicx}
\usepackage{cite}
\usepackage{mathtools}
\usepackage{breqn}
\usepackage{here}
\usepackage{slashed}

\newcommand{\maru}[1]{\raise0.2ex\hbox{\textcircled{\scriptsize{#1}}}}

\begin{document}
\title{
\begin{flushright}
\ \\*[-80pt]
\begin{minipage}{0.2\linewidth}
\normalsize
EPHOU-22-021\\*[50pt]
\end{minipage}
\end{flushright}
{\Large \bf
Number of zero-modes on magnetized $T^4/Z_N$ orbifolds analyzed by modular transformation
\\*[20pt]}}

\author{
~Shota Kikuchi,
~Tatsuo Kobayashi, 
~Kaito Nasu,\\
~Shohei Takada, and
~Hikaru Uchida
\\*[20pt]
\centerline{
\begin{minipage}{\linewidth}
\begin{center}
{\it \normalsize
Department of Physics, Hokkaido University, Sapporo 060-0810, Japan} \\*[5pt]
\end{center}
\end{minipage}}
\\*[50pt]}

\date{
\centerline{\small \bf Abstract}
\begin{minipage}{0.9\linewidth}
\medskip
\medskip
\small
We study fermion zero-mode wavefunctions on $T^4/Z_N$ orbifold with background magnetic fluxes.  
The number of zero-modes is analyzed by use of $Sp(4,\mathbb{Z})$ modular transformation. Conditions needed to realize three generation models are clarified. We also study parity transformation in the compact space which leads to better understanding of relationship between positive and negative chirality wavefunctions.
\end{minipage}
}

\begin{titlepage}
\maketitle
\thispagestyle{empty}
\end{titlepage}

\newpage


\section{Introduction}
\label{Intro}
The Standard Model (SM) of particle physics succeeds in describing particle interactions: electromagnetic, weak, and strong forces. There can be seen good agreements between its predictions and numerous experimental results. However, there are still unsolved puzzles. For example, the theory is incapable of explaining the origin of the flavor structure. This includes the generation number of fermions, hierarchical structures in Yukawa couplings, and so on. Therefore, we are driven to study their origin from a viewpoint of high-energy underlying theory. Superstring theory is a promising candidate of unified theory including quantum gravity. It requires 10 dimensional (10D) space-time which implies that extra 6 dimensional space is compactified. Therefore, we study higher-dimensional theory in the hope of realizing 
the flavor structure of the SM 
through suitable compactification.
See ref. \cite{Ibanez:2012zz} for phenomenological aspects of superstring theory.

Motivated by superstring theory, we start with $\mathcal{N}=1$ super Yang-Mills theory(SYM) in 10 dimensions. Then we consider Kaluza-Klein decomposition of 10D fields. The degeneracy of fermion zero-modes on the compact space is equal to the generation number in the 4D field theory resulting from the dimensional reduction. Yukawa coupling constants can be evaluated by the overlap integral of wavefunctions corresponding to those zero-modes and lightest boson field over the compact space. 
Toroidal compactification with background magnetic fluxes is quite interesting. Thanks to the magnetic field, 4D chiral theory is realized. The degeneracy of the zero-modes is determined by the magnitude of the magnetic flux. Zero-mode wavefunctions are analytically obtained\cite{Cremades:2004wa}. In magnetized $T^2$ models, they are written in terms of Jacobi theta-function. Since their profiles are quasi-localized, nontrivial overlaps may generate hierarchical structure in Yukawa couplings. 

Orbifolding of magnetized tori is 
phenomenologically attractive\cite{Abe:2008fi, Abe:2013bca, Abe:2014noa}. This is because unwanted adjoint matter fields can be projected out which may lead to more realistic models than magnetized torus compactification.
Furthermore, we obtain richer variety of three generation models. Phenomenological aspects of magnetized $T^2/Z_N$, ($N=2,3,4,6$) models were studied in depth, in particular realization of quark and lepton mass matrices \cite{Abe:2014vza,Fujimoto:2016zjs,Kikuchi:2021yog,Hoshiya:2022qvr} and flavor symmetries \cite{Kobayashi:2017dyu,Kobayashi:2018rad,Kobayashi:2018bff,Ohki:2020bpo,Kikuchi:2020frp,Kikuchi:2021ogn,Almumin:2021fbk}. Zero-mode degeneracy was analyzed in various methods. One of the most efficient analyses was done by applying modular transformation\cite{Kobayashi:2017dyu}\footnote{See for analyses on the zero-mode number through 
the index theorem \cite{Sakamoto:2020pev,Sakamoto:2020vdy,Kobayashi:2022tti,Kobayashi:2022xsk}.}. $T^2$ has a discrete geometrical symmetry known as modular symmetry, $SL(2,\mathbb{Z})$. Furthermore, $T^2 \times T^2$ permutation orbifold models were also studied\cite{Kikuchi:2020nxn, Kikuchi: T2T2permutaion}.

It is intriguing to extend such analysis of zero-mode number to magnetized $T^{2n}/Z_N$ orbifolds. 
Magnetized $T^4$ and $T^4/Z_2$ models were previously studied\cite{Cremades:2004wa, Antoniadis:2009bg,Kikuchi:2022_T4, Abe:2014nla}. Zero-mode wavefunctions are written in terms of Riemann-theta function. In $T^4$, there is $Sp(4,\mathbb{Z})$ modular symmetry. Here, we apply this symmetry to zero-mode counting in magnetized $T^4/Z_{N}, (N=3,4,5,6,$ etc.) orbifold models. We perform systematic analyses and clarify conditions needed to realize three generations of fermions. We assume that the 6 dimensional compact space is $T^4/Z_N \times \mathcal{M}_2$ where $\mathcal{M}_2$ is a two dimensional compact space such as two-dimensional orbifolds on which there is a single zero-mode, making no contribution to the overall zero-mode numbers. The D-term condition preserving 4D $\mathcal{N}=1$ supersymmetry (SUSY) depends on
both $T^4/Z_N$ and $\mathcal{M}_2$ sectors.

This paper is organized as follows. In section 2, we give a review of magnetized $T^4$ models.
In section 3, we introduce magnetized $T^4/Z_N$ orbifold models. In section 4, $Sp(4,\mathbb{Z})$ modular transformation of magnetic fluxes as well as zero-mode wavefunctions are explicitly presented. In section 5, we show some algebraic relations satisfied by the generators of $Sp(4,\mathbb{Z})$. In section 6, by use of these algebraic relations, we construct $T^4/Z_N$ orbifold models and systematically analyze zero-mode numbers. In Section 7, we show 
that a parity transformation connects two different magnetized $T^4/Z_6$ orbifold models. There is a duality in the sense that positive chirality modes on one side correspond to negative modes on the other.
Section 8 is our conclusion.
In appendix A, we give a derivation of modular transformation properties of zero-mode wavefunctions. In Appendix B, we provide some details of magnetized $T^4/Z_4$ models. In Appendix C, we define $T^4/Z_2$ permutation orbifolds and show some basic facts. In Appendix D, we show negative chirality wavefunction and its relation to positive chirality mode via parity transformation. In Appendix E, we review the F-term SUSY condition. Some useful properties of the Riemann-theta functions are shown in Appendix F.


\section{Magnetized $T^4$ model}
\label{sec:Revew}
Here, we review zero-mode wavefunctions on magnetized $T^4$ \cite{Cremades:2004wa}, \cite{Antoniadis:2009bg}. 

\subsection{Dirac operator on magnetized $T^4$}
We begin with constructing the Dirac operator on the magnetized four dimensional torus $T^4 \simeq \mathbb{C}^2/\Lambda$, where $\Lambda$ is a lattice spanned by four independent lattice vectors $e_i, (i=1,2,3,4)$ in $\mathbb{C}^2$. In this paper, we are interested in 
cases when 
at least two of the vectors have the same length and perpendicular to each other. This means by a suitable orthogonal rotation of the coordinates of $\mathbb{C}^2 \simeq \mathbb{R}^4$, we can write $e_i$'s of the form
\begin{equation}
\label{eq: lattice_vectors}
    e_1 = 2 \pi R \begin{pmatrix}
   1  \\ 0 
    \end{pmatrix},\ 
    e_2 = 2 \pi R \begin{pmatrix}
    \tau_1 \\ \tau_4
    \end{pmatrix},\ 
    e_3 =  2 \pi R\begin{pmatrix}
    0 \\ 1
    \end{pmatrix},\ 
    e_4=  2 \pi R \begin{pmatrix}
    \tau_3 \\ \tau_2
    \end{pmatrix},
\end{equation}
where $\tau_i, (i=1,2,3,4) \in \mathbb{C}$ 
and $R > 0$. We may further restrict the situation by demanding $\tau_3 = \tau_4$ because we will treat only such cases in later discussions about orbifold models. We introduce 
$x^i, y^i, (i=1,2)$ as 
real coordinates along the lattice vectors of the torus. This means the complex coordinates $(Z_1, Z_2)$ of $\mathbb{C}^2$ are related as
\begin{align}
 \begin{aligned}
 \label{eq: complex_corrdinates_T4}
    \begin{pmatrix}
    Z_1 \\ Z_2
    \end{pmatrix}
    &= 2 \pi R \left[
    \begin{pmatrix}
    x^1 \\ x^2
    \end{pmatrix}
    + 
     \begin{pmatrix}
    \tau_1 & \tau_3 \\
    \tau_3 & \tau_2
    \end{pmatrix}
    \begin{pmatrix}
    y^1 \\ y^2
    \end{pmatrix} 
    \right] 
    = 2 \pi R ( \vec{x} + \Omega \vec{y} )
    = 2 \pi R \vec{z},
    \end{aligned}
\end{align}
where $\vec{z} = \vec{x} + \Omega \vec{y}$ denotes the complex coordinates of the torus. When $T^4/Z_N$ orbifolds are considered, ${\rm Im}\Omega$ can be made positive definite\footnote{When the Coxeter element corresponding to the twist defining the $T^4/Z_N$ orbifold is diagonalized as $\theta= {\rm diag}[exp(2\pi i( \eta^1, \eta^2))]$, both $\eta^1$ and $\eta^2$ can lie between $0$ and $1/2$.
See e.g. refs.\cite{Dixon:1986jc,Markushevich:1986za,Katsuki:1989bf}.}. We parameterize $\Omega$ as
\begin{equation}
\label{eq: moduli}
    \Omega = 
    \begin{pmatrix}
    \tau_1 & \tau_3 \\
    \tau_3 & \tau_2
    \end{pmatrix}.
\end{equation}
The set of such symmetric complex $2 \times 2$ matrices $\Omega$ with positive definite imaginary part is known as Siegel upper-half-space of genus 2, which is denoted by $\mathcal{H}_2$\cite{Mumford:1983}. 
The metric of $\mathbb{C}^2$ is
\begin{equation}
    ds^2 = 2 H_{i \bar{j}} dZ^i d\bar{Z}^j, \quad (i,j = 1,2),
\end{equation}
where
\begin{equation}
    H_{i \bar{j}} = \frac{1}{2}\delta_{i\bar{j}}.
\end{equation}
Therefore, the metric of our $T^4$ is given by
\begin{equation}
    ds^2 = 2 h_{i \bar{j}} dz^i d\bar{z}^j,\quad h_{i \bar{j}} = (2 \pi R)^2 H_{i \bar{j}}.
\end{equation}
Gamma matrices on the complex coordinates of $T^4$ are
\begin{align}
    \begin{aligned}
    \Gamma^{z^i} = \frac{1}{2 \pi R} \Gamma^{Z^i},\ \Gamma^{\bar{z}^i} =
    \frac{1}{2 \pi R}
    \Gamma^{\bar{Z}^i},\quad
    \end{aligned}
\end{align}
satisfying $\{\Gamma^{z^i}, \Gamma^{z^j} \} = 2 h^{i \bar{j}}$, 
where
\begin{align}
\begin{aligned}
    \Gamma^{Z^1} &= \sigma^Z \otimes \sigma^3 = 
    \begin{pmatrix}
    0 & 2 & & \\
    0 & 0 & & \\
    & & 0 & -2 \\
    & & 0 & 0
    \end{pmatrix},\quad 
    \Gamma^{Z^2} =  I \otimes \sigma^Z= 
    \begin{pmatrix}
     &  &2 &0 \\
     & & 0&2 \\
    0& 0&  &  \\
    0& 0&  & 
    \end{pmatrix}, \\
 \Gamma^{\bar{Z}^1} &= \sigma^{\bar{Z}} \otimes \sigma^3 = 
    \begin{pmatrix}
    0 & 0 & & \\
    2 & 0 & & \\
    & & 0 & 0 \\
    & & -2 & 0
    \end{pmatrix},\quad
\Gamma^{\bar{Z}^2} = I \otimes \sigma^{\bar{Z}} =
 \begin{pmatrix}
     &  &0 &0 \\
     & & 0&0 \\
    2& 0&  &  \\
    0& 2&  & 
    \end{pmatrix}.
    \end{aligned}
\end{align}
Here, $\sigma^a\ (a=1,2,3)$ denote the Pauli matrices, and $\sigma^{Z}= \sigma^1 + i \sigma^2,\  \sigma^{\bar{Z}}=\sigma^1 - i \sigma^2$. The chirality matrix $\Gamma^5$ is given by 
\begin{equation}
    \Gamma^5 = 
\begin{pmatrix}
    1 & & & \\ 
      & -1 & & \\
      & & -1 &  \\
      & & & 1
\end{pmatrix}.
\end{equation}
Then, the Dirac operator on $T^4$ is written by,
\begin{equation}
    i \slashed{D} = i  \Gamma^{z^j} D_{z^j} +  i \Gamma^{\bar{z}^j} \bar{D}_{\bar{z}^j}
    =
    \frac{i}{\pi R}
    \begin{pmatrix}
0 & D_{z^1} &  D_{z^2} & 0 \\
\bar{D}_{\bar{z}^1} & 0 & 0 &  D_{{z}^2} \\
\bar{D}_{\bar{z}^2} & 0 & 0 & - D_{z^1} \\
0 &  \bar{D}_{\bar{z}^2} & -  \bar{D}_{\bar{z}^1} & 0
    \end{pmatrix}.
\end{equation}
The covariant derivatives are written by 
\begin{align}
\begin{aligned} 
D_{z^j} &= \partial_{z^j} - i A_{z^j}, \\
\bar{D}_{\bar{z}^j} &= \bar{\partial}_{\bar{z}^j} - i A_{\bar{z}^j},
\end{aligned}
\end{align}
where we assumed that the fermion field coupled to the $U(1)$ vector potential $A = A_{z^j} dz^j + A_{\bar{z}^j} d\bar{z}^j$ is unit charged. 

\subsection{Background magnetic flux}
We introduce uniform magnetic field on $T^4$ written as \cite{Antoniadis:2009bg},
\begin{equation}
F = \frac{1}{2} p_{x^ix^j} dx^i \wedge dx^j + \frac{1}{2} p_{y^i y^j} dy^i \wedge dy^j + p_{x^i y^j} dx^i \wedge dy^j.
\end{equation}
By the complex coordinates the flux is given by the following expression,
\begin{equation}
F = \frac{1}{2} F_{z^i z^j} dz^i \wedge dz^j + \frac{1}{2} F_{\bar{z}^i \bar{z}^j} d\bar{z}^i \wedge d\bar{z}^j + F_{z^i \bar{z}^j} (i dz^i \wedge d\bar{z}^j) ,
\end{equation}
where
\begin{align}
F_{z^i z^j} &= {(\bar{\Omega} - \Omega)^{-1}} (\bar{\Omega} p_{xx} \bar{\Omega} + p_{yy} + p_{xy}^{\rm T} \bar{\Omega} - \bar{\Omega} p_{xy}) {(\bar{\Omega} - \Omega)^{-1}}, \notag\\
F_{\bar{z}^i \bar{z}^j} &= {(\bar{\Omega} - \Omega)^{-1}} ({\Omega} p_{xx} {\Omega} + p_{yy} + p_{xy}^{\rm T} {\Omega} - {\Omega} p_{xy}) {(\bar{\Omega} - \Omega)^{-1}}, \\
F_{z^i \bar{z}^j} &= i {(\bar{\Omega} - \Omega)^{-1}} \left( \bar{\Omega} p_{xx} \Omega + p_{yy}+ p_{xy}^{\rm T} \Omega - \bar{\Omega} p_{xy}  \right) {(\bar{\Omega} - \Omega)^{-1}}. \notag
\end{align}
We constrain the form of flux
to conserve $\mathcal{N}=1$ supersymmetry in the 4D space-time field theory resulting from the dimensional reduction. That is the flux needs to be a $(1,1)$-form corresponding to the F-flat condition \cite{Cremades:2004wa}, which we review in Appendix \ref{appendix: F-flat}. This can be satisfied if,
\begin{equation}
{\Omega} p_{xx} {\Omega} + p_{yy} + p_{xy}^{\rm T} {\Omega} - {\Omega} p_{xy} = 0.
\end{equation}
As a result, $F$ can be written as
\begin{equation}
F = i (p_{xx} \Omega - p_{xy}) (\bar{\Omega} - \Omega)^{-1} (i dz^i \wedge d\bar{z}^j)  = -i {(\bar{\Omega}-\Omega)^{-1}}(\bar{\Omega} p_{xx} + p_{xy}^{\rm T}) (i dz^i \wedge d\bar{z}^j) . 
\end{equation}
From this, we see the hermiticity of the flux: $F_{z^i \bar{z}^j}=F^{\dagger}_{z^i \bar{z}^j}$. 
The expression is further simplified by assuming $p_{xx}=p_{yy}=0$,
\begin{equation}
F_{z^i \bar{z}^j} = i [p_{xy} (\Omega - \bar{\Omega})^{-1}]_{i\bar{j}} .
\end{equation}
The F-flat SUSY condition is also simplified
\begin{equation}
(p_{xy}^{\rm T} \Omega)^{\rm T} = p_{xy}^{\rm T} \Omega.
\end{equation}
The flux must be quantized due to the consistency with the boundary conditions we take. Thus, we can write $p_{xy}$ in terms of a $2 \times 2$ real integer matrix $N$, 
\begin{equation}
p_{xy} = 2 \pi {N}^{\rm T} .
\end{equation}
This is referred to as Dirac's quantization condition. 
Then we obtain
\begin{equation}
F = \pi \left[N^{\rm T} ({\rm Im}\Omega)^{-1} \right]_{i\bar{j}} (i dz^i \wedge d\bar{z}^j),
\end{equation}
with the F-flat condition rewritten as
\begin{equation}
\label{eq: SUSY_condition}
(N \Omega)^{\rm T} = {N} \Omega.
\end{equation}
The corresponding gauge potential is given by
\begin{align}
\begin{aligned}
A(\vec{z}, \vec{\bar{z}}) &= \pi  {\rm Im} \left\{ [{N} (\vec{\bar{z}}  + \vec{\bar{\zeta}})]^{\rm T} ({\rm Im} \Omega)^{-1} d \vec{z}  \right\}  \\
&= - \frac{\pi i}{2} \left\{ [{N} (\vec{\bar{z}}  + \vec{\bar{\zeta}})]^{\rm T} ({\rm Im} \Omega)^{-1} \right\}_i dz^i + \frac{\pi i}{2} \left\{ [{N} (\vec{{z}}  + \vec{{\zeta}})]^{\rm T} ({\rm Im} \Omega)^{-1} \right\}_{\bar{i}} d\bar{z}^i \\
& =: A_{z^i} dz^i + A_{\bar{z}^i} d\bar{z}^i,
\end{aligned}
\end{align}
where $\vec{\zeta}$ is a complex constant 2-component vector known as the Wilson lines. The boundary conditions are
\begin{align}
\begin{aligned}
\label{eq: A_boundary}
A(\vec{z}+\vec{e}_k) &= A(\vec{z}) + d \chi_{\vec{e}_k}(\vec{z}), \\
A(\vec{z} + \Omega \vec{e}_k) &= A(\vec{z}) + d\chi_{\Omega \vec{e}_k} (\vec{z}), 
\end{aligned}
\end{align}
where 
\begin{align}
\begin{aligned}
\chi_{\vec{e}_k} (\vec{z}) := \pi [N^{\rm T} ({\rm Im}\Omega)^{-1} {\rm Im} (\vec{z} + \vec{\zeta})]_k,&\quad \chi_{\Omega \vec{e}_k} (\vec{z}) := \pi {\rm Im}[{N} \bar{\Omega} ({\rm Im}\Omega)^{-1} (\vec{z} + \vec{\zeta})]_k.
\end{aligned}
\end{align}
Here, $\vec{e}_k, (k=1,2)$ denote standard unit vectors.
In order to simplify our discussion, we take vanishing Wilson lines $\vec{\zeta}=0$.
 
\subsection{Zero-mode wavefunction}
Fermion zero-modes satisfy the Dirac equation,
\begin{equation}
\label{eq: Dirac_1}
    i \slashed{D}\Psi(\vec{z}, \vec{\bar{z}}) = 0,
\end{equation}
    where 
\begin{equation}
\Psi(\vec{z}, \vec{\bar{z}}) 
    =
\begin{pmatrix}
\psi^1_+ \\ \psi_-^2 \\ \psi_-^1 \\ \psi_+^2    
\end{pmatrix}.
\end{equation}
Here, $\psi_+^j$ and $\psi_-^j$ denote the positive and negative chirality components respectively. Writing eq.(\ref{eq: Dirac_1}) in components, we get
\begin{align}
\label{eq:1}
    \bar{D}_{\bar{z}^1} \psi_+^1 + D_{z^2} \psi_+^2 &=0, \\
\label{eq:2}
    \bar{D}_{\bar{z}^2} \psi_+^1 - D_{z^1} \psi_+^2 &= 0, \\
\label{eq: 3}
     D_{z^1} \psi_-^2 + D_{z^2} \psi_-^1 &= 0, \\
\label{eq: 4}
   \bar{D}_{\bar{z}^2} \psi_-^2 -  \bar{D}_{\bar{z}^1} \psi_-^1 &= 0,
\end{align}
where
\begin{align}
\begin{aligned}
  D_{z^i} &= \partial_{z^i} - \frac{\pi}{2} ([N \vec{\bar{z}}]^{\rm T}({\rm Im}\Omega)^{-1})_i, \\
  \bar{D}_{\bar{z}^i} & = \bar{\partial}_{\bar{z}^i} + \frac{\pi}{2} ([N \vec{{z}}]^{\rm T}({\rm Im}\Omega)^{-1})_i.
\end{aligned}
\end{align}
Boundary conditions are
\begin{align}
\begin{aligned}
\label{eq: boundaryT4}
    \Psi(\vec{z}+\vec{e}_k) &= e^{i \chi_{\vec{e}_k}(\vec{z})} \Psi(\vec{z}), \\
    \Psi(\vec{z} + \Omega \vec{e}_k) &= e^{i \chi_{\Omega \vec{e}_k (\vec{z})}} \Psi(\vec{z}),
    \end{aligned}
\end{align}
which allow us to solve Dirac equations consistently on $T^4$.

\subsubsection{Solving the Dirac equation}
Commutation relations of covariant derivatives under the SUSY condition are obtained as 
\begin{align}
    [D_{z^i}, D_{z^j}] &= F_{z^i z^j} = 0, \\
    [\bar{D}_{\bar{z}^i}, \bar{D}_{\bar{z}^j}] &= F_{\bar{z}^i \bar{z}^j} = 0, \\
    \label{eq: Fzbz}
    [{D}_{z^i}, \bar{D}_{\bar{z}^j}] &= F_{{z}^i \bar{z}^j}.
\end{align}
We define the Laplace operator $\Delta$ as 
\begin{equation}
\label{eq: laplace}
    \Delta = - \frac{2}{(2\pi R)^2} \sum_{i=1,2} \{D_{z^i}, \bar{D}_{\bar{z}^i}\} .
\end{equation}
Then the positive chirality components satisfy
\begin{align}
\begin{aligned}
    \Delta \psi_+^1 &= 2 (F_{z^1 \bar{z}^1} + F_{z^2 \bar{z}^2}) \psi_+^1, \\
    \Delta \psi_+^2 &= - 2 (F_{z^1 \bar{z}^1} + F_{z^2 \bar{z}^2}) \psi_+^2.
\end{aligned}
\end{align}
Eigenvalues of the Laplace operator in eq.(\ref{eq: laplace}) are non-negative.
This can be checked as follows.
Firstly, consider the following integration of some wavefunctions $\psi_a$ and $\psi_b$ satisfying the same boundary conditions eq.(\ref{eq: boundaryT4}). We find 
\begin{align}
\begin{aligned}
\int_{T^4} d^2z d^2{\bar{z}} (\psi_a)^* (i D_{z^j} \psi_b) =  \int_{T^4} d^2z d^2{\bar{z}} (i \bar{D}_{\bar{z}^j} \psi_a)^* \psi_b,
\end{aligned}
\end{align}
because the surface term vanishes. This shows that $iD_{z^j}$ and $i\bar{D}_{\bar{z}^j}$ are Hermitian conjugate with each other. Therefore, $\Delta$ is a Hermitian operator. 
Secondly, let  us denote an eigenvalue of the Laplacian by $\lambda \in \mathbb{R}$,
\begin{equation}
    \Delta \psi = \lambda \psi,
\end{equation}
and consider 
\begin{equation}
\label{eq: eigenvalue_Delta}
     \int_{T^4} d^2z d^2{\bar{z}} \psi^*    \Delta \psi = \lambda
      \int_{T^4} d^2z d^2{\bar{z}} |\psi|^2.
\end{equation}
The left-hand side of eq.(\ref{eq: eigenvalue_Delta}) is
\begin{align}
\begin{aligned}
    2 \sum_j \int_{T^4} d^2z d^2{\bar{z}} \psi^*     \{i D_{z^j}, i \bar{D}_{\bar{z}^j}\} \psi &= 2 \sum_j   \int_{T^4} d^2z d^2 \bar{z} ( |D_{z^j} \psi|^2 + |\bar{D}_{\bar{z}^j} \psi|^2) \geq 0.
    \end{aligned}
\end{align}
Consequently, we find $\lambda \geq 0$. Thus, if $F_{z^1 \bar{z}^1} + F_{z^2 \bar{z}^2} = {\rm tr}(N^{\rm T} ({\rm Im} \Omega)^{-1} )> 0$, the second positive chirality component is zero, $\psi_+^2 = 0$. This leads to simpler equations for $\psi_+^1$, 
\begin{equation}
\label{eq: psi+1}
    \bar{D}_{\bar{z}^j} \psi_+^1 = 0,\quad (j=1,2).
\end{equation}
On the other hand, if $F_{z^1 \bar{z}^1} + F_{z^2 \bar{z}^2} < 0$, the first positive chirality component is zero, $\psi_+^1 = 0$ leading to 
\begin{equation}
\label{eq: psi+2}
    D_{z^j} \psi_+^2 = 0,\quad (j=1, 2).
\end{equation}

\subsubsection{Solution}
We focus on eq.(\ref{eq: psi+1}). Its solutions are given by\cite{Cremades:2004wa, Antoniadis:2009bg},
\begin{equation}
\label{eq: positive_chirality_zeromode}
    \psi_N^{\vec{J}} = \mathcal{N}_{\vec{J}} \cdot e^{\pi i [N \vec{z}]^{\rm T} \cdot ({\rm Im}\Omega)^{-1} \cdot {\rm Im}\vec{z}} \cdot \vartheta
    \begin{bmatrix}
    {\vec{J}}^{\, \rm T} N^{-1} \\ 0
    \end{bmatrix}({N}\vec{z}, { N}\Omega),
\end{equation}
where 
\begin{equation}
\vartheta 
 \begin{bmatrix}
 \vec{a}^{\, \rm T} \\
 \vec{b}^{\, \rm T}
 \end{bmatrix} (\vec{\nu}, \tilde{\Omega}) = 
 \sum_{\vec{l} \in \mathbb{Z}^2} e^{\pi i (\vec{l} + \vec{a})^{\rm T} \cdot \tilde{\Omega} \cdot (\vec{l} + \vec{a})} e^{2 \pi i (\vec{l} + \vec{a})^{\rm T} \cdot (\vec{\nu} + \vec{b} )},\quad \vec{a}, \vec{b} \in \mathbb{R}^2,\ \vec{\nu} \in \mathbb{C}^2,\   \tilde{\Omega}^{\rm T} = \tilde{\Omega},\ {\rm Im} {\tilde{\Omega}} > 0,
\end{equation}
is known as the Riemann theta-function. 
$\vec{J}$ is a two dimensional integer column vector which lies inside the lattice cell ${\Lambda_N}$ spanned by
\begin{equation}
N^{\rm T}{\vec{e}_n}  ,\quad (n = 1,2),
\end{equation}
where $\vec{e}_n$'s are unit vectors.\footnote{ Normalization of $\vec{J}$ labelling zero-modes degeneracy is different from that in refs. \cite{Cremades:2004wa, Antoniadis:2009bg} where labels $\vec{j}$ satisfying $N^{\rm T} \vec{j} \in \mathbb{Z}^2$ are used. They are related by $\vec{J}=
N^{\rm T} \vec{j} \pmod{N^{\rm T} \vec{e}_n}$. }
$\mathcal{N}_J$ is a normalization constant,
\begin{equation}
\label{eq: normalization}
    \mathcal{N}_{\vec{J}} = 
    [{\rm Vol}(T^4)]^{-1/2} ({\rm det}N)^{1/4},
\end{equation}
which follows from normalization condition,
\begin{equation}
    \int_{T^4} d^2z d^2{\bar{z}}\  \psi^{\vec{J}}_N
      (\psi^{\vec{K}}_N)^* = (2^2 {\rm det}({\rm Im}\Omega))^{-1/2} \delta_{\vec{J},\vec{K}}.
\end{equation}
By noting 
\begin{equation}
\label{eq: identify}
\psi_N^{\vec{J} + N^{\rm T} \vec{e}_n} = \psi_N^{\vec{J}},
\end{equation}
the degeneracy of zero-modes is ${\rm det}N$. In general, $\psi_+^1$ is written by their linear combinations,
\begin{equation}
\label{eq: linear_combination}
    \psi_+^1 = \sum_{\vec{J} \in \Lambda_N} a_{\vec{J}} \cdot \psi^{\vec{J}}_N,\quad a_{\vec{J}} \in \mathbb{C}.
\end{equation}
Note that wavefunctions in eq.(\ref{eq: positive_chirality_zeromode}) are normalizable when the following positive definite condition,
\begin{equation}
\label{eq: positive_definite_conditon}
    N {\rm Im}\Omega > 0,
\end{equation}
is satisfied\cite{Cremades:2004wa}.\footnote{Eq.(\ref{eq: positive_definite_conditon}) is equivalent to $N^{\rm T} ({\rm Im}\Omega)^{-1}>0$ under the SUSY condition and $\Omega \in \mathcal{H}_2$.}
This is equivalent to 
\begin{align}
\begin{aligned}
    {\rm det}N > 0, \\
    {\rm tr}(N {\rm Im}\Omega) > 0.
\end{aligned}
\end{align}
When these conditions are not met, other components of the spinor have non-zero solution. In the case of ${\rm det}N>0$ and $ {\rm tr}(N {\rm Im}\Omega)<0$, 
\begin{equation}
\label{eq: negative_definite}
    \psi_N^{\vec{J}} = \mathcal{N}_{\vec{J}} \cdot e^{\pi i [N \vec{\bar{z}}]^{\rm T} \cdot ({\rm Im}\bar{\Omega})^{-1} \cdot {\rm Im}\vec{\bar{z}}} \cdot \vartheta
    \begin{bmatrix}
    {\vec{J}}^{\, \rm T} N^{-1} \\ 0
    \end{bmatrix}({N}\vec{\bar{z}}, { N}\bar{\Omega}),
\end{equation}
are the solutions of eq.(\ref{eq: psi+2}).
If ${\rm det}N < 0$, negative chirality components become non-zero. In Appendix \ref{appendix: negative}, we discuss negative chirality wavefunctions.

\section{Magnetized $T^4/Z_N$ model}
Here, we briefly summarize how to construct magnetized $T^4/Z_N$ twisted orbifold models. Our discussion is along  ref.\cite{Abe:2013bca} where magnetized $T^2/Z_N$ twisted orbifold models are studied. 

$T^4/Z_N$ orbifolds are defined by imposing following identification,
\begin{equation}
\label{eq: twist_identify}
    \vec{z} \sim \Omega_{\rm twist} \vec{z},
\end{equation}
on $T^4 \simeq \mathbb{C}^2/\Lambda$. Here, we take $\Omega_{\rm twist}$ as a $2 \times 2$ unitary matrix satisfying
$(\Omega_{\rm twist})^N = I_2$, representing a $Z_N$-twist. We define the boundary condition on $T^4/Z_N$ as 
\begin{equation}
    \Psi(
    \Omega_{\rm twist} \vec{z},
    \bar{\Omega}_{\rm twist} \vec{\bar{z}}
    ) = \mathcal{S}V(\vec{z},\vec{\bar{z}})\Psi(\vec{z},\vec{\bar{z}}),
\end{equation}
in addition to eq.(\ref{eq: boundaryT4}). Here, $\mathcal{S}$ denotes the spinor representation of the $Z_N$-twist and $V(\vec{z},\vec{\bar{z}})$ is a transformation function. As in the preceding study ref.\cite{Abe:2013bca}, we conventionally define $V(\vec{z},\vec{\bar{z}})$ so that $\mathcal{S}$ acts trivially on $\psi_+^1$. For example, if we consider the twist 
\begin{equation}
\begin{pmatrix}
z_1 \\ z_2
\end{pmatrix}
\rightarrow 
\begin{pmatrix}
    e^{2 \pi i k_1/N} z_1 \\ e^{2 \pi i k_2/N} z_2
\end{pmatrix},\quad k_1, k_2 \in \mathbb{Z},
\end{equation}
we take  
\begin{align}
\begin{aligned}
    \mathcal{S} &= \frac{1}{4}(\Gamma^{Z^1}\Gamma^{\bar{Z}^1} + e^{2 \pi i k_1/N} \Gamma^{\bar{Z}^1} \Gamma^{Z^1} + \Gamma^{Z^2}\Gamma^{\bar{Z}^2} + e^{2 \pi i k_2/N} \Gamma^{\bar{Z}^2} \Gamma^{Z^2}) \\
    &= {\rm diag}(1,e^{2 \pi i k_1/N},e^{2 \pi i k_2/N},e^{2 \pi i (k_1 + k_2)/N}),
\end{aligned}
\end{align}
so that $\psi_+^1$ is unchanged.
In some other $Z_N$-twists, $\Omega_{\rm twist}$ may not be diagonal, so $z_1$ and $z_2$ are mixed. Then,  $\psi_+^1$ and $\psi_+^2$ will be mixed by the action of $\mathcal{S}$, so the situation seems more complicated. However, it still makes sense to define $V(\vec{z},\vec{\bar{z}})$ so that on-shell $\psi_+^1$ is invariant under $\mathcal{S}$. Suppose we start with conditions eq.(\ref{eq: positive_definite_conditon}), so that only $\psi_+^1$ is non-zero, and $\psi_+^2 = 0$.
As we will see in the following sections, both $N$ and $\Omega$ will be invariant under our twists, therefore $\psi_+^2=0$ is maintained. Thus, $\mathcal{S}$ would act to on-shell $\psi_+^1$
up to $U(1)$ phase which can be removed by the redefinition of $V(\vec{z},\vec{\bar{z}})$.
The transformation function $V(\vec{z},\vec{\bar{z}})$ is written as 
\begin{equation}
    V(\vec{z},\vec{\bar{z}}) = e^{2\pi i \beta}, 
\end{equation}
where $\beta$ is a real number. Since $N$ repeated $Z_N$-twists need to be identity, we must have
\begin{equation}
    \beta N \equiv 0 \pmod1.
\end{equation}
This shows that there are $N$ different sectors with different $Z_N$-twist charges, $e^{2\pi i \frac{k}{N}}, (k=0,1,...,N-1)$. 
 
The eigenstates on $T^4/Z_N$ are written by the linear combination of states on $T^4$ as in  eq.(\ref{eq: linear_combination}).
However, $a_{\vec{J}}$ are no longer arbitrary. They are chosen to satisfy the additional $Z_N$-twist boundary condition.

\section{Modular transformation}
Under the modular transformation $\gamma \in Sp(4, \mathbb{Z})$, the complex coordinates and the complex structure $\Omega \in \mathcal{H}_2$ transform as \cite{Mumford:1983},
\begin{equation}
    (\vec{z}, \Omega) \rightarrow ({(C \Omega + D )^{-1}}^{\rm  T} \vec{z}, (A \Omega + B)(C \Omega + D )^{-1}),
\end{equation}
where $\gamma$ is given by $2 \times 2$ matrices, $A, B, C$, and $D$ as
\begin{equation}
    \gamma = 
    \begin{pmatrix}
    A & B \\ C & D
    \end{pmatrix},
\end{equation}
By the definition of $Sp(4,\mathbb{Z})$, $\gamma$ must satisfy $\gamma^{\rm T} J \gamma = J$ where 
\begin{equation}
    J = 
    \begin{pmatrix}
    0 & I_2 \\
    -I_2 & 0
    \end{pmatrix}.
\end{equation}
We will use following generators of  $Sp(4,\mathbb{Z})$,
\begin{align}
S  = 
  \begin{pmatrix}
  0 & I_2 \\
  - I_2 & 0
  \end{pmatrix},\quad
T_i =
  \begin{pmatrix}
  I_2 & B_i \\
  0 & I_2
  \end{pmatrix},\ (i=1,2,3),
\end{align}
as in ref.\cite{Ding:2021}. Here, $B_i$'s are symmetric $2 \times 2$ integer matrices,
\begin{equation}
    B_1 
    = 
    \begin{pmatrix}
    1 & 0 \\ 0 & 0 
    \end{pmatrix},\quad
    B_2 
    = 
    \begin{pmatrix}
    0 & 0 \\ 0 & 1 
    \end{pmatrix},\quad 
    B_3
    = 
    \begin{pmatrix}
    0 & 1 \\ 1 & 0 
    \end{pmatrix}.  
\end{equation}

\subsection{$S$ transformation}
Under the $S$ transformation, we obtain
\begin{align}
\begin{aligned}
\label{eq: S_trans_complex}
    \vec{z} &\xrightarrow{S} \vec{z}_{(S)} = - \Omega^{-1} \vec{z}, \\
    \Omega &\xrightarrow{S} \Omega_{(S)} = - \Omega^{-1}.
\end{aligned}
\end{align}
where $\vec{z}_{(S)} = \vec{x}_{(S)} -\Omega^{-1} \vec{y}_{(S)}$ denotes the coordinates after the $S$ transformation.
From eq.(\ref{eq: S_trans_complex}), we find,
\begin{align}
    \begin{aligned}
\vec{x}_{(S)} &= - \vec{y}, \\
\vec{y}_{(S)} &= \vec{x}.
    \end{aligned}
\end{align}

\subsubsection{$S$ transformation of the magnetic flux}
 We study the $S$ transformation of the magnetic flux, 
\begin{align}
 \begin{aligned}
    F 
    &= \frac{1}{2} (p_{xx}^{(S)})_{ij} dx_{(S)}^i \wedge dx^j_{(S)} + \frac{1}{2} (p_{yy}^{(S)})_{ij} dy^i_{(S)} \wedge dy^j_{(S)} + (p_{xy}^{(S)})_{ij} dx^i_{(S)} \wedge dy^j_{(S)} \\
    &=  \frac{1}{2} (p_{xx}^{(S)})_{ij} dy^i \wedge dy^j + \frac{1}{2} (p_{yy}^{(S)})_{ij} dx^i \wedge dx^j + (p_{xy}^{(S)})^{\rm T}_{ij} dx^i \wedge dy^j.
 \end{aligned}
\end{align}
Thus, we find
\begin{align}
    \begin{aligned}
    p_{xx}^{(S)} &= p_{yy},\\ 
    p_{yy}^{(S)} &= p_{xx},\\
    p_{xy}^{(S)} &= (p_{xy})^{\rm T}.
    \end{aligned}
\end{align}
We clearly see that the condition $p_{xx}=p_{yy}=0$ is consistent with $S$ transformation. The flux matrix $N$ is transformed as
\begin{equation}
\label{eq: N_S_trans}
    N \xrightarrow{S} N_{(S)} = N^{\rm T}.
\end{equation}
At last, we should check the F-flatness SUSY condition eq.(\ref{eq: SUSY_condition}). It can be easily verified that 
\begin{align}
(N_{(S)} \Omega_{(S)})^{\rm T} 
&= N_{(S)} \Omega_{(S)},
\end{align}
holds if $(N\Omega)^{\rm T} = N \Omega$ is satisfied. This shows that the F-flat condition is consistent with $S$ transformation.

\subsubsection{$S$ transformation of zero-modes}
We concentrate on the case when ${N}^{\rm T}=N$, so that it is invariant under $S$. As we will see later, this symmetry follows automatically from the F-flat SUSY condition in orbifolds we have studied. Then, the zero-modes in eq.(\ref{eq: positive_chirality_zeromode}) transform as
\begin{equation}
\label{eq: S_zero-mode}
    \psi_N^{\vec{J}}(-\Omega^{-1}\vec{z},-\Omega^{-1}) = \sqrt{{\rm det}(N^{-1}\Omega/i)} \sum_{\vec{K} \in \Lambda_N} e^{2 \pi i \vec{J}^{\rm T} \cdot N^{-1}\cdot \vec{K}} \psi_N^{\vec{K}}(\vec{z}, \Omega),
\end{equation}
where the branch of the square root is chosen so that it takes positive value when $\Omega$ is purely imaginary. We give a proof of eq.(\ref{eq: S_zero-mode}) in Appendix \ref{appendix: S-zero}.

\subsection{$T_i$ transformation}
Under the $T_i$ transformation, we obtain 
\begin{align}
\begin{aligned}
\label{eq: T_trans_complex}
    \vec{z} &\xrightarrow{T_i} \vec{z}_{(T_i)} = \vec{z}, \\
    \Omega &\xrightarrow{T_i} \Omega + B_i,
\end{aligned}
\end{align}
where $\vec{z}_{(T_i)} = \vec{x}_{(T_i)} + (\Omega + B_i) \vec{y}_{(T_i)}$ denotes the coordinates after the $T_i$ transformation.
From eq.(\ref{eq: T_trans_complex}), we find,
\begin{align}
    \begin{aligned}
\vec{x}_{(T)} &= \vec{x} - B \vec{y}, \\
\vec{y}_{(T)} &= \vec{y}.
    \end{aligned}
\end{align}
Here, we omitted the subscript $i(=1,2,3)$. This is because when we consider the composite of $T_i$, we still obtain the same form of transformations as in eq.(\ref{eq: T_trans_complex}). We refer to them by just saying $T$ transformation. This means we may take $B$ as any symmetric $2 \times 2$ integer matrices including $B_i$.

\subsubsection{$T_i$ transformation of the magnetic flux}
 We study the $T$ transformation of the magnetic flux,
\begin{align}
 \begin{aligned}
    F &= \frac{1}{2} (p_{xx}^{(T)})_{ij} dx_{(T)}^i \wedge dx^j_{(T)} + \frac{1}{2} (p_{yy}^{(T)})_{ij} dy^i_{(T)} \wedge dy^j_{(T)} + (p_{xy}^{(T)})_{ij} dx^i_{(T)} \wedge dy^j_{(T)} \\
      &= \frac{1}{2} (p_{xx}^{(T)})_{ij} dx^i \wedge dx^j + \frac{1}{2} [(p_{yy}^{(T)}) + (B p_{xx}^{(T)} B) - (B p_{xy}^{(T)}) + (Bp_{xy}^{(T)})^{\rm T} ]_{ij} dy^i \wedge dy^j \\
      & + [(p_{xy}^{(T)}) - (p_{xx}^{(T)} B)]_{ij} dx^i \wedge dy^j. 
 \end{aligned}
\end{align}
Thus, we find
\begin{align}
\begin{aligned}
    p_{xx}^{(T)} &= p_{xx}, \\
    p_{yy}^{(T)} &= p_{yy} + (B p_{xx} B ) + (B p_{xy}) - (Bp_{xy})^{\rm T}, \\
    p_{xy}^{(T)} &= p_{xy} + (p_{xx} B).
\end{aligned}
\end{align}
We clearly see that the condition $p_{xx}=p_{yy}=0$ is no longer maintained when $T$ transformations are considered, unless
\begin{equation}
\label{eq: T_consistency}
    (B p_{xy})^{\rm T} = B p_{xy},
\end{equation}
is a symmetric matrix. 
 Under this condition, the magnetic flux $N$ is invariant,
\begin{equation}
\label{eq: N_T_trans}
    N \xrightarrow{T} N_{(T)} = N.
\end{equation}
At last, we should check the F-flat SUSY condition eq.(\ref{eq: SUSY_condition}). It can be easily verified that 
\begin{align}
(N_{(T)} \Omega_{(T)})^{\rm T} 
&= N_{(T)} \Omega_{(T)},
\end{align}
holds if $(N\Omega)^{\rm T} = N \Omega$ is satisfied. This shows that the F-flat condition is consistent with the $T$ transformation given eq.(\ref{eq: T_consistency}).

\subsubsection{$T$ transformation of zero-modes}
Here, we assume that the magnetic flux $N$ satisfies 
\begin{equation}
\label{eq: consistent_T}
    (N B)^{\rm T} = NB,
\end{equation}
so the corresponding $T$ transformation is consistent with $p_{xx}=p_{yy}=0$.
Furthermore, we demand that the diagonal matrix element of $NB$ be all even.
Then the zero-modes in eq.(\ref{eq: positive_chirality_zeromode}) transform as,
\begin{equation}
\label{eq: T_zero-mode}
    \psi_N^{\vec{J}}(\vec{z},\Omega+B) =  e^{ \pi i \vec{J}^{\rm T} \cdot (N^{-1}B)\cdot \vec{J}} \psi_N^{\vec{J}}(\vec{z}, \Omega).
\end{equation}
We give a proof of eq.(\ref{eq: T_zero-mode}) in Appendix \ref{appendix: T-zero}.

\subsection{$A \in GL(2,\mathbb{Z})$ transformation}
For later discussion, consider matrices of the form
\begin{equation}
\label{eq: gamma_A}
    \gamma =
    \begin{pmatrix}
    A & 0 \\ 0 & (A^{-1})^{\rm T}
    \end{pmatrix},
\end{equation}
where $A \in GL(2,\mathbb{Z})$. Then, $\gamma$ is an element of $Sp(4,\mathbb{Z})$.
Under the $A\in GL(2,\mathbb{Z})$ transformation, we obtain
\begin{align}
\begin{aligned}
\label{eq: A_P}
    \vec{z} &\xrightarrow{A} \vec{z}_{(A)} = A \vec{z}, \\
    \Omega &\xrightarrow{A} \Omega_{(A)} = A \Omega A^{\rm T}.
\end{aligned}
\end{align}
\subsubsection{$A \in GL(2,\mathbb{Z})$ transformation of the magnetic flux}
 The flux matrix $N$ is transformed as,
\begin{equation}
\label{eq: N_A_trans}
    N \xrightarrow{A} A N A^{-1}.
\end{equation}
It is straightforward to verify the consistency of $A$ transformation with the 
condition $p_{xx}=p_{yy}=0$.
\subsubsection{$A \in GL(2,\mathbb{Z})$ transformation of zero-modes}
We concentrate on the case when $N$ is invariant under a specific $A \in GL(2,\mathbb{Z})$ transformation,
\begin{equation}
\label{eq: A_inv_N}
    A N A^{-1} = N.
\end{equation}
Then, the zero-modes in eq.(\ref{eq: positive_chirality_zeromode}) transform as 
\begin{equation}
\label{eq: A_trans_zero}
 \psi_N^{\vec{J}}(A \vec{z}, A \Omega A^{\rm T}) = \psi_N^{A^{\rm T}\vec{J}}(\vec{z}, \Omega).
\end{equation}
We give a proof of eq.(\ref{eq: A_trans_zero}) in Appendix \ref{appendix: A-zero}.

\section{Algebraic relations}
We defined $T^4/Z_N$ orbifold by identifying different positions on $T^4$ related by the $Z_N$-twist as shown in eq.(\ref{eq: twist_identify}). In fact, the twist can be written as a modular transformation as we will see. In ref. \cite{Kobayashi:2017dyu}, magnetized $T^2/Z_N$ orbifolds were studied based on this fact. Thus, we  extend it
to $T^4/Z_N$. Our starting point is to look for a modular transformation  $\gamma \in Sp(4,\mathbb{Z})$ satisfying the algebraic relation $\gamma^N = I_4$, so it may represent a $Z_N$-twist.\footnote{$\gamma^k \neq I_4$ is necessary where $k \not\equiv 0 \pmod{N}$.}

Thus, we list up algebraic relations satisfied by the elements of $Sp(4,\mathbb{Z})$ which we have used to study magnetized $T^4/Z_N$ orbifold models,
\begin{align}
    \begin{aligned}
    & S^4 = I_4, \\
    &(ST_1T_2)^3 = I_4, \\
    &(ST_1T_3)^5 = I_4,\  (ST_2T_3)^5 = I_4, \\
    &(ST_3)^6 = I_4,\ (S T_1 T_2^{-1})^6 = I_4,\ ((T_1T_2)^{-1}S)^6 = I_4,\ (ST_1T_2\gamma_P)^6 = I_4,  \\ 
    &(S T_1 T_2^{-1} T_3^{-1})^8 = I_4, \\
    & ((T_1T_3)^{-1}S)^{10} = I_4,\ 
    ((T_2T_3)^{-1}S)^{10} = I_4, \\
    &(ST_1)^{12} = I_4,\ (ST_2)^{12} = I_4,
    \end{aligned}
\end{align}
    where $\gamma_P$ is an $A \in GL(2,\mathbb{Z})$ transformation with 
    \begin{equation}
A = \begin{pmatrix}
            0 & 1 \\
            1 & 0
        \end{pmatrix}.
    \end{equation}

\section{Degeneracy of zero-modes}
\subsection{$T^4/Z_4$}
We focus on the following algebraic relation,
\begin{equation}
    S^4 = I_4.
\end{equation}
This allows us to relate $S$ to the $Z_4$-twist which then is used for the construction of $T^4/Z_4$ orbifold,
as we will see. The complex coordinate and the complex structure moduli are transformed as eq.(\ref{eq: S_trans_complex}). We parameterize the moduli as eq.(\ref{eq: moduli}).
Let us look for the $S$-invariant $\Omega \in \mathcal{H}_2$. Then it must satisfy,
\begin{equation}
    \Omega^2 
    =
    \begin{pmatrix}
    \tau_1^2 + \tau_3^2 & \tau_3 (\tau_1 + \tau_2) \\
    \tau_3 (\tau_1 + \tau_2) & \tau_2^2 + \tau_3^2
    \end{pmatrix} = - I_2.
\end{equation}
From the off-diagonal elements, $\tau_3 = 0$ or $\tau_1=-\tau_2$ must hold. The latter case is discarded, because it contradicts to the requirement $\Omega \in \mathcal{H}_2$. Therefore, we have $\tau_3 = 0$. Then from the diagonal elements, the moduli parameters are fixed as, $\tau_1^2 = -1$, $\tau_2^2 = -1$. The relevant solution is unique,
\begin{equation}
\label{eq: S_inv_omega}
    \Omega_{(S)} =
    \begin{pmatrix}
    i & 0 \\ 0 & i
    \end{pmatrix},
\end{equation}
where we denoted the $S$-invariant moduli as $\Omega_{(S)}$.
As a result, the complex coordinate of $T^4$ is transformed under $S$ as
\begin{equation}
    \vec{z} \xrightarrow{S} i \vec{z}.
\end{equation}
This is nothing but a $Z_4$-twist.  
We are interested in the $T^4/Z_4$ orbifold which is defined by imposing the following identification 
\begin{equation}
    \vec{z} \sim i \vec{z},
\end{equation}
on the complex coordinate of $T^4$.  
This means that the moduli of $T^4$ must be $S$-invariant leading to the fixing of moduli as shown in  eq.(\ref{eq: S_inv_omega}). 

\subsubsection{Lattice vectors}
Here, we explicitly show the lattice vectors defining the $T^4/Z_4$ orbifold. They are given by,
\begin{equation}
    e_1 = 2 \pi R \begin{pmatrix}
   1  \\ 0 
    \end{pmatrix},\ 
    e_2 = 2 \pi R \begin{pmatrix}
    i \\ 0
    \end{pmatrix},\ 
    e_3 =  2 \pi R\begin{pmatrix}
    0 \\ 1
    \end{pmatrix},\ 
    e_4=  2 \pi R \begin{pmatrix}
    0 \\ i
    \end{pmatrix}.
\end{equation}
Lattice vectors are perpendicular with each other, namely they form a root lattice which corresponds to $SU(2)^4 \simeq SO(4)^2$. Both $(e_1, e_2)$ and $(e_3, e_4)$ are simple roots of $SO(4)$.
Under the $Z_4$-twist realized by the $S$ transformation, they behave as
\begin{align}
    \begin{pmatrix}
    e_2^{(S)}\\
    e_4^{(S)}\\
    e_1^{(S)}\\
    e_3^{(S)}
    \end{pmatrix}
    =
    \begin{pmatrix}
    0 & 0 & 1 & 0 \\
    0 & 0 & 0 & 1 \\
    -1 & 0 & 0 & 0 \\
    0 & -1 & 0 & 0 
    \end{pmatrix}
        \begin{pmatrix}
    e_2 \\
    e_4 \\
    e_1 \\
    e_3
    \end{pmatrix}.
\end{align}
The $Z_4$-twist is the generalized Coxeter element of $SO(4)^2$ including the $Z_2$ outer automorphism exchanging $e_1 \leftrightarrow e_2$ and $e_3 \leftrightarrow e_4$ \cite{Markushevich:1986za, Katsuki:1989bf}.

\subsubsection{Flux}
When the complex structure moduli are $\Omega_{(S)}$, the F-flat SUSY condition shown in eq.(\ref{eq: SUSY_condition}) is satisfied only if $N$ is a symmetric matrix,
\begin{equation} 
\label{eq: Z_4_flux}
    N = 
    \begin{pmatrix}
    n_1 & m \\ m & n_2
    \end{pmatrix}, \quad n_1, n_2, m \in \mathbb{Z}.
\end{equation}
Note that $N$ in eq.(\ref{eq: Z_4_flux}) is $S$-invariant according to eq.(\ref{eq: N_S_trans}). We study the case when the positive definite condition eq.(\ref{eq: positive_definite_conditon}) is satisfied so that only the first positive chirality component $\psi_+^1$ has non-zero solution. This means that we further restrict $N$ by 
\begin{equation}
    {\rm det}N > 0,\quad  {\rm tr}N >0.
\end{equation}

\subsubsection{Zero-mode counting method}
We have seen that zero-mode wavefunctions behave as eq.(\ref{eq: S_zero-mode}) under the $S$ transformation or we could say $Z_4$-twist instead. To analyze the degeneracy of zero-modes, it would be helpful to know the trace of the transformation matrix $\rho(S)$, 
\begin{align}
\begin{aligned}
\label{eq: trace_S_T4Z4}
    {\rm tr}\rho(S)&= \frac{1}{\sqrt{{\rm det}N}} \sum_{\vec{K} \in \Lambda_N} e^{2 \pi i \vec{K}^{\rm T} . N^{-1}. \vec{K}}\\
    &= 
    \frac{1}{\sqrt{{\rm det}N}} \sum_{\vec{K} \in \Lambda_N} e^{\frac{2\pi i}{{\rm det}N}(n_2K_1^2 + n_1 K_2^2 - 2m K_1 K_2)}.
\end{aligned}
\end{align}
Let us denote the number of zero-modes as $D_{(\pm 1)}, D_{(\pm i)}$ which correspond to the $Z_4$-twist eigenvalues $\pm 1, \pm i$. Then we obtain following relations,
\begin{align}
\label{eq: total_Z4}
    D_{(+1)} + D_{(+i)} + D_{(-1)} + D_{(-i)} = {\rm det}N, \\
\label{eq: trS_Z4}
    D_{(+1)} + iD_{(+i)} - D_{(-1)} -i D_{(-i)} = {\rm tr}\rho(S).
\end{align}
Eqs.(\ref{eq: total_Z4}) 
and (\ref{eq: trS_Z4}) together constrain three real degrees of freedom. One more independent constraint is needed.
The information of zero-modes' number on magnetized $T^4/Z_2$ provides it. This is due to the fact that two consecutive $Z_4$-twists are equivalent to $Z_2$-twist. Thus, $Z_4$ eigenstates with eigenvalues $\pm1$ are $Z_2$ even. $Z_4$ eigenstates with eigenvalues $\pm i$ are $Z_2$ odd. The number of $Z_2$ eigenstates was studied in refs.  \cite{Kikuchi:2022_T4, Abe:2014nla}. We have
\begin{align}
\begin{aligned}
\label{eq: Z2even}
  N_{Z_2+}-N_{Z_2-} &= D_{(+1)} - D_{(+i)} + D_{(-1)} - D_{(-i)} \\
 &= 
\begin{cases}
1: {\rm if}\ {\rm det}N \equiv 1 \pmod{2},\\
2: {\rm if}\ {\rm det}N
    \equiv 0 \pmod{2},\ {\rm gcd}(n_i,m) \equiv 1 \pmod{2}\ {\rm for}\ i = 1\ {\rm or}\ 2, \\
4:  {\rm gcd}(n_i,m) \equiv 0   \pmod{2}\ {\rm for}\ i = 1\ {\rm and}\ 2,
\end{cases}
\end{aligned}
\end{align}
where $N_{Z_2+}$ and $N_{Z_2-}$ denote the number of $Z_2$ even($+$) and odd($-$) modes respectively.

As we saw in eq.(\ref{eq: trS_Z4}), the trace of $\rho(S)$ is important in determining the zero-mode degeneracy. Even when we fixed the value of ${\rm det}N$, there are a large number of possible $N$.
This means we have a variety of $\Lambda_N$, characterizing the range of summation variables $\vec{K}$ in eq.(\ref{eq: trace_S_T4Z4}), so one might expect that we must treat individual $N$'s separately. However, it is possible to simplify the analysis greatly, if one makes rearrangements of $\vec{K}$. 
Firstly, consider when ${\rm det}N=p$ is a prime number and $N$ is written by eq.(\ref{eq: Z_4_flux}).
There are three cases,  
\begin{enumerate}
    \item  ${\rm gcd}(n_1, m) = p$ and ${\rm gcd}(n_2, m) = 1$,
    \item ${\rm gcd}(n_1, m) = 1$  and ${\rm gcd}(n_2, m) = p$,
    \item ${\rm gcd}(n_1, m) = {\rm gcd}(n_2, m) = 1$. 
\end{enumerate}
For case 1, $\vec{K}$ can be rearranged as $\vec{K}^{\rm T}=(0,0), (1,0), ..., (p-1, 0)$. This is because two $\vec{K}$s' are identified, if they differ by 
\begin{equation}
\label{eq: identify_v}
    \vec{v} =
\begin{pmatrix}
v_1 \\ v_2
\end{pmatrix}
=
    \alpha \begin{pmatrix}
    n_1 \\ m
    \end{pmatrix}
    + \beta 
    \begin{pmatrix}
    m \\ n_2
    \end{pmatrix},\ (\alpha, \beta \in \mathbb{Z}),
\end{equation}
according to eq.(\ref{eq: identify}).
Let us take $\alpha$ and $\beta$ to satisfy
\begin{equation}
    v_2 = \alpha m + \beta n_2 =0 .
\end{equation}
Since $n_2$ and $m$ are relatively prime, general solutions are \begin{equation}
    \alpha = n_2 l, \
    \beta = - m l,\quad (l \in \mathbb{Z}).
\end{equation}
Then, we get
\begin{equation}
    v_1 = \alpha n_1 + \beta m = p \cdot l.
\end{equation}
If we choose $l=1$, we obtain
\begin{equation}
    \vec{v} = \begin{pmatrix}
    p \\ 0 
    \end{pmatrix},
\end{equation}
which is the smallest unit of identification. All of $p$ integer points on this $\vec{v}$ are not equivalent, so our claim is verified. For case 2, we may rearrange as $\vec{K}^{\rm T} = (0,0), (0,1), ..., (0, p-1)$. For case 3, $\vec{K}^{\rm T}$ can be taken as $(n,0)$ or $(0,n)$ where $ n = 0,1,\cdots ,p -1$.

Secondly, consider general cases, when ${\rm det}N$ is prime factorized as 
\begin{equation}
    {\rm det}N=\prod_{i=1}^n p_i^{a_i}.
\end{equation}
Here, $p_i$ are prime numbers and $a_i$ are non-negative integers.
Then, ${\rm gcd}(n_2,m)=\prod_{i=1}^{n}p_i^{b_i}$ where $0 \leq b_i \leq a_i$. As before, $\vec{K}$s' are identical if they differ by Eq.(\ref{eq: identify_v}). Now, let us select $\alpha, \beta$ to satisfy $v_2=0$, so we have
 \begin{equation}
     - \frac{m}{{\rm gcd}(m,n_2)} \alpha = \frac{n_2}{{\rm gcd}(m,n_2)} \beta.
 \end{equation}
 General solutions are given by
 \begin{equation}
     \alpha = \frac{n_2}{{\rm gcd}(m,n_2)}l,\ 
     \beta = -
     \frac{m}{{\rm gcd}(m,n_2)} l,\quad l\in \mathbb{Z}. 
 \end{equation} 
 If we take $l=1$, we obtain
 \begin{equation}
     \vec{v} =
     \begin{pmatrix}
     \prod_{i=1}^{n} p_i^{a_i - b_i} \\
     0
     \end{pmatrix}. 
 \end{equation}
 This means following $ \prod_{i=1}^{n} p_i^{a_i - b_i}$ points are independent
 \begin{equation}
 \label{eq: J_1st}
 (0,0),\ (1,0),\ \cdots ,(\prod_{i=1}^{n} p_i^{a_i - b_i} -1,0).
 \end{equation}
We know that $v_2 = \alpha m + \beta n_2$ is always 0 modulo $\prod_{i=1}^{n}p_i^{b_i}$. Thus, following points are inequivalent with each other and not identified with any of those in Eq.(\ref{eq: J_1st}),
\begin{align}
    \begin{aligned}
(0,1),\ (1,1),\ &\cdots ,(\prod_{i=1}^{n} p_i^{a_i - b_i} -1,1), \\
(0,2),\ (1,2),\ &\cdots ,(\prod_{i=1}^{n} p_i^{a_i - b_i} -1,2),         \\
& \cdots \\
(0, \prod_{i=1}^{n}p_i^{b_i}-1), &\cdots , (\prod_{i=1}^{n} p_i^{a_i - b_i} -1,\prod_{i=1}^{n}p_i^{b_i}-1).
    \end{aligned}
\end{align}
 Altogether we classified alignment of all $\vec{K}$.
 
Consequently, ${\rm tr}\rho(S)$ can be evaluated systematically for each value of ${\rm det}N$ without individual treatment of $N$.

\subsubsection{Result}
Here, we show the results showing how the number of zero-modes is dependent on the assignments of the flux $N$. They are summarized in Tables \ref{tb: T4/Z4_1-19},\ref{tb: T4/Z4_20-23}, and \ref{tb: T4/Z4_24}. It is convenient to use the Legendre and the Jacobi symbols for the presentation of our results. The Legendre symbol is defined for an odd prime $p$ and an integer $a$ as \cite{Vinogradov:1954},
\begin{align}
    \begin{aligned}
        \left( \frac{a}{p} \right)_L :=
        \begin{cases}
            +1:\  {\rm if}\   a\ {\rm  is\  a\  quadratic\  residue\ modulo}\ p{\rm\ and}\  a\not\equiv0
            \pmod{p}, \\
            -1:\  {\rm if}\   a\ {\rm  is\ not\ a\  quadratic\  residue\ modulo\ }p, \\
            0: {\rm if}\ a \equiv 0 \pmod{p}.
        \end{cases}
    \end{aligned}
\end{align}
The Jacobi symbol is an extension of the Legendre symbol. For a positive integer $a$ and a positive odd integer $P$, the Jacobi symbol $\left( \frac{a}{P} \right)_J$ is defined in terms of the Legendre symbol,
\begin{equation}
    \left( \frac{a}{P}  \right)_J := \left( \frac{a}{p_1} \right)_L^{k_1} \left( \frac{a}{p_2} \right)_L^{k_2}\cdots \left( \frac{a}{p_r} \right)_L^{k_r},
\end{equation}
where $P$ is prime factorized as $P = p_1^{k_1}p_2^{k_2}\cdots p_r^{k_r}$.
It is worth noting that the Jacobi symbol has the following property \cite{Vinogradov:1954},
    \begin{equation}
\label{eq: Legendre_prod}
    \left(\frac{a}{P} \right)_J \left(\frac{b}{P} \right)_J= \left(\frac{ab}{P} \right)_J.
\end{equation}

\begin{table}[H]
\begin{center}
\begin{tabular}{|c|c||c|c|c|c|} \hline
${\rm det}N$ & conditions & $+1$ & $-1$ & $+i$ & $-i$      \\ \hline \hline 
1 & & 1 & 0 & 0 & 0   \\ \hline
2 & & 1 & 1 & 0 & 0  \\ \hline
3 & 
\begin{tabular}{c}
$ (\frac{n_i}{3})_L=1,$\ 
$ (i = 1\  {\rm or}\ 2)$
\end{tabular} 
    & 1 & 1 & 1 & 0   \\ \hline
3 & 
\begin{tabular}{c}
$ (\frac{n_i}{3})_L=-1,$\ 
$ (i = 1\  {\rm or}\ 2)$
\end{tabular} 
    & 1 & 1 & 0 & 1  \\\hline 
4 &
\begin{tabular}{c}
${\rm gcd}(n_i,m)=1,$\  
$ (i = 1\  {\rm or}\ 2)$
\end{tabular}
    & 2 & 1 & 1 & 0  \\ \hline
4 &
\begin{tabular}{c}
${\rm gcd}(n_i,m)=2,$\  
$ (i = 1\  {\rm and}\ 2)$
\end{tabular}
    & 2 & 2 & 0 & 0 \\ \hline
5 & 
\begin{tabular}{c}
$ (\frac{n_i}{5})_L=1,$\ 
$ (i = 1\  {\rm or}\ 2)$ 
\end{tabular}
    & 2 & 1 & 1 & 1 \\ \hline
5 & 
\begin{tabular}{c}
$ (\frac{n_i}{5})_L=-1,$\ 
$ (i = 1\  {\rm or}\ 2)$ 
\end{tabular}
    & 1 & 2 & 1 & 1 \\ \hline    
6 & & 2 & 2 & 1 & 1 \\ \hline
7 & & 2 & 2 & 2 & 1 \\ \hline
8 & 
\begin{tabular}{c}
$n_i \stackrel{\rm{mod}8}{\equiv} 1,$\ 
$ (i = 1\  {\rm or}\ 2)$
\end{tabular} 
    & 3 & 2 & 2 & 1 \\ \hline 
8 & 
\begin{tabular}{c}
$n_i \stackrel{\rm{mod}8}{\equiv} 3,$\ 
$ (i = 1\  {\rm or}\ 2)$
\end{tabular} 
    & 2 & 3 & 2 & 1 \\ \hline 
8 & 
\begin{tabular}{c}
${\rm gcd}(n_i,m)\stackrel{\rm{mod}2}{\equiv}0,$ \
$ (i = 1\ {\rm and}\ 2)$
\end{tabular} 
    & 3 & 3 & 1 & 1 \\ \hline
9 & 
 \begin{tabular}{c}
${\rm gcd}(n_i,m)=1,$ \
$ (i = 1\  {\rm or}\ 2)$
\end{tabular}
    & 3 & 2 & 2 & 2 \\ \hline
9 & 
 \begin{tabular}{c}
${\rm gcd}(n_i,m)=3,$ \
$ (i = 1\  {\rm and}\ 2)$
\end{tabular}
    & 2 & 3 & 2 & 2 \\ \hline
10 && 3 & 3 & 2 & 2 \\ \hline
11&
\begin{tabular}{c}
$ (\frac{n_i}{11})_L=1,$\
$ (i = 1\  {\rm or}\ 2)$
\end{tabular} 
    & 3 & 3 & 3 & 2 \\ \hline
11&
\begin{tabular}{c}
$ (\frac{n_i}{11})_L=-1,$\
$ (i = 1\  {\rm or}\ 2)$
\end{tabular} 
    & 3 & 3 & 2 & 3 \\ \hline
12& 
\begin{tabular}{c}
$n_i \stackrel{\rm{mod}4}{\equiv} 1,$\
$ (i = 1\  {\rm or}\ 2)$
\end{tabular} 
    & 4 & 3 & 3 & 2 \\ \hline
12& 
\begin{tabular}{c}
$n_i \stackrel{\rm{mod}4}{\equiv} -1,$\
$ (i = 1\  {\rm or}\ 2)$
\end{tabular} 
    & 3 & 4 & 3 & 2 \\ \hline
12& 
\begin{tabular}{c}
$n_i \stackrel{\rm{mod}4}{\equiv} 2,$\
$ (i = 1\  {\rm or}\ 2)$
\end{tabular} 
    & 4 & 4 & 2 & 2 \\ \hline
12& 
\begin{tabular}{c}
$n_i \stackrel{\rm{mod}4}{\equiv} 0,$\
$ (i = 1\  {\rm and}\ 2)$
\end{tabular} 
    & 4 & 4 & 3 & 1 \\ \hline
13& 
\begin{tabular}{c}
$ (\frac{n_i}{13})_L=1,$\
$ (i = 1\  {\rm or}\ 2)$
\end{tabular}
    & 4 & 3 & 3 & 3 \\ \hline
13& 
\begin{tabular}{c}
$ (\frac{n_i}{13})_L=-1,$\
$ (i = 1\  {\rm or}\ 2)$
\end{tabular}
    & 3 & 4 & 3 & 3 \\ \hline  
14& & 4 & 4 & 3 & 3 \\ \hline 
15& & 4 & 4 & 4 & 3 \\ \hline
16& 
\begin{tabular}{c}
${\rm gcd}(n_i,m)=1,$ \
$ (i = 1\ {\rm or}\ 2)$
\end{tabular} 
    & 5 & 4 & 4 & 3 \\ \hline
16& 
\begin{tabular}{c}
${\rm gcd}(n_i,m)=2,$ \
$ (i = 1\ {\rm or}\ 2)$
\end{tabular} 
    & 5 & 5 & 3 & 3 \\ \hline
16& 
\begin{tabular}{c}
${\rm gcd}(n_i,m)=4,$ \
$ (i = 1\ {\rm and}\ 2)$
\end{tabular} 
    & 5 & 5 & 4 & 2 \\ \hline    
17&
\begin{tabular}{c}
$ (\frac{n_i}{17})_L=1,$\
$ (i = 1\  {\rm or}\ 2)$
\end{tabular}
    & 5 & 4 & 4 & 4 \\ \hline 
17&
\begin{tabular}{c}
$ (\frac{n_i}{17})_L=-1,$\
$ (i = 1\  {\rm or}\ 2)$
\end{tabular}
    & 4 & 5 & 4 & 4 \\ \hline 
18& & 5 & 5 & 4 & 4 \\ \hline
19& 
\begin{tabular}{c}
$ (\frac{n_i}{19})_L=1,$\
$ (i = 1\  {\rm or}\ 2)$
\end{tabular}
    & 5 & 5 & 5 & 4 \\ \hline
19 & 
\begin{tabular}{c}
$ (\frac{n_i}{19})_L=-1,$\
$ (i = 1\  {\rm or}\ 2)$
\end{tabular}
    & 5 & 5 & 4 & 5 \\ \hline
\end{tabular}  
\end{center}
\caption{$T^4/Z_4, (1 \leq  {\rm det }N \leq 19$)}
\label{tb: T4/Z4_1-19}
\end{table}

\begin{table}[H]
\begin{center}
\begin{tabular}{|c|c||c|c|c|c|} \hline
${\rm det}N$ & conditions & $+1$ & $-1$ & $+i$ & $-i$      \\ \hline \hline 
20 & 
\begin{tabular}{c}
$n_i \stackrel{\rm{mod}4}{\equiv} 1,\ (\frac{n_i}{5})_L=1 ,(i = 1\  {\rm or}\ 2)$\\
 {\rm or} \\
$n_1 \stackrel{\rm{mod}4}{\equiv} 1,\ (\frac{n_2}{5})_L=1$ \\
 {\rm or} \\
$n_2 \stackrel{\rm{mod}4}{\equiv} 1,\ (\frac{n_1}{5})_L=1$
\end{tabular} 
    & 6 & 5 & 5 & 4 \\ \hline 
20 & 
\begin{tabular}{c}
$n_i \stackrel{\rm{mod}4}{\equiv} -1,\ (\frac{n_i}{5})_L=-1 ,(i = 1\  {\rm or}\ 2)$\\
 {\rm or} \\
$n_1 \stackrel{\rm{mod}4}{\equiv} -1,\ (\frac{n_2}{5})_L=-1$ \\
 {\rm or} \\
$n_2 \stackrel{\rm{mod}4}{\equiv} -1,\ (\frac{n_1}{5})_L=-1$
\end{tabular} 
    & 5 & 6 & 5 & 4    \\ \hline
20 & 
\begin{tabular}{c}
${\rm gcd}(n_i,m)\stackrel{\rm{mod}2}{\equiv}0,$ \
$ (i = 1\ {\rm and}\ 2)$
\end{tabular} 
    & 6 & 6 & 4 & 4 \\ \hline
21 & 
\begin{tabular}{c}
$(\frac{n_i}{21})_J=1 ,(i = 1\  {\rm or}\ 2)$\\
 {\rm or} \\
$(\frac{n_1}{7})_L=\pm1,\ (\frac{n_2}{3})_L=\pm1$ (signs same order)\\
 {\rm or} \\
$(\frac{n_1}{3})_L=\pm1,\ (\frac{n_2}{7})_L=\pm1$
(signs same order)
\end{tabular}  
    & 6 & 5 & 5 & 5  \\ \hline
21 & 
\begin{tabular}{c}
$(\frac{n_i}{21})_J=-1 ,(i = 1\  {\rm or}\ 2)$\\
 {\rm or} \\
$(\frac{n_1}{7})_L=\pm1,\ (\frac{n_2}{3})_L=\mp1$ (signs same order)\\
 {\rm or} \\
$(\frac{n_1}{3})_L=\pm1,\ (\frac{n_2}{7})_L=\mp1$
    (signs same order)
\end{tabular}
    & 5 & 6 & 5 & 5 \\ \hline
22& & 6 & 6 & 5 & 5  \\ \hline
23& & 6 & 6 & 6 & 5 \\ \hline
\end{tabular} 
\end{center}
\caption{$T^4/Z_4, (20 \leq  {\rm det }N \leq 23$)}
\label{tb: T4/Z4_20-23}
\end{table}

\begin{table}[H]
\begin{center}
\begin{tabular}{|c|c||c|c|c|c|} \hline
${\rm det}N$ & conditions & $+1$ & $-1$ & $+i$ & $-i$      \\ \hline \hline 
24 & 
\begin{tabular}{c}
$ n_i \stackrel{\rm{mod}8}{\equiv} 1,\ (\frac{n_i}{3})_L=1\  (i = 1\  {\rm or}\ 2)$\\
 {\rm or} \\
$ n_1 \stackrel{\rm{mod}8}{\equiv} 1,\ (\frac{n_2}{3})_L=1$ \\
 {\rm or} \\
 $n_2 \stackrel{\rm{mod}8}{\equiv} 1,\ (\frac{n_1}{3})_L=1$ \\
 {\rm or} \\
$ n_i \stackrel{\rm{mod}8}{\equiv} -3,\ (\frac{n_i}{3})_L=-1\  (i = 1\  {\rm or}\ 2)$\\
 {\rm or} \\
$ n_1 \stackrel{\rm{mod}8}{\equiv} -3,\ (\frac{n_2}{3})_L=-1$ \\
 {\rm or} \\
 $n_2 \stackrel{\rm{mod}8}{\equiv} -3,\ (\frac{n_1}{3})_L=-1$
\end{tabular}
    & 7 & 6 & 6 & 5 \\ \hline 
24 & 
\begin{tabular}{c}
$ n_i \stackrel{\rm{mod}8}{\equiv} -1,\ (\frac{n_i}{3})_L=1\  (i = 1\  {\rm or}\ 2)$\\
 {\rm or} \\
$ n_1 \stackrel{\rm{mod}8}{\equiv} -1,\ (\frac{n_2}{3})_L=1$ \\
 {\rm or} \\
 $n_2 \stackrel{\rm{mod}8}{\equiv} -1,\ (\frac{n_1}{3})_L=1$ \\
 {\rm or} \\
$ n_i \stackrel{\rm{mod}8}{\equiv} 3,\ (\frac{n_i}{3})_L=-1\  (i = 1\  {\rm or}\ 2)$\\
 {\rm or} \\
$ n_1 \stackrel{\rm{mod}8}{\equiv} 3,\ (\frac{n_2}{3})_L=-1$ \\
 {\rm or} \\
 $n_2 \stackrel{\rm{mod}8}{\equiv} 3,\ (\frac{n_1}{3})_L=-1$
\end{tabular} 
    & 6 & 7 & 6 & 5  \\ \hline
24 & 
\begin{tabular}{c}
${\rm gcd}(n_i,m)\stackrel{\rm{mod}2}{\equiv}0,$ \
$ (i = 1\ {\rm and}\ 2)$ 
\end{tabular}
    & 7 & 7 & 5 & 5 \\ \hline
\end{tabular}  
\end{center}
\caption{$T^4/Z_4, ({\rm det}N=24)$}
\label{tb: T4/Z4_24}
\end{table}

\subsubsection{Derivation}
Here, we show the derivation of some of our results. The reason of the appearance of Legendre and Jacobi symbols will become clear.

\paragraph{When ${\rm det}N = p$ ($p$ in an odd prime):}\ \\
We have ${\rm gcd}(n_i,m)=1,\ (i=1{\ \rm or\ }2)$. Thus, the trace is given by
\begin{equation}
\label{eq:trace_Z4}
      {\rm tr}\rho(S)= 
         \frac{1}{\sqrt{p}} \sum_{K = 0}^{p-1} e^{\frac{2\pi i}{p} n_i K^2}.
\end{equation}
This can be evaluated by use of results in number theory \cite{Vinogradov:1954, Moore: 2022} as,
\begin{align}
\begin{aligned}
    {\rm tr}\rho(S)&= 
    \begin{cases}
    \left( \frac{n_i}{p}
    \right)_L&: p \equiv 1 \pmod{4},\\
    i \left( \frac{n_i}{p}
    \right)_L&: p \equiv 3 \pmod{4}.
    \end{cases}
\end{aligned}
\end{align}
Then, one might wonder why we have only ${\rm tr}\rho(S)= i$
when ${\rm det}N=7$ in Table\ref{tb: T4/Z4_1-19}.
It is simply explained because there is no positive definite $2 \times 2$ symmetric integer matrix $N$ such that $n_i$ is non-quadratic residue modulo $7$. 
We prove this fact in Appendix \ref{appendix: existence_of_N}. 
Due to the same reason, we do not have ${\rm tr}\rho(S)=-i$ when ${\rm det}N = 23$ in Table \ref{tb: T4/Z4_20-23}.

\paragraph{When ${\rm det}N = pq$ ($p, q$ are odd primes where $p \neq q$):}\ \\
If ${\rm gcd}(n_i,m)=1$, we obtain 
\begin{align}
\begin{aligned}
{\rm tr}\rho(S)
&=
\frac{1}{\sqrt{pq}} \sum_{K = 0}^{pq-1} e^{\frac{2\pi i}{pq} n_i K^2} \\
&=
\frac{1}{\sqrt{pq}}\sum_{L=0}^{q-1}
e^{\frac{2 \pi i}{q}pn_iL^2} \sum_{M=0}^{p-1} e^{\frac{2 \pi i}{p} qn_i M^2} \\
&=
\begin{cases}
\left(\frac{n_i}{pq} \right)_J &:
{\rm if}\ p \equiv q \equiv 1\ {\rm or}\ p \equiv q \equiv 3 \pmod4,\\
i \left(\frac{n_i}{pq} \right)_J&:
{\rm if}\ p \equiv 1, q \equiv 3 \pmod{4}\ {\rm or\ vice\ versa},
\end{cases}
\end{aligned}
\end{align}
where we transformed the summation variable as $K = pL + q M,\ (L\in \mathbb{Z}/q), (M \in \mathbb{Z}/p)$ in the second equality. We have used the law of quadratic reciprocity,
\begin{align}
    \begin{aligned}
    \left(\frac{p}{q} \right)_L \left(\frac{q}{p} \right)_L   
    = 
    (-1)^{\frac{p-1}{2} \frac{q-1}{2}},
    \end{aligned}
\end{align}
in the third equality.
If ${\rm gcd}(n_1,m)=p,\ {\rm gcd}(n_2,m)=q$, we obtain
\begin{align}
\begin{aligned}
{\rm tr}\rho(S)
&=
\frac{1}{\sqrt{p}} \sum_{K_1 = 0}^{p-1} e^{\frac{2\pi i}{p}\cdot \frac{n_2}{q}K_1^2}
\frac{1}{\sqrt{q}} \sum_{K_2=0}^{q-1} e^{\frac{2\pi i}{q}\cdot \frac{n_1}{p}K_2^2}
\\
    &=
\begin{cases}
\left(\frac{n_2}{p} \right)_L \left(\frac{n_1}{q} \right)_L &:
{\rm if}\ p \equiv q \equiv 1\ {\rm or}\ p \equiv q \equiv 3 \pmod4,\\
i \left(\frac{n_2}{p} \right)_L
\left(\frac{n_1}{q} \right)_L&:
{\rm if}\ p \equiv 1, q \equiv 3 \pmod{4}\ {\rm or\ vice\ versa}.
\end{cases}
\end{aligned}
\end{align}
We have only ${\rm tr}\rho(S)= i$
when ${\rm det}N=15$ in Table\ref{tb: T4/Z4_1-19}. This is because there is no positive definite $2 \times 2$ symmetric integer matrix $N$ which would correspond to ${\rm tr}\rho(S)= -i$.

\paragraph{When ${\rm det}N = 2p$ ($p$ is an odd prime):}\ \\
It is straightforward to verify
\begin{equation}
        {\rm tr}\rho(S) = 0.
\end{equation}

\paragraph{When ${\rm det}N \equiv 0 \pmod{4}$:}\ \\
We do not obtain a general formula. However, the trace can be evaluated analytically if desired. As an example, we show our derivation when ${\rm det}N=12$ in Appendix \ref{appendix: 12}.

\paragraph{When ${\rm det}N = p^2$ ($p$ is an odd prime):}\ \\
If ${\rm gcd}(n_i,m)=1$, we obtain
\begin{align}
\begin{aligned}
\label{eq: p^2_1}
{\rm tr}\rho(S)
&=
\frac{1}{p} \sum_{K = 0}^{p^2-1} e^{\frac{2\pi i}{p^2} n_i K^2} 
  &=1,
\end{aligned}
\end{align}
as in ref.\cite{Moore: 2022}.

If ${\rm gcd}(n_1,m)={\rm gcd}(n_2,m)=p$, we can show
\begin{align}
\label{eq: p^2}
    {\rm tr}\rho(S) = \left(\frac{-1}{p} \right)_L,
\end{align}
as in Appendix \ref{appendix: p^2}.
 In fact, 
\begin{equation}
\label{eq: (-1/p)}
    \left(\frac{-1}{p} \right)_L = (-1)^{\frac{p-1}{2}},
\end{equation}
is a well-known result\cite{Vinogradov:1954}. 
It may be interesting to construct a trace evaluating formula which is applicable when ${\rm det}N$ is an arbitrary integer.  

\subsubsection{Observation and discussion}
We find that three generation models appear when the size of the flux is $8 \leq {\rm det}N \leq 16$. It may be questioned the possibility of three degeneracy in larger ${\rm det}N$, so that we have not captured all three generation models within Tables \ref{tb: T4/Z4_1-19}, \ref{tb: T4/Z4_20-23}, and {\ref{tb: T4/Z4_24}}. Following analysis will clarify that our results are in fact enough.
It is obvious from eq.(\ref{eq:trace_Z4}) that \begin{equation}
\label{eq: trS<sqrtN}
    \left|{\rm tr}\rho(S) \right| \leq \sqrt{{\rm det}N},
\end{equation}
holds. 
Firstly, let us assume that $D_{(+i)}$ is the smallest.
From eqs.({\ref{eq: total_Z4}}) and (\ref{eq: Z2even}), 
\begin{equation}
    D_{(+i)} + D_{(-i)} = \frac{{\rm det}N - (N_{Z_2+}-N_{Z_2-})}{2}.
\end{equation}
From eqs.(\ref{eq: trS_Z4}) and (\ref{eq: trS<sqrtN}), one finds
\begin{align}
\begin{aligned}
     D_{(-i)} - D_{(+i)} \leq \sqrt{{\rm det}N}.
\end{aligned}
\end{align}
Then one obtains
\begin{equation}
\label{eq: necessary_Z4}
    2 D_{(+i)} \geq \frac{{\rm det}N - (N_{Z_2+}-N_{Z_2-})}{2} - \sqrt{{\rm det}N}.
\end{equation}
We find that the necessary condition to obtain $D_{(+i)}=3$ is ${\rm det}N \leq 24$. Similarly, we consider the cases when $D_{(\pm1)}$ or $ D_{(-i)}$ is the smallest. It can be checked that ${\rm det}N \leq 24$ is still necessary to generate $3$ degeneracy of zero-modes. 

\subsection{$T^4/Z_3^{(a)}$}
\label{subsec: ST1T2}
We focus on the following algebraic relation,
\begin{equation}
    (ST_1T_2)^3 = I_4.
\end{equation}
This allows us to relate $ST_1T_2$ to a $Z_3$-twist which then is used to construct a $T^4/Z_3$ orbifold.
The complex structure moduli behave under the $ST_1T_2$ as,
\begin{equation}
    \Omega \xrightarrow{T_2}
    \Omega + B_2
    \xrightarrow{T_1}
    \Omega + B_1 + B_2
    \xrightarrow{S}
    -(\Omega + I_2)^{-1}.
\end{equation}
$ST_1T_2$ invariant $\Omega \in \mathcal{H}_2$ is uniquely determined,
\begin{equation}
\label{eq: ST1T2_inv_omega}
    \Omega_{(ST_1T_2)} =
    \begin{pmatrix}
    \omega & 0 \\ 0 & \omega
    \end{pmatrix},\quad \omega = e^{\frac{2\pi i}{3}}.
\end{equation}
Then the complex coordinate is transformed as,
\begin{equation}
\label{eq: z3-twist}
    \vec{z} \xrightarrow{ST_1T_2} \omega \vec{z}.
\end{equation}
This is nothing but a $Z_3$-twist.
We define $T^4/Z_3^{(a)}$ orbifold by imposing the following $Z_3$ identification
\begin{equation}
    \vec{z} \sim \omega \vec{z},
\end{equation}
on the complex coordinate of $T^4$. This means that the moduli of $T^4$ must be $ST_1T_2$ invariant leading to the fixing of moduli as shown in eq.(\ref{eq: ST1T2_inv_omega}). 

\subsubsection{Lattice vectors}
Here, we explicitly show the lattice vectors defining the $T^4/Z_3^{(a)}$ orbifold,
\begin{equation}
\label{eq: latticeT4/Z3a}
    e_1 = 2 \pi R \begin{pmatrix}
   1  \\ 0 
    \end{pmatrix},\ 
    e_2 = 2 \pi R \begin{pmatrix}
    \omega \\ 0
    \end{pmatrix},\ 
    e_3 =  2 \pi R\begin{pmatrix}
    0 \\ 1
    \end{pmatrix},\ 
    e_4=  2 \pi R \begin{pmatrix}
    0 \\ \omega
    \end{pmatrix}.
\end{equation}
Notice that this is identical to the root lattice of $SU(3) \times SU(3)$.
Under the $Z_3$-twist realized by the $ST_1T_2$ transformation, they behave as
\begin{align}
    \begin{pmatrix}
    e_2^{(ST_1T_2)}\\
    e_4^{(ST_1T_2)}\\
    e_1^{(ST_1T_2)}\\
    e_3^{(ST_1T_2)}
    \end{pmatrix}
    =
    \begin{pmatrix}
    0 & 0 & 1 & 0 \\
    0 & 0 & 0 & 1 \\
    -1 & 0 & -1 & 0 \\
    0 & -1 & 0 & -1 
    \end{pmatrix}
        \begin{pmatrix}
    e_2 \\
    e_4 \\
    e_1 \\
    e_3
    \end{pmatrix}.
\end{align}
This $Z_3$-twist is the Coxeter element of $SU(3)$ \cite{Markushevich:1986za, Katsuki:1989bf}.

\subsubsection{Flux}
    When the complex structure moduli are $\Omega_{(ST_1T_2)}$, the F-flat condition shown in eq.(\ref{eq: SUSY_condition}) restricts $N$ to be a symmetric matrix. 
    Then we notice that $T_1T_2$ transformation is consistent with the the condition $p_{xx}=p_{yy}=0$ according to eq.(\ref{eq: consistent_T}).
    Moreover, in order to write down the $T_1T_2$ transformation of the wavefunctions as in eq.(\ref{eq: T_zero-mode}), diagonal elements of $N$ must be even. Thus,
\begin{equation}
\label{eq: Z3a_flux}
    N = 
    \begin{pmatrix}
    n_1 & m \\
    m & n_2
    \end{pmatrix}
    =
    \begin{pmatrix}
    2n_{1}' & m \\
    m & 2n_{2}'
    \end{pmatrix},\quad n'_{1},  n'_{2}, m \in \mathbb{Z} .
\end{equation}
Note that $N$ in eq.(\ref{eq: Z3a_flux}) is $ST_1T_2$ invariant.
We will consider the case when the positive definite condition eq.(\ref{eq: positive_definite_conditon}) is satisfied, so that $N$ is positive definite.

\subsubsection{Zero-mode counting method}
In order to analyze the number of zero-modes, we evaluate the trace of the transformation matrix,  $\rho(ST_1T_2)$,
\begin{align}
\begin{aligned}
    {\rm tr}\rho(ST_1T_2)&= \frac{e^{-\frac{\pi i}{6}}}{\sqrt{{\rm det}N}} \sum_{\vec{K} \in \Lambda_N} e^{3 \pi i \vec{K}^{\rm T} . N^{-1}. \vec{K}}.
\end{aligned}
\end{align}
Let us denote the number of degeneracy in zero-modes by $D_{(\omega^n)},\ (n=0,1,2)$. Then we have
\begin{align}
\label{eq: total_Z3}
    D_{(+1)} + D_{(\omega)} + D_{(\omega^2)} &= {\rm det}N, \\
\label{eq: trS_Z3}
    D_{(+1)} + \omega D_{(\omega)} + \omega^2 D_{(\omega^2)} &= {\rm tr}\rho(ST_1T_2).
\end{align}
Eqs.(\ref{eq: total_Z3}) 
and (\ref{eq: trS_Z3}) constrain three real degrees of freedom. Thus, we have sufficient information to determine $D_{(+1)}, D_{(\omega)}$, and $D_{(\omega^2)}$.
Results are shown in Tables \ref{tb: T4/Z3_1} and \ref{tb: T4/Z3_48}.

\begin{table}[H]
\begin{center}
\begin{tabular}{|c|c||c|c|c|} \hline
${\rm det}N$ & conditions & $+1$ & $\omega$ & $\omega^2$      \\ \hline \hline 
3 & - & 2 & 0 & 1   \\ \hline
4 & - & 1 & 1 & 2   \\ \hline
7 & - & 2 & 2 & 3   \\ \hline
8 & - & 3 & 3 & 2   \\ \hline
11& - & 4 & 4 & 3   \\ \hline
12 &
\begin{tabular}{c}
$n_i' \stackrel{\rm{mod}2}{\equiv} 1,$\
$ (i = 1\  {\rm or}\ 2)$
\end{tabular} 
& 5 & 3 & 4   \\ \hline
12 &
\begin{tabular}{c}
$n_i' \stackrel{\rm{mod}2}{\equiv} 0,$\
$ (i = 1\  {\rm and}\ 2)$
\end{tabular} 
& 3 & 5 & 4   \\ \hline
15 & 
\begin{tabular}{c}
$ (\frac{n_i'}{5})_L=1,$\
$ (i = 1\  {\rm or}\ 2)$
\end{tabular} 
    & 6 & 4 & 5 \\ \hline
15 & 
\begin{tabular}{c}
$ (\frac{n_i'}{5})_L=-1,$\
$ (i = 1\  {\rm or}\ 2)$
\end{tabular} 
    & 4 & 6 & 5 \\ \hline  
16& - & 5 & 5 & 6 \\ \hline
19& - & 6 & 6 & 7 \\ \hline
20& - & 7 & 7 & 6 \\ \hline
23& - & 8 & 8 & 7 \\ \hline
24& 
\begin{tabular}{c}
$n_i' \stackrel{\rm{mod}8}{\equiv} \pm 1$, \
($i=1\ {\rm or}\ 2$)
\end{tabular}
    & 9 & 7 & 8 \\ \hline
24& 
\begin{tabular}{c}
$n_i' \stackrel{\rm{mod}8}{\equiv} \pm 3$, \
($i=1\ {\rm or}\ 2$)
\end{tabular}
    & 7 & 9 & 8 \\ \hline   
27& 
\begin{tabular}{c}
${\rm gcd}(n_i',m) \stackrel{\rm{mod}3}{\equiv} 1,$\
($i=1\ {\rm or}\ 2$)
\end{tabular} 
    & 10 & 8 & 9 \\ \hline
27& 
    \begin{tabular}{c}
${\rm gcd}(n_i',m) \stackrel{\rm{mod}3}{\equiv} 0,$\
($i=1\ {\rm and}\ 2$)
\end{tabular} 
    & 10 & 10 & 7 \\ \hline
28 & - & 9 & 9 & 10 \\ \hline
31 & - & 10 & 10 & 11 \\ \hline
32 & - & 11 & 11 & 10 \\ \hline
35 & - & 12 & 12 & 11 \\ \hline
\end{tabular}
\end{center}
\caption{$T^4/Z_3^{(a)}, (3 \leq {\rm det}N \leq 35)$}
\label{tb: T4/Z3_1}
\end{table}

\begin{table}[H]
\begin{center}
\begin{tabular}{|c|c||c|c|c|} \hline
${\rm det}N$ & conditions & $+1$ & $\omega$ & $\omega^2$      \\ \hline \hline 
36 & 
\begin{tabular}{c}
$n_i'\stackrel{\rm{mod}12}{\equiv} \pm 1,$\
($i=1\ {\rm or}\ 2$) \\
{\rm or} \\
$n_1'\stackrel{\rm{mod}3}{\equiv} 0$,\ $n_2' \stackrel{\rm{mod}4}{\equiv}-2$ \\
{\rm or} \\
$n_2'\stackrel{\rm{mod}3}{\equiv} 0$,\ $n_1' \stackrel{\rm{mod}4}{\equiv}-2$ 
\end{tabular} 
   & 13 & 11 & 12 \\ \hline
36 & 
\begin{tabular}{c}
$n_i'\stackrel{\rm{mod}12}{\equiv} \pm 5,$\
($i=1\ {\rm or}\ 2$) \\
{\rm or} \\
$n_1'\stackrel{\rm{mod}3}{\equiv} 0$,\ $n_2' \stackrel{\rm{mod}4}{\equiv}2$ \\
{\rm or} \\
$n_2'\stackrel{\rm{mod}3}{\equiv} 0$,\ $n_1' \stackrel{\rm{mod}4}{\equiv}2$ 
\end{tabular} 
    & 11 & 13 & 12 \\ \hline
36 & 
\begin{tabular}{c}
${\rm gcd}(n_i',m')=3,$ \
$ (i = 1\  {\rm and}\ 2)$
\end{tabular}
    & 13 & 13 & 10 \\ \hline
39 & 
\begin{tabular}{c}
$ (\frac{n_i'}{13})_L=1,$\
$ (i = 1\  {\rm or}\ 2)$
\end{tabular}
    & 14 & 12 & 13 \\ \hline
39 & 
\begin{tabular}{c}
$ (\frac{n_i'}{13})_L=-1,$\
$ (i = 1\  {\rm or}\ 2)$
\end{tabular}
    & 12 & 14 & 13 \\ \hline
40 & - & 13 & 13 & 14 \\ \hline
43 & - & 14 & 14 & 15 \\ \hline
44 & - & 15 & 15 & 14 \\ \hline
47 & - & 16 & 16 & 15 \\ \hline 
48 & 
\begin{tabular}{c}
${\rm gcd}(n_i',m')
\stackrel{\rm{mod}2}{\equiv}1
,$ \
$ (i = 1\ {\rm or}\ 2)$
\end{tabular}
    & 17 & 15 & 16  \\ \hline
48 & 
\begin{tabular}{c}
$n_i' \stackrel{\rm{mod}4}{\equiv} 2,$\
$ (i = 1\  {\rm or}\ 2)$
\end{tabular} 
    & 15 & 17 & 16 \\ \hline
48 & 
\begin{tabular}{c}
$n_i' \stackrel{\rm{mod}4}{\equiv} 0,$\
$ (i = 1\  {\rm and}\ 2)$
\end{tabular}
    & 17 & 15 & 16  \\ \hline
\end{tabular}
\end{center}
\caption{$T^4/Z_3^{(a)}, (36 \leq {\rm det}N \leq 48)$}
\label{tb: T4/Z3_48}
\end{table}

\subsubsection{Observation and discussion}
We find that three generation models appear when the size of the flux is ${\rm det}N = 7, 11, 12$. It can be shown that three degeneracy of zero-modes is not produced for other values of ${\rm det}N$. 
It is obvious that 
\begin{equation}
\label{eq: trST1T2<sqrtN}
    \left|{\rm tr}\rho(ST_1T_2) \right| \leq \sqrt{{\rm det}N},
\end{equation}
holds. 
Firstly, let us assume $D_{(+1)} \geq D_{(\omega)} \geq D_{(\omega^2)}$ and carry out our analysis.
From eqs.(\ref{eq: trS_Z3}) and (\ref{eq: trST1T2<sqrtN}), we get 
\begin{equation}
    {(D_{(+1)}-D_{(\omega^2)})^2 - (D_{(+1)}-D_{(\omega^2)}) (D_{(\omega)}-D_{(\omega^2)}) + 
(D_{(\omega)} -  D_{(\omega^2)})^2} \leq {{\rm det}N}.
\end{equation}
We substitute $D_{(\omega^2)}=3$, and define $x, y$ as $x := D_{(+1)}-3,\ y:= D_{(\omega)}-3,\ (x \geq y \geq 0)$. Then, $f(x,y)=x^2 - xy + y^2$ satisfies 
\begin{equation}
    f(x,y) \leq {\rm det}N,\quad (x\geq y \geq 0).
\end{equation}
From eq.(\ref{eq: total_Z3}), the variables $x$ and $y$ are related by $x + y = {\rm det}N -9$. Thus, we obtain
\begin{align}
\begin{aligned}
f(x(y),y) = 3 \left( y-\frac{{\rm det}N - 9}{2} \right)^2 + \frac{1}{4}({\rm det}N - 9)^2 \leq {\rm det}N, \quad (y \geq 0).
\end{aligned}
\end{align}
If ${\rm det}N \geq 9$, the minimum of $f(x(y),y),\ (y\geq0)$ is $\frac{1}{4}({\rm det}N - 9)^2$. This means 
\begin{equation}
\label{eq: inequality}
    \frac{1}{4}({\rm det}N - 9)^2 \leq {\rm det}N.
\end{equation}
needs to be satisfied as a necessary condition for the realization of three generation models. Eq.(\ref{eq: inequality}) tells us that ${\rm det}N \leq 17$ is necessary. Similarly, we can consider cases when $D_{(+1)} \geq D_{(\omega)} \geq D_{(\omega^2)}$ is not satisfied. In any cases, eq.(\ref{eq: inequality}) is obtained. Consequently, we do not need to search for three  generation models in the region, ${\rm det}N > 17$.

\subsection{$T^4/Z_3^{(b)}$}
\label{subsec: (ST3)^2}
We focus on the following algebraic relation,
\begin{equation}
    (ST_3)^6 = I_4.
\end{equation}
This allows us to relate $(ST_3)^2$ to a $Z_3$-twist which then is used to construct a $T^4/Z_3$ orbifold.
The complex structure moduli behave under $(ST_3)^2$ as.
\begin{equation}
    \Omega \xrightarrow{ST_3} - (\Omega + B_3)^{-1} \xrightarrow{ST_3} -(-(\Omega + B_3)^{-1} + B_3)^{-1}.
\end{equation}
To find a $(ST_3)^2$ invariant $\Omega \in \mathcal{H}_2$, we search $ST_3$ invariant moduli, $\Omega_{(ST_3)}$. It is shown in eq.(\ref{eq: ST3_moduli}). Then, the complex coordinate is transformed as,
\begin{align}
\label{eq: z3-twist-b}
\begin{pmatrix}
    z_1 \\ z_2 
\end{pmatrix}
\xrightarrow{(ST_3)^2}
\begin{pmatrix}
    -\frac{1}{2} & - i \frac{\sqrt{3}}{2} \\
    - i \frac{\sqrt{3}}{2} & -\frac{1}{2}
\end{pmatrix}
\begin{pmatrix}
    z_1 \\ z_2 
\end{pmatrix}.
\end{align}
We define $T^4/Z_3^{(b)}$ orbifold by identifying the $Z_3$-twist in eq.($\ref{eq: z3-twist-b}$).

The lattice vectors of $T^4/Z_3^{(b)}$ correspond to the root lattice of $SU(3) \times SU(3)$. 
The flux can be introduced if it is of the form as shown in eq.(\ref{eq: N_ST3}). For details, see subsection \ref{eq: Z6_ST3}. 
\subsubsection{Zero-mode counting method}
To determine the number of zero-modes, we  evaluate the transformation matrix $\rho((ST_3)^2)$ as,
\begin{align}
    {\rm tr}\rho((ST_3)^2) = \frac{1}{{\rm det}N}
    \sum_{\vec{K}\in \Lambda_N}  \sum_{\vec{L}\in \Lambda_N} e^{\pi i \vec{K}^{\rm T} (N^{-1}B_3) \vec{K}} e^{\pi i \vec{L}^{\rm T} (N^{-1}B_3) \vec{L}} e^{4 \pi i \vec{K}^{\rm T} N^{-1} \vec{L}}.
\end{align}
Then we solve equations of the form eqs.(\ref{eq: total_Z3}) and (\ref{eq: trS_Z3}).
Results are shown in Table \ref{tb: T4/Z3_2}.

\begin{table}[H]
\begin{center}
\begin{tabular}{|c|c||c|c|c|} \hline
${\rm det}N$ & $(n,m)$ & $+1$ & $\omega$ & $\omega^2$   \\ \hline \hline 
1 & (1,0) & 1 & 0 & 0    \\ \hline
4 & (2,0) & 2 & 1 & 1    \\ \hline
5 & (3,$\pm$2) & 1 & 2 & 2   \\ \hline
9& (5,4) & 3 & 2 & 4  \\ \hline
9 & (5,$-4$) & 3 & 4 & 2   \\ \hline
9 & (3,0) & 5 & 2 & 2    \\ \hline
12 & (4,2) & 4 & 5 & 3 \\ \hline
12 & (4,$-2$) & 4 & 3 & 5  \\ \hline 
13& (7,$\pm 6$) & 5 & 4 & 4  \\ \hline
16& (4,0) & 6 & 5 & 5  \\ \hline
17& (9,$\pm8$) & 5 & 6 & 6  \\ \hline
20& (6,$\pm4$) & 6 & 7 & 7 \\ \hline 
21& (11,10) or (5,$-2$) & 7 & 6 & 8  \\ \hline
21 & (11,$-10$) or (5,2) & 7 & 8 & 6  \\ \hline
25 & (13,$\pm$12) or (5,0) & 9 & 8 & 8 \\ \hline
28 & (8,$\pm6$) & 10 & 9 & 9  \\ \hline
29 & (15,$\pm14$) & 9 & 10 & 10  \\ \hline
32 & (6,$\pm 2$) & 10 & 11 & 11 \\ \hline
33 & (17,16) or (7,4) & 11 & 10 & 12 \\  \hline
33 & (17,$-16$) or (7,$-4$) & 11 & 12 & 10 \\  \hline
36 & (10,$8$) & 12 & 13 & 11 \\ \hline
36 & (10,$-8$) & 12 & 11 & 13 \\ \hline
36 & (6,0) & 14 & 11 & 11 \\ \hline
37 & (19, $\pm 18$) & 13 & 12 & 12 \\ \hline
41 & (21, $\pm 20$) & 13 & 14 & 14 \\ \hline
44 & (12, $\pm 10$) & 14 & 15 & 15 \\ \hline
45 & (23,22) or (7,$-2$) & 15 & 14 & 16 \\  \hline
45 & (23,$-22$) or (7,2) & 15 & 16 & 14 \\  \hline
45 & (9,$\pm 6$) & 17 & 14 & 14 \\  \hline
48 & (8,4) & 16 & 15 & 17 \\  \hline
48 & (8,$-4$) & 16 & 17 & 15 \\  \hline
48 & (25,$\pm 24$) or (7,0) & 17 & 16 & 16 \\  \hline
\end{tabular}
\end{center}
\caption{$T^4/Z_3^{(b)}, (1 \leq {\rm det}N \leq 48)$}
\label{tb: T4/Z3_2}
\end{table}

We find that three generation models appear when ${\rm det}N$ is $9$ or $12$.
\subsection{$T^4/Z_5$}
We focus on the following algebraic relation,
\begin{equation}
\label{eq: Z5_algebra}
    (ST_1T_3)^5 = I_4.
\end{equation}
This allows us to relate $ST_1T_3$ to a $Z_5$-twist which then is used to construct a $T^4/Z_5$ orbifold.
The complex structure moduli behave under the $ST_1T_3$ as,
\begin{equation}
    \Omega \xrightarrow{T_3}
    \Omega + B_3
    \xrightarrow{T_1}
    \Omega + B_1 + B_3
    \xrightarrow{S}
    -(\Omega + B_1 + B_3)^{-1}.
\end{equation}
$ST_1T_3$ invariant $\Omega \in \mathcal{H}_2$ is determined as,
\begin{equation}
\label{eq: ST1T3_inv_omega}
    \Omega_{(ST_1T_3)} =
    \begin{pmatrix}
     -\frac{1}{2} + \frac{i}{2} \sqrt{\frac{5+2\sqrt{5}}{5}} & 
     - \frac{1}{2} + \frac{i}{2}\sqrt{\frac{5+2\sqrt{5}}{5}} - i \sqrt{\frac{5+\sqrt{5}}{10}} \\
     - \frac{1}{2} + \frac{i}{2}\sqrt{\frac{5+2\sqrt{5}}{5}} - i \sqrt{\frac{5+\sqrt{5}}{10}} & 
    i \sqrt{\frac{5+\sqrt{5}}{10}}
    \end{pmatrix}.
\end{equation}
Then the complex coordinate is transformed as 
\begin{equation}
\label{eq: Z5-twist}
    \vec{z} \xrightarrow{ST_1T_3}
    \Omega_{(ST_1T_3)} \vec{z}.
\end{equation}
Since $\Omega_{(ST_1T_3)}^5 = I_2$,
this is nothing but a $Z_5$-twist. We define a $T^4/Z_5$ orbifold by imposing the following $Z_5$ identification
\begin{equation}
    \vec{z} \sim \Omega_{(ST_1T_3)} \vec{z},
\end{equation}
on the complex coordinate of $T^4$. This means that the moduli of $T^4$ must be $ST_1T_3$ invariant leading to the fixing of moduli as shown in eq.(\ref{eq: ST1T3_inv_omega}).

\subsubsection{Lattice vectors}
Here, we explicitly show the lattice vectors defining the $T^4/Z_5$ orbifold, 
\begin{align}
\begin{aligned}
    e_1 &= 2 \pi R
    \begin{pmatrix}
    1 \\ 0 
    \end{pmatrix},\ 
    e_2 = 2 \pi R
    \begin{pmatrix}
      -\frac{1}{2} + \frac{i}{2} \sqrt{\frac{5+2\sqrt{5}}{5}} \\ 
       - \frac{1}{2} + \frac{i}{2}\sqrt{\frac{5+2\sqrt{5}}{5}} - i \sqrt{\frac{5+\sqrt{5}}{10}}
    \end{pmatrix},\\
    e_3 &= 2 \pi R
    \begin{pmatrix}
    0 \\ 1 
    \end{pmatrix},\
    e_4 = 2 \pi R
    \begin{pmatrix}
         - \frac{1}{2} + \frac{i}{2}\sqrt{\frac{5+2\sqrt{5}}{5}} - i \sqrt{\frac{5+\sqrt{5}}{10}}
         \\ 
          i \sqrt{\frac{5+\sqrt{5}}{10}}
    \end{pmatrix}.
    \end{aligned}
\end{align}
Notice that this is identical to the root lattice of $SU(5)$.
Under the $Z_5$-twist realized by the $ST_1T_3$ transformation, they behave as
\begin{align}
    \begin{pmatrix}
    e_2^{(ST_1T_3)}\\
    e_4^{(ST_1T_3)}\\
    e_1^{(ST_1T_3)}\\
    e_3^{(ST_1T_3)}
    \end{pmatrix}
    =
    \begin{pmatrix}
    0 & 0 & 1 & 0 \\
    0 & 0 & 0 & 1 \\
    -1 & 0 & -1 & -1 \\
    0 & -1 & -1 & 0 
    \end{pmatrix}
        \begin{pmatrix}
    e_2 \\
    e_4 \\
    e_1 \\
    e_3
    \end{pmatrix}.
\end{align}
This $Z_5$-twist is the Coxeter element of $SU(5)$ root lattice.
\subsubsection{Flux}
 When the complex structure moduli are $\Omega_{(ST_1T_3)}$, the F-flat condition shown in eq.(\ref{eq: SUSY_condition}) restricts $N$ to be of the form 
\begin{equation}
    N = 
    \begin{pmatrix}
    n_1 & n_1 - n_2 \\
    n_1 - n_2 & n_2
    \end{pmatrix},\ 
    n_{1},  n_{2} \in \mathbb{Z} .
\end{equation}
Then, we notice that $T_1T_3$ transformation is consistent with the the condition $p_{xx}=p_{yy}=0$ according to eq.(\ref{eq: consistent_T}).
Moreover, in order to write down the $T_1T_3$ transformation of the wavefunctions as in eq.(\ref{eq: T_zero-mode}), diagonal elements of $N(B_1 + B_3)$ must be even. This further restricts $N$ to satisfy $n_1 \equiv n_2 \equiv 0 \pmod{2}$. Thus, we obtain 
\begin{equation}
\label{eq: flux_Z5}
    N = 2
    \begin{pmatrix}
    n_1' & n_1' - n_2' \\
    n_1' - n_2' & n_2'
    \end{pmatrix},\ 
    n_{1}',  n_{2}' \in \mathbb{Z}.
\end{equation}
As a result, ${\rm det}N \equiv 0 \pmod{4}$ is necessary. We will consider the case when the positive definite condition eq.(\ref{eq: positive_definite_conditon}) is satisfied.

\subsubsection{Zero-mode counting method}
    To analyze the number of zero-modes, we  evaluate the transformation matrix $\rho(ST_1T_3)$ as,

\begin{align}
\begin{aligned}
    {\rm tr}\rho(ST_1T_3)&= \frac{e^{-\frac{\pi i}{10}}}{\sqrt{{\rm det}N}} \sum_{\vec{K} \in \Lambda_N} e^{2 \pi i \vec{K}^{\rm T} N^{-1} \vec{K}}e^{\pi i \vec{K}^{\rm T} [N^{-1} (B_1 + B_3)] \vec{K}}.
\end{aligned}
\end{align}
Let us denote the number of degeneracy in zero-modes by $D_{(\zeta^k)},\ (k=0,1,2,3,4)$, where $\zeta = e^{\frac{2 \pi i}{5}}$. 
\begin{align}
\label{eq: total_Z5}
    \sum_{k=0}^4 D_{(\zeta^k)} &= {\rm det}N, \\
\label{eq: trS_Z5}
   \sum_{k=0}^4 \zeta^k D_{(\zeta^k)}  &= {\rm tr}\rho(ST_1T_3).
\end{align}
Once the trace is evaluated, Eq.(\ref{eq: trS_Z5}) constrain $D_{(\zeta^k)}$ up to  $D_{(1)}=D_{(\zeta)}=\cdots = D_{(\zeta^4)}=1$.\footnote{Suppose $a_0 + a_1 \zeta + a_2 \zeta^2 + a_3 \zeta^3 + a_4 \zeta^4 = 0$ where $a_n$'s are non-negative integers. Then we find that $a_0 = a_1 = \cdots a_4$ needs to be satisfied. The arbitrariness is due to $\sum_{k=0}^4 {\zeta^k}=0$.} This arbitrariness can be eliminated once we fixed ${\rm det}N$. Therefore, eqs.(\ref{eq: total_Z5}) and (\ref{eq: trS_Z5}) provide sufficient information to determine $D_{(\zeta^k)}$. 
The results are shown in Table \ref{tb: T4/Z5}.

\begin{table}[H]
\begin{center}
\begin{tabular}{|c||c|c|c|c|c|} \hline
${\rm det}N$ & $+1$ & $\zeta$ & $\zeta^2$ & $\zeta^3$ & $\zeta^4$      \\ \hline \hline 
4 & 1 & 0 & 1 & 1 & 1 \\ \hline
16 & 3 & 4 & 3 & 3 & 3 \\ \hline
20 & 5 & 4 & 5 & 3 & 3 \\ \hline
36 & 7 & 8 & 7 & 7 & 7 \\ \hline
\end{tabular}
\end{center}
\caption{$T^4/Z_5$}
\label{tb: T4/Z5}
\end{table}

We find that three generation models appear when the size of the flux is ${\rm det}N = 16, 20$.
It is obvious that even if we used the algebraic relation $(ST_2T_3)^5=I_4$ instead of eq.(\ref{eq: Z5_algebra}), we obtain essentially the same result. This simply corresponds to the interchange of two complex coordinates $z_1$ and $z_2$. 

\subsection{$T^4/Z_6^{(1)}$}
We focus on the following algebraic relation,
\begin{equation}
    (ST_1T_2^{-1})^6 = I_4.
\end{equation}
$ST_1T_2^{-1}$ invariant $\Omega \in \mathcal{H}_2$ is 
\begin{equation}
\Omega_{(ST_1T_2^{-1})} = 
    \begin{pmatrix}
    \omega & 0 \\ 
    0 & \kappa
    \end{pmatrix},\quad \omega = e^{\frac{2 \pi i}{3}},\ \kappa = e^{\frac{\pi i}{3}}.
\end{equation}
Note that $\Omega_{(ST_1T_2^{-1})}$ is connected to $\Omega_{(ST_1T_2)}$ by $T_2$ transformation. This means that
lattice points are identical between $T^4/Z_6^{(1)}$ and $T^4/Z_3^{(a)}$.
However, the complex coordinate is transformed differently 
\begin{equation}
\label{eq: Z6^1-twist}
    \vec{z} = 
    \begin{pmatrix}
    z_1 \\ z_2
    \end{pmatrix}
    \xrightarrow{ST_1T_2^{-1}} \Omega_{(ST_1T_2^{-1})} \vec{z} = 
    \begin{pmatrix}
    \omega z_1 \\ \kappa z_2
    \end{pmatrix}.
\end{equation}
This is a $Z_6$-twist. 
We define $T^4/Z_6^{(1)}$ orbifold by identifying this twist. 
The $Z_6$-twist is the generalized Coxeter element of $SU(3)$ including the outer automorphism of two simple roots \cite{Markushevich:1986za, Katsuki:1989bf}.
\subsubsection{Flux}
When the complex structure moduli are $\Omega_{(ST_1T_2^{-1})}$, the F-flat condition shown in eq.(\ref{eq: SUSY_condition}) restricts $N$ to be a diagonal matrix. 
Then we notice that $T_1T_2^{-1}$ transformation is consistent with the the condition $p_{xx}=p_{yy}=0$ according to eq.(\ref{eq: consistent_T}).
Moreover, in order to write down the $T_1T_2^{-1}$ transformation of the wavefunctions as in eq.(\ref{eq: T_zero-mode}), matrix elements of $N$ must be even. Thus, we obtain
\begin{equation}
    N = 
    \begin{pmatrix}
    M_1 & 0 \\
    0 & M_2
    \end{pmatrix}
    =
    \begin{pmatrix}
    2n_1' & 0 \\ 0 & 2n_2'
    \end{pmatrix}.
\end{equation}
where $n_1', n_2' \in \mathbb{Z}^+$. Then, $N$ is $ST_1T_2^{-1}$ invariant.\footnote{We still end up with diagonal flux even if we have chosen a coordinate system with moduli $\Omega_{(ST_1T_2)}$. This follows from the requirement that the flux be invariant under the $Z_6$-twist shown in eq.(\ref{eq: Z6^1-twist}).}
Notice that we are led to analyze magnetized $T^2/Z_3 \times T^2/Z_6$ where the flux size is $M_1=2n_1'$ and $M_2=2n_2'$ respectively. Number of zero-modes on them are already studied in ref.\cite{Kobayashi:2017dyu}. Tables \ref{tb: T2/Z3} and \ref{tb: T2/Z6} summarize the results.

\begin{table}[H]
\begin{center}
\begin{tabular}{|c||c|c|c|} \hline
 & $n_1'=3m$ & $n_1'=3m+1$ & $n_1'=3m+2$       \\ \hline
1  & $2m+1$ & $2m+1$ & $2m+1$           \\ \hline
$\omega$  & $2m$ & $2m$ & $2m+2$     \\ \hline
$\omega^2$  & $2m-1$ & $2m+1$ & $2m+1$   \\ \hline
\end{tabular}
\end{center}
\caption{ $T^2/Z_3$ with flux $M_1=2n_1'$}
\label{tb: T2/Z3}
\end{table}

\begin{table}[H]
\begin{center}
\begin{tabular}{|c||c|c|c|} \hline
 & $n_2'=3n$ & $n_2'=3n+1$ & $n_2'=3n+2$       \\ \hline
1                & $n+1$ & $n+1$ & $n+1$           \\ \hline
$\kappa$  & $n$ & $n$ & $n+1$     \\ \hline
$\kappa^2$  & $n$ & $n+1$ & $n+1$   \\ \hline
$\kappa^3$             & $n$ & $n$ & $n$
\\ \hline
$\kappa^4$ &  $n$ & $n$ & $n+1$
\\ \hline
$\kappa^5$ & $n-1$ & $n$ & $n$ 
\\ \hline
\end{tabular}
\end{center}
\caption{$T^2/Z_6$ with flux $M_2=2n_2'$}
\label{tb: T2/Z6}
\end{table}

Zero-mode wavefunctions on magnetized $T^2/Z_3 \times T^2/Z_6$ are given by the product of those on $T^2/Z_3$ and $T^2/Z_6$, namely
\begin{equation}
    \psi^{\vec{J}}_{N} (\vec{z}) = \psi^{J_1,M_1}_{T^2/Z_3}(z_1) \otimes \psi^{J_2,M_2}_{T^2/Z_6}(z_2).
\end{equation}
Under the $Z_6$-twist in eq.(\ref{eq: Z6^1-twist}), we find
\begin{align} 
\begin{aligned}
\psi^{\vec{J}}_{N} (\Omega_{(ST_1T_2^{-1})}\vec{z}) &= \psi^{J_1,M_1}_{T^2/Z_3}(\omega z_1) \otimes \psi^{J_2,M_2}_{T^2/Z_6}(\kappa z_2) \\
&= \kappa^{2k+l} \psi^{J_1,M_1}_{T^2/Z_3}(z_1) \otimes \psi^{J_2,M_2}_{T^2/Z_6}(z_2) \\
&= \kappa^{2k+l} \psi^{\vec{J}}_{N}(\vec{z}).
\end{aligned}
\end{align}
This shows that the $Z_6$ charge of 
$\psi^{\vec{J}}_{N}(\vec{z})$ is given by the product of $Z_3$ and $Z_6$ charges of $\psi^{J_1,M_1}_{T^2/Z_3}(z_1)$ and $ \psi^{J_2,M_2}_{T^2/Z_6}(z_2)$ respectively.
Results of the number of zero-modes on $T^4/Z_6^{(1)} = T^2/Z_3 \times T^2/Z_6 $ are shown in Table \ref{tb: T4/Z6_1}.

\begin{table}[H]
\begin{center}
\begin{tabular}{|c|c||c|c|c|c|c|c|} \hline
${\rm det}N$ & $N={\rm diag}(M_1,M_2)$ & $+1$ & $\kappa$ & $\kappa^2$ & $\kappa^3$ &  $\kappa^4$ & $\kappa^5$  \\ \hline \hline 
4 & (2,2) & 2 & 0 & 1 & 0 & 1 & 0   \\ \hline
8 & (2,4) & 2 & 1 & 2 & 0 & 2 & 1   \\ \hline
8 & (4,2) & 2 & 0 & 3 & 0 & 3 & 0   \\ \hline
12 & (2,6) & 3 & 2 & 2 & 1 & 3 & 1  \\
\hline
12& (6,2) & 4 & 0 & 5 & 0 & 3 & 0  \\ \hline
16 & (2,8) & 4 & 2 & 3 & 2 & 3 & 2   \\ \hline
16 & (4,4) & 4 & 1 & 4 & 2 & 4 & 1   \\ \hline
16 & (8,2) & 6 & 0 & 5 & 0 & 5 & 0 \\ \hline
20 & (2,10) & 4 & 3 & 4 & 2 & 4 & 3 \\ \hline 
20& (10,2) & 6 & 0 & 7 & 0 & 7 & 0 \\ \hline
24& (2,12) & 5 & 4 & 4 & 3 & 4 & 3 \\ \hline
24& (4,6) & 5 & 2 & 6 & 3 & 5 & 3  \\ \hline
24& (6,4) & 6 & 3 & 6 & 2 & 6 & 1 \\ \hline
24& (12,2) & 8 & 0 & 9 & 0 & 7 & 0 \\ \hline
28& (2,14) & 6 & 4 & 5 & 4 & 5 & 4 \\ \hline 
28& (14,2) & 10 & 0 & 9 & 0 & 9 & 0 \\ \hline
32& (2,16) & 6 & 4 & 6 & 5 & 6 & 5 \\ \hline
32 & (4,8) & 6 & 4 & 7 & 4 & 7 & 4 \\ \hline
32 & (8,4) & 8 & 3 & 8 & 2 & 8 & 3 \\ \hline
32 & (16,2) & 10 & 0 & 11 & 0 & 11 & 0 \\ \hline
36 & (2,18) & 7 & 6 & 6 & 5 & 7 & 5 \\ \hline
36 & (6,6) & 9 & 4 & 8 & 5 & 7 & 3 \\ \hline
36 & (18,2) & 12 & 0 & 13 & 0 & 11 & 0 \\ \hline
40 & (2,20) & 8 & 6 & 7 & 6 & 7 & 6 \\ \hline
40 & (4,10) & 8 & 5 & 8 & 6 & 8 & 5 \\ \hline
40 & (10,4) & 10 & 3 & 10 & 4 & 10 & 3 \\ \hline
40 & (20,2) & 14 & 0 & 13 & 0 & 13 & 0 \\ \hline
44 & (2,22) & 8 & 7 & 8 & 6 & 8 & 7 \\ \hline
44 & (22,2) & 14 & 0 & 15 & 0 & 15 & 0 \\ \hline
48 & (2,24) & 9 & 8 & 8 & 7 & 9 & 7 \\ \hline
48 & (4,12) & 9 & 6 & 10 & 7 & 9 & 7 \\ \hline
48 & (6,8) & 10 & 6 & 11 & 6 & 9 & 6 \\ \hline
48 & (8,6) & 11 & 6 & 10 & 5 & 11 & 5 \\ \hline
48 & (12,4) & 12 & 5 & 12 & 4 & 12 & 3 \\ \hline
48 & (24,2) & 16 & 0 & 17 & 0 & 15 & 0 \\ \hline
\end{tabular}
\end{center}
\caption{$T^4/Z_6^{(1)}$}
\label{tb: T4/Z6_1}
\end{table}

We find that three generation models appear when ${\rm det}N \leq 48$ except ${\rm det}N = 4,28$, and $44$. 

\subsection{$T^4/Z_6^{(2)}$}
We focus on the following algebraic relation,
\begin{equation}
    ((T_1 T_2)^{-1} S)^6 = I_4.
\end{equation}
$(T_1 T_2)^{-1} S$ invariant $\Omega \in \mathcal{H}_2$ is uniquely determined as,
\begin{equation}
    \Omega_{((T_1T_2)^{-1}S)} =
    \begin{pmatrix}
    \omega & 0 \\ 0 & \omega
    \end{pmatrix},\quad \omega = e^{\frac{2\pi i}{3}},
\end{equation}
which is identical to $\Omega_{(ST_1T_2)}$ in eq.(\ref{eq: ST1T2_inv_omega}).
The complex coordinate is transformed as 
\begin{equation}
    \vec{z} \xrightarrow{(T_1T_2)^{-1}S} \kappa \vec{z},\quad \kappa =  e^{\frac{\pi i}{3}}.
\end{equation}
This is nothing but a $Z_6$-twist which we identify to define $T^4/Z_6^{(2)}$ orbifold. Note that two consecutive $Z_6$-twists are equivalent to the $Z_3$-twist in eq.(\ref{eq: z3-twist}). Three consecutive $Z_6$-twists are equivalent to the $Z_2$-twist.

\subsubsection{Lattice vectors and flux}
The lattice vectors defining the $T^4/Z_6^{(2)}$ orbifold are shown in eq.(\ref{eq: latticeT4/Z3a}).
Under the $Z_6$-twist realized by $(T_1T_2)^{-1}S$ transformation, they behave as
\begin{align}
    \begin{pmatrix}
    e_2^{((T_1T_2)^{-1}S)}\\
    e_4^{((T_1T_2)^{-1}S)}\\
    e_1^{((T_1T_2)^{-1}S)}\\
    e_3^{((T_1T_2)^{-1}S)}
    \end{pmatrix}
    =
    \begin{pmatrix}
    1 & 0 & 1 & 0 \\
    0 & 1 & 0 & 1 \\
    -1 & 0 & 0 & 0 \\
    0 & -1 & 0 & 0 
    \end{pmatrix}
        \begin{pmatrix}
    e_2 \\
    e_4 \\
    e_1 \\
    e_3
    \end{pmatrix}.
\end{align}

The consistent flux $N$ on this $T^4/Z_6$ orbifold is shown in eq.(\ref{eq: Z3a_flux}). 
Then $(T_1T_2)^{-1}S$ invariance of the flux is satisfied.

\subsubsection{Zero-mode counting method}
To analyze the number of zero-modes, we evaluate the trace of $\rho((T_1T_2)^{-1}S)$ as,
\begin{align}
\begin{aligned}
    {\rm tr}\rho((T_1T_2)^{-1}S)&= \frac{e^{\frac{\pi i}{6}}}{\sqrt{{\rm det}N}} \sum_{\vec{K} \in \Lambda_N} e^{ \pi i \vec{K}^{\rm T} . N^{-1}. \vec{K}}.
\end{aligned}
\end{align}
Let us denote the number of degeneracy in zero-modes by $D_{(\kappa^k)},\ (k=0,1,..,5)$. We have the following information,
\begin{align}
\label{eq: total_Z6}
    \sum_{k=0}^5 D_{(\kappa^k)} &= {\rm det}N, \\
\label{eq: trS_Z6}
   \sum_{k=0}^5 \kappa^k D_{(\kappa^k)} &= {\rm tr}\rho((T_1T_2)^{-1}S).
\end{align}
Eqs.(\ref{eq: total_Z6}) 
and (\ref{eq: trS_Z6}) constrain only three real degrees of freedom. Thus, we need more constraint equations to determine $D$'s. From the fact that $Z_6$ eigenstates with eigenvalues $1, \kappa^2$, and $\kappa^4$ are $Z_2$ even, we have
\begin{equation}
    D_{(+1)} + D_{(\kappa^2)} + D_{(\kappa^4)} = N_{Z_2,+}.
\end{equation}
Here, $N_{Z_2,+}$ denotes the number of even modes on $T^4/Z_2$. 
Likewise from the relationship between $Z_6$ and $Z_3$ eigenvalues, we have
\begin{align}
\begin{aligned}
D_{(+1)} + D_{(\kappa^3)} &= D_{Z_3,(+1)}, \\
D_{(\kappa)} + D_{(\kappa^4)} &= D_{Z_3,(\omega)}.
\end{aligned}
\end{align}
In the above, $D_{Z_3, (\omega^k)}$ represent the number of zero-modes with $Z_3$ charges $\omega^k$ on the $T^4/Z_3^{(a)}$ orbifold.
Results are shown in Tables \ref{tb: T4/Z6_2} and \ref{tb: T4/Z6_2_2}.

\begin{table}[H]
\begin{center}
\begin{tabular}{|c|c||c|c|c|c|c|c|} \hline
${\rm det}N$ & conditions & $+1$ & $\kappa$ & $\kappa^2$ & $\kappa^3$ &  $\kappa^4$ & $\kappa^5$  \\ \hline \hline 
3 & - & 1 & 0 & 1 & 1 & 0 & 0   \\ \hline
4 & - & 1 & 0 & 2 & 0 & 1 & 0   \\ \hline
7 & - & 1 & 1 & 2 & 1 & 1 & 1   \\ \hline
8 & - & 2 & 1 & 2 & 1 & 2 & 0  \\
\hline
11& - & 2 & 2 & 2 & 2 & 2 & 1  \\ \hline
12 &
\begin{tabular}{c}
$n_i' \stackrel{\rm{mod}2}{\equiv} 1,$\
$ (i = 1\  {\rm or}\ 2)$
\end{tabular} 
& 3 & 1 & 3 & 2 & 2 & 1   \\ \hline
12 &
\begin{tabular}{c}
$n_i' \stackrel{\rm{mod}2}{\equiv} 0,$\
$ (i = 1\  {\rm and}\ 2)$
\end{tabular} 
& 2 & 2 & 3 & 1 & 3 & 1   \\ \hline
15 & 
\begin{tabular}{c}
$ (\frac{n_i'}{5})_L=1,$\
$ (i = 1\  {\rm or}\ 2)$
\end{tabular} 
    & 3 & 2 & 3 & 3 & 2 & 2 \\ \hline
15 & 
\begin{tabular}{c}
$ (\frac{n_i'}{5})_L=-1,$\
$ (i = 1\  {\rm or}\ 2)$
\end{tabular} 
    & 2 & 3 & 3 & 2 & 3 & 2 \\ \hline 
16& - & 3 & 2 & 4 & 2 & 3 & 2 \\ \hline
19& - & 3 & 3 & 4 & 3 & 3 & 3 \\ \hline
20& - & 4 & 3 & 4 & 3 & 4 & 2  \\ \hline
23& - & 4 & 4 & 4 & 4 & 4 & 3 \\ \hline
24& 
\begin{tabular}{c}
$n_i' \stackrel{\rm{mod}8}{\equiv} \pm 1$, \
($i=1\ {\rm or}\ 2$)
\end{tabular}
    & 5 & 3 & 5 & 4 & 4 & 3 \\ \hline
24& 
\begin{tabular}{c}
$n_i' \stackrel{\rm{mod}8}{\equiv} \pm 3$, \
($i=1\ {\rm or}\ 2$)
\end{tabular}
    & 4 & 4 & 5 & 3 & 5 & 3 \\ \hline   
27& 
\begin{tabular}{c}
${\rm gcd}(n_i',m) \stackrel{\rm{mod}3}{\equiv} 1,$\
($i=1\ {\rm or}\ 2$)
\end{tabular} 
    & 5 & 4 & 5 & 5 & 4 & 4 \\ \hline
27& 
    \begin{tabular}{c}
${\rm gcd}(n_i',m) \stackrel{\rm{mod}3}{\equiv} 0,$\
($i=1\ {\rm and}\ 2$)
\end{tabular} 
    & 5 & 5 & 4 & 5 & 5 & 3 \\ \hline
28 & - & 5 & 4 & 6 & 4 & 5 & 4 \\ \hline
31 & - & 5 & 5 & 6 & 5 & 5 & 5 \\ \hline
32 & - & 6 & 5 & 6 & 5 & 6 & 4 \\ \hline
35 & - & 6 & 6 & 6 & 6 & 6 & 5 \\ \hline
\end{tabular}
\end{center}
\caption{$T^4/Z_6^{(2)}, (3 \leq {\rm det}N \leq 35)$}
\label{tb: T4/Z6_2}
\end{table}

\begin{table}[H]
\begin{center}
\begin{tabular}{|c|c||c|c|c|c|c|c|} \hline
${\rm det}N$ & conditions & $+1$ & $\kappa$ & $\kappa^2$ & $\kappa^3$ & $\kappa^4$ & $\kappa^5$     \\ \hline \hline 
36 & 
\begin{tabular}{c}
$n_i'\stackrel{\rm{mod}12}{\equiv} \pm 1,$\
($i=1\ {\rm or}\ 2$) \\
{\rm or} \\
$n_1'\stackrel{\rm{mod}3}{\equiv} 0$,\ $n_2' \stackrel{\rm{mod}4}{\equiv}-2$ \\
{\rm or} \\
$n_2'\stackrel{\rm{mod}3}{\equiv} 0$,\ $n_1' \stackrel{\rm{mod}4}{\equiv}-2$ 
\end{tabular} 
   & 7 & 5 & 7 & 6 & 6 & 5 \\ \hline
36 & 
\begin{tabular}{c}
$n_i'\stackrel{\rm{mod}12}{\equiv} \pm 5,$\
($i=1\ {\rm or}\ 2$) \\
{\rm or} \\
$n_1'\stackrel{\rm{mod}3}{\equiv} 0$,\ $n_2' \stackrel{\rm{mod}4}{\equiv}2$ \\
{\rm or} \\
$n_2'\stackrel{\rm{mod}3}{\equiv} 0$,\ $n_1' \stackrel{\rm{mod}4}{\equiv}2$ 
\end{tabular} 
    & 6 & 6 & 7 & 5 & 7 & 5 \\ \hline
36 & 
\begin{tabular}{c}
${\rm gcd}(n_i',m')=3,$ \
$ (i = 1\  {\rm and}\ 2)$
\end{tabular}
& 7 & 6 & 6 & 6 & 7 & 4 \\ \hline
39 & 
\begin{tabular}{c}
$ (\frac{n_i'}{13})_L=1,$\
$ (i = 1\  {\rm or}\ 2)$
\end{tabular}
    & 7 & 6 & 7 & 7 & 6 & 6 \\ \hline
39 & 
\begin{tabular}{c}
$ (\frac{n_i'}{13})_L=-1,$\
$ (i = 1\  {\rm or}\ 2)$
\end{tabular}
    & 6 & 7 & 7 & 6 & 7 & 6 \\ \hline
40 & - & 7 & 6 & 8 & 6 & 7 & 6 \\ \hline
43 & - & 7 & 7 & 8 & 7 & 7 & 7 \\ \hline
44 & - & 8 & 7 & 8 & 7 & 8 & 6 \\ \hline
47 & - & 8 & 8 & 8 & 8 & 8 & 7 \\ \hline 
48 & 
\begin{tabular}{c}
${\rm gcd}(n_i',m')
\stackrel{\rm{mod}2}{\equiv}1
,$ \
$ (i = 1\ {\rm or}\ 2)$
\end{tabular}
    & 9 & 7 & 9 & 8 & 8 & 7  \\ \hline
48 & 
\begin{tabular}{c}
$n_i' \stackrel{\rm{mod}4}{\equiv} 2,$\
$ (i = 1\  {\rm or}\ 2)$
\end{tabular} 
    & 8 & 8 & 9 & 7 & 9 & 7 \\ \hline
48 & 
\begin{tabular}{c}
$n_i' \stackrel{\rm{mod}4}{\equiv} 0,$\
$ (i = 1\  {\rm and}\ 2)$
\end{tabular}
    & 9 & 7 & 9 & 8 & 8 & 7  \\ \hline
\end{tabular}
\end{center}
\caption{$T^4/Z_6^{(2)}, 36 \leq {\rm det}N \leq 48$}
\label{tb: T4/Z6_2_2}
\end{table}

We find that three generation models appear when $ 12 \leq {\rm det}N \leq 27$. 
Although we can not give a complete proof, it is observed that the trace, ${\rm tr}\rho((T_1T_2)^{-1}S)$ is always equal to $\omega=e^{\frac{2 \pi i}{3}}$.

\subsection{$T^4/Z_6^{(3)}$}
\label{eq: Z6_ST3}
We focus on the following algebraic relation,
\begin{equation}
    (ST_3)^6 = I_4.
\end{equation}
$ST_3$ invariant $\Omega \in \mathcal{H}_2$ is uniquely determined as,
\begin{equation}
\label{eq: ST3_moduli}
    \Omega_{(ST_3)} = \begin{pmatrix}
    i \frac{\sqrt{3}}{2} & - \frac{1}{2} \\
    - \frac{1}{2} & 
    i \frac{\sqrt{3}}{2}
    \end{pmatrix}.
\end{equation}
The complex coordinates are transformed as,
\begin{equation}
\label{eq: Z6-twist(3)}
    \vec{z} 
    \xrightarrow{ST_3}  \Omega_{(ST_3)} \vec{z}. 
\end{equation}
Since $\Omega_{(ST_3)}^6 = I_2$, this is a $Z_6$-twist which we identify to define $T^4/Z_6^{(3)}$ orbifold. Note that the composition of three $Z_6$-twists is equivalent to 
\begin{equation}
    \begin{pmatrix}
    z_1 \\ z_2
    \end{pmatrix}
     \xrightarrow{(ST_3)^3} 
    \begin{pmatrix}
    z_2 \\ z_1
    \end{pmatrix},
\end{equation}
which is the $Z_2$ permutation. Composition of two $Z_6$-twists is equivalent to the $Z_3$-twist in eq.(\ref{eq: z3-twist-b}). 

\subsubsection{Lattice vectors}
\begin{figure}[H]
\centering
\includegraphics[width=6cm]{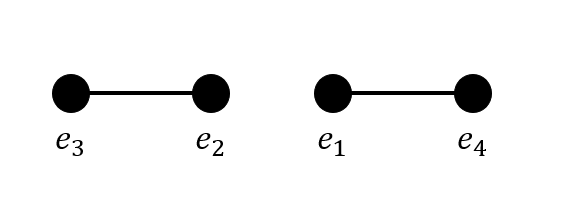}
\caption{Lattice vectors of $T^4/Z_6^{(3)}$}
\label{fig: t4/z6(3)}
\end{figure}

Here, we explicitly show the lattice vectors defining the $T^4/Z_6^{(3)}$ orbifold, 
\begin{equation}
\label{eq: T4Z6(3)-lattice}
e_1 = 2 \pi R
\begin{pmatrix}
    1 \\ 0
\end{pmatrix},\ 
e_2 = 2 \pi R
\begin{pmatrix}
    i \frac{\sqrt{3}}{2} \\ -\frac{1}{2}
\end{pmatrix},\ 
e_3 = 2 \pi R
\begin{pmatrix}
    0 \\ 1
\end{pmatrix},\ 
e_4 = 2 \pi R
\begin{pmatrix}
    -\frac{1}{2} \\ i \frac{\sqrt{3}}{2} 
\end{pmatrix}.\ 
\end{equation}
Notice that this is identical to the root lattice of $SU(3)\times SU(3)$ with the  Dynkin diagram shown in Fig.\ref{fig: t4/z6(3)}.
Under the $Z_6$-twist realized by $ST_3$ transformation, they behave as
\begin{align}
\label{eq: ST3_e_behave}
    \begin{pmatrix}
    e_2^{(ST_3)}\\
    e_4^{(ST_3)}\\
    e_1^{(ST_3)}\\
    e_3^{(ST_3)}
    \end{pmatrix}
    =
    \begin{pmatrix}
    0 & 0 & 1 & 0 \\
    0 & 0 & 0 & 1 \\
    -1 & 0 & 0 & -1 \\
    0 & -1 & -1 & 0 
    \end{pmatrix}
        \begin{pmatrix}
    e_2 \\
    e_4 \\
    e_1 \\
    e_3
    \end{pmatrix}.
\end{align}
\subsubsection{Flux}
Magnetic fluxes on the $T^4/Z_6^{(3)}$ orbifold can be consistently introduced if $N$ is of the form,
\begin{equation}
\label{eq: N_ST3}
    N = 
    \begin{pmatrix}
    n &  m \\  m & n
    \end{pmatrix}
    =
    \begin{pmatrix}
    n & 2 m' \\ 2 m' & n
    \end{pmatrix},\ n, m' \in \mathbb{Z}.
\end{equation}
We see that the F-flat condition shown in eq.(\ref{eq: SUSY_condition}) is satisfied. 
We also notice that $T_3$ transformation is consistent with $p_{xx}=p_{yy}=0$ according to eq.(\ref{eq: consistent_T}).
Moreover, since off-diagonal elements of $N$ are even, $T_3$ transformation of the wavefunctions can be  written by eq.(\ref{eq: T_zero-mode}). Note that $N$ in eq.(\ref{eq: N_ST3}) is $ST_3$ invariant. We will study the case when positive definite condition eq.(\ref{eq: positive_definite_conditon}) is satisfied, so that $N$ is positive definite.

\subsubsection{Zero-mode counting method}
To analyze the number of zero-modes, we evaluate the trace of $\rho(ST_3)$ as,
\begin{align}
\begin{aligned}
   {\rm tr}\rho(ST_3) 
&=
\frac{1}{\sqrt{{\rm det}N}} \sum_{\vec{K} \in \Lambda_N} e^{\pi i \vec{K}^{\rm T}. N^{-1} (2I_2 + B_3) . \vec{K}} \\ 
&=
\frac{1}{\sqrt{{\rm det}N}} \sum_{\vec{K} \in \Lambda_N} e^{\frac{2 \pi i}{{\rm det}N}[(n-m')K_1^2 + (n-m')K_2^2 + (n-4m')K_1K_2]}.
\end{aligned}
\end{align}
Let us denote the number of zero-modes by $D_{(\kappa^k)}, (k=0,1,...,5)$. We have equations of the form eqs.(\ref{eq: total_Z6}) and (\ref{eq: trS_Z6}). From the fact that $Z_6$ eigenstates with eigenvalues $1, \kappa^2$ and $\kappa^4$ are even-modes of the $Z_2$-permutation, we have
\begin{equation}
    D_{(+1)} + D_{(\kappa^2)} + D_{(\kappa^4)} = \frac{{\rm det}N + (n+m)}{2}.
\end{equation}
For details of $T^4/Z_2$ permutation orbifold models, see Appendix \ref{appendix: z2_per}. Moreover, from the relationship between $Z_6$ and $Z_3$ eigenvalues, we have
\begin{align}
\begin{aligned}
D_{(+1)} + D_{(\kappa^3)} &= D_{Z_3,(+1)}, \\
D_{(\kappa)} + D_{(\kappa^4)} &= D_{Z_3,(\omega)}.
\end{aligned}
\end{align}
In the above, $D_{Z_3, (\omega^k)}$ represent the number of zero-modes with $Z_3$ charges $\omega^k$ on the $T^4/Z_3^{(b)}$ orbifold. Results are shown in Tables \ref{tb: T4/Z6_3} and \ref{tb: T4/Z6_3_2}.

\begin{table}[H]
\begin{center}
\begin{tabular}{|c|c||c|c|c|c|c|c|} \hline
${\rm det}N$ & $(n,m)$ & $+1$ & $\kappa$ & $\kappa^2$ & $\kappa^3$ &  $\kappa^4$ & $\kappa^5$  \\ \hline \hline 
1 & (1,0) & 1 & 0 & 0 & 0 & 0 & 0   \\ \hline
4 & (2,0) & 1 & 0 & 1 & 1 & 1 & 0   \\ \hline
5 & (3,2) & 1 & 0 & 2 & 0 & 2 & 0   \\ \hline
5 & (3,$-2$) & 1 & 1 & 1 & 0 & 1 & 1  \\
\hline
9& (5,4) & 3 & 0 & 4 & 0 & 2 & 0  \\ \hline
9 & (5,$-4$) & 2 & 2 & 1 & 1 & 2 & 1   \\ \hline
9 & (3,0) & 3 & 1 & 2 & 2 & 1 & 0   \\ \hline
12 & (4,2) & 3 & 2 & 3 & 1 & 3 & 0 \\ \hline
12 & (4,$-2$) & 2 & 1 & 3 & 2 & 2 & 2 \\ \hline 
13& (7,6) & 5 & 0 & 4 & 0 & 4 & 0 \\ \hline
13& (7,$-6$) & 3 & 2 & 2 & 2 & 2 & 2 \\ \hline
16& (4,0) & 4 & 2 & 3 & 2 & 3 & 2  \\ \hline
17& (9,8) & 5 & 0 & 6 & 0 & 6 & 0 \\ \hline
17& (9,$-8$) & 3 & 3 & 3 & 2 & 3 & 3 \\ \hline
20& (6,4) & 5 & 2 & 5 & 1 & 5 & 2 \\ \hline 
20& (6,$-4$) & 3 & 3 & 4 & 3 & 4 & 3 \\ \hline
21& (11,10) & 7 & 0 & 8 & 0 & 6 & 0 \\ \hline
21 & (11,$-10$) & 4 & 4 & 3 & 3 & 4 & 3 \\ \hline
21 & (5,2) & 5 & 3 & 4 & 2 & 5 & 2 \\ \hline
21 & (5,$-2$) & 4 & 3 & 5 & 3 & 3 & 3 \\ \hline
25 & (13,12) & 9 & 0 & 8 & 0 & 8 & 0 \\ \hline
25 & (13,$-12$) & 5 & 4 & 4 & 4 & 4 & 4 \\ \hline
25 & (5,0) & 5 & 3 & 5 & 4 & 5 & 3 \\ \hline
28 & (8,6) & 7 & 2 & 7 & 3 & 7 & 2 \\ \hline
28 & (8,$-6$) & 5 & 4 & 5 & 5 & 5 & 4 \\ \hline
29 & (15,14) & 9 & 0 & 10 & 0 & 10 & 0 \\ \hline
29 & (15,$-14$) & 5 & 5 & 5 & 4 & 5 & 5 \\ \hline
\end{tabular}
\end{center}
\caption{$T^4/Z_6^{(3)}, (1 \leq {\rm det}N \leq 29)$}
\label{tb: T4/Z6_3}
\end{table}

\begin{table}[H]
\begin{center}
\begin{tabular}{|c|c||c|c|c|c|c|c|} \hline
${\rm det}N$ & $(n,m)$ & $+1$ & $\kappa$ & $\kappa^2$ & $\kappa^3$ &  $\kappa^4$ & $\kappa^5$  \\ \hline \hline 
32 & (6,2) & 6 & 4 & 7 & 4 & 7 & 4   \\ \hline
32 & $(6,-2)$ & 6 & 5 & 6 & 4 & 6 & 5   \\ \hline
33 & (17,16) & 11 & 0 & 12 & 0 & 10 & 0   \\ \hline
33 & $(17,-16)$ & 6 & 6 & 5 & 5 & 6 & 5  \\
\hline
33 & (7,4) & 7 & 3 & 8 & 4 & 7 & 4  \\ \hline
33 & $(7,-4)$ & 6 & 6 & 6 & 5 & 6 & 4   \\ \hline
36 & (10,8) & 9 & 4 & 9 & 3 & 9 & 2   \\ \hline
36 & $(10,-8)$ & 6 & 5 & 7 & 6 & 6 & 6 \\ \hline
36 & (6,0) & 8 & 5 & 7 & 6 & 6 & 4 \\ \hline 
37 & (19,18) & 13 & 0 & 12 & 0 & 12 & 0 \\ \hline
37 & $(19,-18)$ & 7 & 6 & 6 & 6 & 6 & 6 \\ \hline
41 & (21,20) & 13 & 0 & 15 & 0 & 15 & 0  \\ \hline
41 & $(21,-20)$ & 7 & 7 & 7 & 6 & 7 & 7 \\ \hline
44 & (12,10) & 11 & 4 & 11 & 3 & 11 & 4 \\ \hline
44 & $(12,-10)$ & 7 & 7 & 8 & 7 & 8 & 7 \\ \hline 
45& (23,22) & 15 & 0 & 16 & 0 & 14 & 0 \\ \hline
45& $(23,-22)$ & 8 & 8 & 7 & 7 & 8 & 7 \\ \hline
45 & (9,6) & 11 & 5 & 10 & 6 & 9 & 4 \\ \hline
45 & $(9,-6)$ & 9 & 7 & 8 & 8 & 7 & 6 \\ \hline
45 & (7,2) & 9 & 7 & 9 & 6 & 9 & 5 \\ \hline
45 & $(7,-2)$ & 8 & 6 & 9 & 7 & 8 & 7 \\ \hline
45 & (8,4) & 10 & 6 & 11 & 6 & 9 & 6 \\ \hline
48 & $(8,-4)$ & 9 & 8 & 8 & 7 & 9 & 7 \\ \hline
49 & (25,24) & 17 & 0 & 16 & 0 & 16 & 0 \\ \hline
49 & $(25,-24)$ & 9 & 8 & 8 & 8 & 8 & 8 \\ \hline
49 & (7,0) & 10 & 7 & 9 & 7 & 9 & 7 \\ \hline
\end{tabular}
\end{center}
\caption{$T^4/Z_6^{(3)}, (32 \leq {\rm det}N \leq 49)$}
\label{tb: T4/Z6_3_2}
\end{table}

We find that three generation models appear when $9 \leq {\rm det}N \leq 28$ or ${\rm det}N = 33, 36, 44$.

\subsection{$T^4/Z_6^{(4)}$}
We focus on the following algebraic relation,
\begin{equation}
\label{eq: permutation_trans}
    (ST_1T_2 \gamma_P)^6 = I_4,
\end{equation}
where $\gamma_P \in Sp(4,\mathbb{Z})$ is defined as
\begin{equation}
    \gamma_P =
    \begin{pmatrix}
    A_P & 0 \\ 0 & (A_P^{-1})^{\rm T}
    \end{pmatrix},\quad 
    A_P =
    \begin{pmatrix}
    0 & 1 \\ 1 & 0
    \end{pmatrix} \in GL(2,\mathbb{Z}).
\end{equation}
This is a $Z_6$-twist which we identify to define $T^4/Z_6^{(4)}$ orbifold.
From eq.(\ref{eq: A_P}), we understand $\gamma_P$
as $Z_2$ permutation which interchanges the two complex coordinates $z_1$ and $z_2$. 
The complex coordinates are transformed as,
\begin{equation}
 \begin{pmatrix}
     z_1 \\
     z_2
 \end{pmatrix}
 \xrightarrow{ST_1T_2\gamma_P}
  \begin{pmatrix}
      \omega z_2 \\
      \omega z_1
  \end{pmatrix}.
\end{equation}
Note that the composition of three $Z_6$-twists is equivalent to the $Z_2$ permutation. Composition of two $Z_6$-twists is equivalent to the inverse of the $Z_3$-twist we studied in eq.(\ref{eq: z3-twist}).
\subsubsection{Lattice vectors}
\begin{figure}[H]
\centering
\includegraphics[width=6cm]{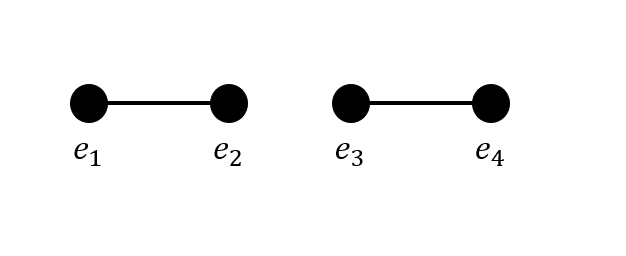}
\caption{Lattice vectors of $T^4/Z_6^{(4)}$}
\label{fig: t4/z6(4)}
\end{figure}

The lattice vectors defining the $T^4/Z_6^{(4)}$ orbifold are shown in eq.(\ref{eq: latticeT4/Z3a}). It is identical to the root lattice of $SU(3)\times SU(3)$ with a Dynkin diagram in Fig.\ref{fig: t4/z6(4)}.
Under the $Z_6$-twist realized by $ST_1T_2\gamma_P$ transformation, they behave as
\begin{align}
\label{eq: ST1T2P_e_behave}
    \begin{pmatrix}
    e_2^{(ST_1T_2\gamma_P)}\\
    e_4^{(ST_1T_2\gamma_P)}\\
    e_1^{(ST_1T_2\gamma_P)}\\
    e_3^{(ST_1T_2\gamma_P)}
    \end{pmatrix}
    =
    \begin{pmatrix}
    0 & 0 & 0 & 1 \\
    0 & 0 & 1 & 0 \\
    0 & -1 & 0 & -1 \\
    -1 & 0 & -1 & 0 
    \end{pmatrix}
        \begin{pmatrix}
    e_2 \\
    e_4 \\
    e_1 \\
    e_3
    \end{pmatrix}.
\end{align}

\subsubsection{Flux}
Magnetic fluxes on the $T^4/Z_6^{(4)}$ orbifold can be consistently introduced if $N$ is of the form,
\begin{equation}
\label{eq: N-ST1T2P}
N =
\begin{pmatrix}
2n' & m \\
m & 2n'
\end{pmatrix},\ 
n', m \in \mathbb{Z}.
\end{equation}
This is a special case of the flux on  $T^4/Z_3^{(a)}$ orbifold as shown in eq.(\ref{eq: Z3a_flux}). This is because we have constructed $T^4/Z_6^{(4)}$ orbifold by an additional identification of $z_1$ and $z_2$ on $T^4/Z_3^{(a)}$ orbifold via the $Z_2$ permutation, $\gamma_P \in Sp(4,\mathbb{Z})$. From eq.(\ref{eq: N_A_trans}), we find fluxes $N$ of the form in eq.(\ref{eq: Z3a_flux}) are transformed as,
\begin{equation}
\begin{pmatrix}
2n_1' & m \\
m & 2n_2'
\end{pmatrix}
\xrightarrow{\gamma_P} 
\begin{pmatrix}
2n_2' & m \\
m & 2n_1'
\end{pmatrix}.
\end{equation}
Thus, further identification by $Z_2$ permutation requires $n_1'=n_2'$.

\subsubsection{Zero-mode counting method}
To analyze the number of zero-modes, we evaluate the trace of $\rho(ST_1T_2\gamma_P)$ as,
\begin{align}
\begin{aligned}
{\rm tr}\rho(ST_1T_2\gamma_P)
&=
\frac{e^{-\frac{\pi i}{6}}}{\sqrt{{\rm det}N}} \sum_{\vec{K} \in \Lambda_N} e^{2 \pi i (A_P \vec{K})^{\rm T}. N^{-1}. \vec{K}} e^{\pi i \vec{K}^{\rm T}.N^{-1}.\vec{K}}  \\
&=
\frac{e^{-\frac{\pi i}{6}}}{\sqrt{{\rm det}N}} \sum_{\vec{K} \in \Lambda_N} e^{\frac{2 \pi i}{{\rm det}N} [(n-m')K_1^2 + (n-m')K_2^2 + (n-4m')K_1 K_2]}.
\end{aligned}
\end{align}
Let us denote the number of zero-modes by $D_{(\kappa^k)}, (k=0,1,...,5)$. We have equations of the form eqs.(\ref{eq: total_Z6}) and (\ref{eq: trS_Z6}). From the fact that $Z_6$ eigenstates with eigenvalues $1, \kappa^2$ and $\kappa^4$ are even-modes of the $Z_2$-permutation, we have
\begin{equation}
    D_{(+1)} + D_{(\kappa^2)} + D_{(\kappa^4)} = \frac{{\rm det}N + (n+m)}{2}.
\end{equation}
For details of $T^4/Z_2$ permutation orbifold models, see Appendix \ref{appendix: z2_per}. Moreover, from the relationship between $Z_6$ and the inverse of $Z_3$ eigenvalues, we have
\begin{align}
\begin{aligned}
D_{(+1)} + D_{(\kappa^3)} &= D_{Z_3,(+1)}, \\
D_{(\kappa)} + D_{(\kappa^4)} &= D_{Z_3,(+\omega^2)}.
\end{aligned}
\end{align}
In the above, $D_{Z_3, (+\omega^k)}$ represent the number of zero-modes with $Z_3$ charges $\omega^k$ on the $T^4/Z_3^{(a)}$ orbifolds. Results are shown in Table \ref{tb: T4/Z6_4} and \ref{tb: T4/Z6_4_2}.

\begin{table}[H]
\begin{center}
\begin{tabular}{|c|c||c|c|c|c|c|c|} \hline
${\rm det}N$ & $(n,m)$ & $+1$ & $\kappa$ & $\kappa^2$ & $\kappa^3$ &  $\kappa^4$ & $\kappa^5$  \\ \hline \hline 
3 & (2,1) & 2 & 0 & 0 & 0 & 1 & 0   \\ \hline
3 & (2,$-1$) & 1 & 0 & 0 & 1 & 1 & 0   \\ \hline
4 & (2,0) & 1 & 1 & 1 & 0 & 1 & 0   \\ \hline
7 & (4,3) & 2 & 0 & 2 & 0 & 3 & 0  \\
\hline
7& (4,$-3$) & 1 & 1 & 1 & 1 & 2 & 1  \\ \hline
11 & (6,5) & 4 & 0 & 4 & 0 & 3 & 0   \\ \hline
11 & (6,$-5$) & 2 & 1 & 2 & 2 & 2 & 2   \\ \hline
12 & (4,2) & 3 & 1 & 3 & 0 & 3 & 2 \\ \hline
12 & (4,$-2$) & 2 & 2 & 3 & 1 & 2 & 2 \\ \hline 
15& (8,7) & 6 & 0 & 4 & 0 & 5 & 0 \\ \hline
15& (8,$-7$) & 3 & 2 & 2 & 3 & 3 & 2 \\ \hline
15& (4,1) & 3 & 2 & 4 & 1 & 3 & 2  \\ \hline
15& (4,$-1$) & 3 & 2 & 3 & 1 & 3 & 3 \\ \hline
16& (4,0) & 3 & 2 & 3 & 2 & 4 & 2 \\ \hline
19& (10,9) & 6 & 0 & 6 & 0 & 7 & 0 \\ \hline 
19& (10,$-9$) & 3 & 3 & 3 & 3 & 4 & 3 \\ \hline
20& (8,6) & 5 & 1 & 5 & 2 & 5 & 2 \\ \hline
20 & (8,$-6$) & 4 & 3 & 4 & 3 & 3 & 3 \\ \hline
23 & (12,11) & 8 & 0 & 8 & 0 & 7 & 0 \\ \hline
23 & (12,$-11$) & 4 & 3 & 4 & 4 & 4 & 4 \\ \hline
27 & (14,13) & 10 & 0 & 8 & 0 & 9 & 0 \\ \hline
27 & (14,$-13$) & 5 & 4 & 4 & 5 & 5 & 4 \\ \hline
27 & (6,3) & 7 & 2 & 6 & 3 & 5 & 4 \\ \hline
27 & (6,$-3$) & 6 & 3 & 5 & 4 & 4 & 5 \\ \hline
28 & (8,6) & 7 & 3 & 7 & 2 & 7 & 2 \\ \hline
28 & (8,$-6$) & 5 & 5 & 5 & 4 & 5 & 4 \\ \hline
\end{tabular}
\end{center}
\caption{$T^4/Z_6^{(4)}, (3 \leq {\rm det}N \leq 28)$}
\label{tb: T4/Z6_4}
\end{table}

\begin{table}[H]
\begin{center}
\begin{tabular}{|c|c||c|c|c|c|c|c|} \hline
${\rm det}N$ & $(n,m)$ & $+1$ & $\kappa$ & $\kappa^2$ & $\kappa^3$ &  $\kappa^4$ & $\kappa^5$  \\ \hline \hline 
31 & (16,15) & 10 & 0 & 10 & 0 & 11 & 0   \\ \hline
31 & (16,$-15$) & 5 & 5 & 5 & 5 & 6 & 5   \\ \hline
32 & (6,2) & 7 & 4 & 7 & 4 & 6 & 4   \\ \hline
32 & (6,$-2$) & 6 & 4 & 6 & 5 & 6 & 5  \\
\hline
35& (18,17) & 12 & 0 & 12 & 0 & 11 & 0  \\ \hline
35 & (18,$-17$) & 6 & 5 & 6 & 6 & 6 & 6   \\ \hline
35 & (6,1) & 7 & 4 & 7 & 5 & 7 & 5   \\ \hline
35 & (6,$-1$) & 7 & 5 & 7 & 5 & 6 & 5 \\ \hline
36 & (10,8) & 9 & 3 & 9 & 2 & 9 & 4 \\ \hline 
36& (10,$-8$) & 6 & 6 & 7 & 5 & 6 & 6 \\ \hline
36& (6,0) & 8 & 4 & 7 & 5 & 6 & 6 \\ \hline
39& (20,19) & 14 & 0 & 12 & 0 & 13 & 0  \\ \hline
39& (20,$-19$) & 7 & 6 & 6 & 7 & 7 & 6 \\ \hline
39& (8,5) & 9 & 4 & 8 & 5 & 9 & 4 \\ \hline
39& (8,$-5$) & 8 & 6 & 6 & 6 & 7 & 6 \\ \hline 
43& (22,21) & 14 & 0 & 14 & 0 & 15 & 0 \\ \hline
43& (22,$-21$) & 7 & 7 & 7 & 7 & 8 & 7 \\ \hline
44 & (12,10) & 11 & 3 & 11 & 4 & 11 & 4 \\ \hline
44 & (12,$-10$) & 8 & 7 & 8 & 7 & 7 & 7 \\ \hline
47 & (24,23) & 16 & 0 & 16 & 0 & 15 & 0 \\ \hline
47 & (24,$-23$) & 8 & 7 & 8 & 8 & 8 & 8 \\ \hline
48 & (8,4) & 11 & 6 & 9 & 6 & 10 & 6 \\ \hline
48 & (8,$-4$) & 9 & 7 & 8 & 8 & 9 & 7 \\ \hline
\end{tabular}
\end{center}
\caption{$T^4/Z_6^{(4)}, (31 \leq {\rm det}N \leq 48)$}
\label{tb: T4/Z6_4_2}
\end{table}

We find that three generation models appear when $7 \leq {\rm det}N \leq 28$ or ${\rm det}N = 36, 44$.

\subsection{$T^4/Z_8$}
We focus on the following algebraic relation,
\begin{equation}
    (ST_1T_2^{-1}T_3^{-1})^8 = I_4.
\end{equation}
This suggests that we can construct a $T^4/Z_{8}$ orbifold with $ST_{1}T_2^{-1}T_{3}^{-1}$ invariant complex structure moduli $\Omega \in \mathcal{H}_2$ which is uniquely determined as,
\begin{equation}
\label{eq: omegaZ8}
    \Omega_{(ST_1T_2^{-1}T_3^{-1})} =
    \begin{pmatrix}
    -\frac{1}{2} + \frac{i}{\sqrt{2}} & \frac{1}{2} \\
    \frac{1}{2} & \frac{1}{2} + \frac{i}{\sqrt{2}}
    \end{pmatrix}. 
\end{equation}
The complex coordinates are transformed as,
\begin{equation}
\label{eq: Z8_twist}
    \vec{z}
 \xrightarrow{ST_1T_2^{-1}T_3^{-1}}
 \Omega_{(ST_1T_2^{-1}T_3^{-1})} \vec{z}.
\end{equation}
Since $\Omega_{(ST_1T_2^{-1}T_3^{-1})}^8 = I_2$, this is a $Z_8$-twist. By identifying coordinates of $T^4$ by this twist, we obtain the $T^4/Z_{8}$ orbifold.

\subsubsection{Lattice vectors}
\begin{figure}[H]
\centering
\includegraphics[width=6cm]{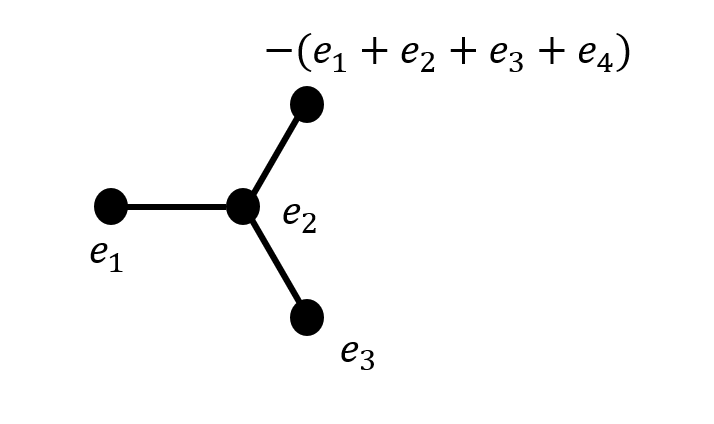}
\caption{Lattice vectors of $T^4/Z_8$}
\label{fig: t4/z8}
\end{figure}

Here, we study the lattice vectors defining the $T^4/Z_8$ orbifold. For this purpose, it is convenient to consider  moduli $\Omega$ related to $\Omega_{(ST_1T_2^{-1}T_3^{-1})}$ via $T_2T_3$ transformation. Although basis vectors of the lattice are different between the two, lattice points are invariant under the $Sp(4,\mathbb{Z})$ transformation. Thus, it makes sense to move to 
\begin{equation}
\Omega' \equiv 
\Omega_{(ST_1T_2^{-1}T_3^{-1})} - B_2 - B_3 
=   
\begin{pmatrix}
-\frac{1}{2} + \frac{i}{\sqrt{2}} & - \frac{1}{2} \\
- \frac{1}{2} & - \frac{1}{2} + \frac{i}{\sqrt{2}}
\end{pmatrix}.
\end{equation}
Lattice vectors which correspond to $\Omega'$ are 
\begin{equation}
    e_1 = 2 \pi R \begin{pmatrix}
   1  \\ 0 
    \end{pmatrix},\ 
    e_2 = 2 \pi R \begin{pmatrix}
    -\frac{1}{2} + \frac{i}{\sqrt{2}} \\ -\frac{1}{2}
    \end{pmatrix},\ 
    e_3 =  2 \pi R\begin{pmatrix}
    0 \\ 1
    \end{pmatrix},\ 
    e_4=  2 \pi R \begin{pmatrix}
    -\frac{1}{2}  \\ 
     -\frac{1}{2} + \frac{i}{\sqrt{2}}
    \end{pmatrix}.
\end{equation}
Their orientation corresponds to the root lattice of $SO(8)$ as in Fig.\ref{fig: t4/z8}.
The $Z_8$-twist in eq.(\ref{eq: Z8_twist}) is the general Coxeter element of $SO(8)$ including the $Z_2$ outer automorphism \cite{Markushevich:1986za, Katsuki:1989bf}.

\subsubsection{Flux}
Magnetic fluxes on the $T^4/Z_{8}$ orbifold with complex structure moduli $\Omega_{(ST_1T_2^{-1}T_3^{-1})}$ can be consistently introduced if $N$ is of the form, 
\begin{equation}
\label{eq: Z8_flux}
    N = 
    \begin{pmatrix}
    n_1 & \frac{n_2-n_1}{2} \\
    \frac{n_2-n_1}{2} & n_2
    \end{pmatrix}, \quad n_1, n_2 \in \mathbb{Z},
\end{equation}
where $n_1 \equiv n_2 \pmod{2}$. 
Then the flux is $ST_1T_2^{-1}T_3^{-1}$ invariant.

\subsubsection{Discussion}
Degeneracy of zero-modes $D_{(\xi^0)},..,D_{(\xi^7)}\ (\xi = e^{ \pi i/4 })$ can be analyzed as before by use of modular transformation.
Note that two and four consecutive $Z_8$-twists are $Z_4$- and $Z_2$-twists respectively. 

\subsection{$T^4/Z_{10}$}
We focus on the following algebraic relation,
\begin{equation}
\label{eq: Z10}
    ((T_1T_3)^{-1}S)^{10} = I_4.
\end{equation}
This suggests that we can construct a $T^4/Z_{10}$ orbifold with $(T_{1}T_3)^{-1}S$ invariant complex structure moduli $\Omega \in \mathcal{H}_2$. We find $(T_1T_3)^{-1}S=(ST_1T_3)^{-1}S^2$ where $S^2$ is the $Z_2$-twist, $\vec{z} \rightarrow -\vec{z}$. Obviously, $\forall \Omega \in \mathcal{H}_2$ are invariant under $S^2$. We also note that $ST_1T_3$-invariant moduli are $(ST_1T_3)^{-1}$-invariant and vice versa. Therefore, $ST_1T_3$-invariant moduli in eq.(\ref{eq: ST1T3_inv_omega}) are $(T_1T_3)^{-1}S$-invariant and vice versa. 
This means that the lattice vectors of the $T^4/Z_{10}$ orbifold are equivalent to the root lattice of $SU(5)$.
Then, magnetic fluxes can be consistently introduced if $N$ is of the form in eq.(\ref{eq: flux_Z5}).

The complex coordinates are transformed as,
\begin{equation}
\vec{z}
 \xrightarrow{(T_1T_3)^{-1}S}
 - \Omega^{-1}_{(ST_1T_3)}
\vec{z},
\end{equation}
under the $Z_{10}$-twist. This is the general Coxeter element of $SU(5)$ including the $Z_2$ outer automorphism. Note that the composition of five $Z_{10}$-twists is equivalent to the $Z_2$-twist. Composition of four $Z_{10}$-twists is equivalent to the $Z_{5}$-twist by the $ST_1T_3$ transformation as in eq.(\ref{eq: Z5-twist})\footnote{We can check $[(T_1T_3)^{-1}S]^4 = ST_1T_3$.}. Degeneracy of zero-modes $D_{(\eta^0)},...,D_{(\eta^9)} (\eta = e^{\pi i/5})$ can be analyzed as before by use of modular transformation.
It seems clear that we obtain essentially the same result when we consider the algebraic relation $((T_2T_3)^{-1}S)^{10}=I_4$ instead of eq.(\ref{eq: Z10}) as this corresponds to the interchange of two complex coordinates $z_1$ and $z_2$. 

\subsection{$T^4/Z_{12}$}
We focus on the following algebraic relation,
\begin{equation}
\label{eq: Z12_algebra}
    (ST_1)^{12} = I_4.
\end{equation}
This suggests that we can construct a $T^4/Z_{12}$ orbifold with $ST_{1}$ invariant complex structure moduli $\Omega \in \mathcal{H}_2$ which is uniquely determined as,
\begin{equation}
\label{eq: omegaZ12}
    \Omega_{(ST_1)}=
    \begin{pmatrix}
    \omega & 0 \\
    0 & i
    \end{pmatrix}.
\end{equation}
The complex coordinates are transformed as,
\begin{equation}
\begin{pmatrix}
    z_1 \\ z_2
\end{pmatrix}
 \xrightarrow{ST_1}
 \begin{pmatrix}
 \omega z_1 \\i z_2
 \end{pmatrix},
\end{equation}
which is clearly a $Z_{12}(\cong Z_3 \times Z_4)$-twist. By identifying coordinates of $T^4$ by this twist, we obtain the $T^4/Z_{12}$ orbifold. The lattice vectors are equivalent to the root lattice of $SU(3)\times SU(2) \times SU(2)$. 

\subsubsection{Flux}
Magnetic fluxes on the $T^4/Z_{12}$ orbifold can be consistently introduced if $N$ is of the form, 
\begin{equation}
\label{eq: Z12_flux}
    N = 
    \begin{pmatrix}
    2 n_1' & 0 \\ 0 & n_2
    \end{pmatrix}, \quad n_1', n_2\in \mathbb{Z}.
\end{equation}
The F-flat condition shown in eq.(\ref{eq: SUSY_condition}) is satisfied only if $N$ is diagonal. Then, it follows that flux is $ST_1$ invariant. The  $(1,1)$-component of $N$ needs to be even for the
$T_1$ transformation
of zero-modes written by eq.(\ref{eq: T_zero-mode})\footnote{It was also pointed out that the flux needs to be even for the invariance of boundary conditions under the $T$ transformation in magnetized $T^2$ models\cite{Kikuchi:2020frp, Kikuchi:2020nxn}.}.

\subsubsection{Discussion}
Both complex structure $\Omega$
and flux $N$ are diagonal as shown in eqs.(\ref{eq: omegaZ12}) and (\ref{eq: Z12_flux}). Thus, we are led to analyze magnetized $T^2/Z_3 \times T^2/Z_4$ where the flux size is $M_1 = 2 n_1'$ and $M_2=n_2$ respectively. Zero-modes on them are already studied in ref.\cite{Kobayashi:2017dyu}. Thus, zero-mode number can be analyzed easily by just noting the fact that zero-mode wavefunctions on magnetized $T^2/Z_3 \times T^2/Z_4$ are given by the product of those on each orbifold.

It is obvious that even if we used the algebraic relation $(ST_2)^2=I_4$ instead of eq.(\ref{eq: Z12_algebra}), we obtain essentially the same result. This simply corresponds to the interchange of two complex coordinates $z_1$ and $z_2$. 

\section{Duality of $T^4/Z_6^{(3)}$ and $T^4/Z_6^{(4)}$ by parity transformation}
\label{section: duality}
We notice that behaviours of lattice vectors in $T^4/Z_6^{(3)}$ and $T^4/Z_6^{(4)}$ are similar when we compare eqs.(\ref{eq: ST3_e_behave}) and (\ref{eq: ST1T2P_e_behave}). Furthermore, we saw that both cases correspond to the root lattice of $SU(3) \times SU(3)$ as in Figs.\ref{fig: t4/z6(3)} and \ref{fig: t4/z6(4)}.
 This implies there is a nontrivial relation between the two orbifolds.
In fact, they are related by a parity transformation in the extra dimension. We can find a duality that positive chirality zero-mode wavefunctions on $T^4/Z_6^{(3)}$ correspond to negative chirality wavefunctions on $T^4/Z_6^{(4)}$ and vice versa.

A parity transformation,
\begin{align}
\begin{aligned}
\label{eq: P_2}
    {\rm Re}z_1 &= {\rm Re}z^{(p)}_{2}, \\
    {\rm Re}z_2 &= {\rm Re}z^{(p)}_1, \\
    {\rm Im}{z}_i &=
    {\rm Im}{z}^{(p)}_i,\ (i=1,2),
\end{aligned}
\end{align}
relates the two orbifolds.
If $\vec{z}$ denotes complex coordinate of $T^4/Z_6^{(3)}$,   $\vec{z}^{\, (p)}$ correspond to that of $T^4/Z_6^{(4)}$.

Firstly, we check the complex structure moduli. Lattice vectors of $T^4/Z_6^{(3)}$ orbifolds in eq.(\ref{eq: T4Z6(3)-lattice}) are transformed as 
\begin{align}
\begin{aligned}
 e_1 = 2 \pi R
\begin{pmatrix}
1 \\ 0 
\end{pmatrix}
&\xrightarrow{parity}
e_3^{(p)} =2 \pi R
\begin{pmatrix}
0 \\ 1
\end{pmatrix}, \\
 e_2 =2 \pi R
\begin{pmatrix}
\frac{\sqrt{3}}{2}i \\ -\frac{1}{2} 
\end{pmatrix}
&\xrightarrow{parity}
e_2^{(p)} = 2 \pi R
\begin{pmatrix}
\omega \\ 0
\end{pmatrix}, \\
 e_3 =2 \pi R
\begin{pmatrix}
0 \\ 1 
\end{pmatrix}
&\xrightarrow{parity}
e_1^{(p)} =2 \pi R
\begin{pmatrix}
1 \\ 0
\end{pmatrix}, \\
 e_4 =2 \pi R
\begin{pmatrix}
-\frac{1}{2} \\ \frac{\sqrt{3}}{2}i 
\end{pmatrix}
&\xrightarrow{parity}
e_4^{(p)} =2 \pi R
\begin{pmatrix}
0 \\ \omega
\end{pmatrix}.
\end{aligned}
\end{align}
We observe that $e_i^{(p)}$ 
is exactly the same as lattice vectors of $T^4/Z_6^{(4)}$ orbifold. Therefore, their complex structure moduli are connected by eq.(\ref{eq: P_2}).

Secondly, we look at the background fluxes on them.
$N$ on $T^4/Z_6^{(3)}$ and $T^4/Z_6^{(4)}$ are given by eqs.(\ref{eq: N_ST3}) and (\ref{eq: N-ST1T2P}) respectively. We find
\begin{align}
    \begin{aligned}
F &=\pi [N_{(ST_3)}^{\rm T}. ({\rm Im}\Omega_{(ST_3)})^{-1}]_{ij} (i {dz}_i \wedge d\bar{z}_j) \\
 &= \pi [N_{(p)}^{\rm T}. ({\rm Im}\Omega_{(ST_1T_2\gamma_P)})^{-1}]_{ij} (i {dz^{(p)}_i} \wedge d\bar{z}^{(p)}_j),
    \end{aligned}
\end{align}
where
\begin{align}
    N_{(ST_3)}=
    \begin{pmatrix}
        n & 2m' \\
        2m' & n
    \end{pmatrix},\quad
    N_{(p)}=
    \begin{pmatrix}
        2m' & n \\
        n & 2m'
    \end{pmatrix},
\end{align}
showing that $N_{(p)}$ is consistently identified with $N_{(ST_1T_2\gamma_P)}$ in eq.(\ref{eq: N-ST1T2P}).
Notice that the determinant changes sign as ${\rm det}N_{(ST_3)} = -{\rm det}N_{(p)}$ implying that positive (negative) chirality on $T^4/Z_6^{(3)}$ and negative (positive) chirality on $T^4/Z_6^{(4)}$ are related. For more detailed discussions, see Appendix \ref{appendix: negative}.

Lastly, we confirm that $Z_6$-twist generated by $ST_3$ on $\vec{z}$ coordinate system is identified as $Z_6$-twist by $ST_1T_2\gamma_P$ on $\vec{z}^{\, (p)}$ system. It is straightforward to check that
\begin{align}
\begin{pmatrix}
    z_1 \\ z_2
\end{pmatrix} \xrightarrow{(ST_3)}
\begin{pmatrix}
i \frac{\sqrt{3}}{2} & -\frac{1}{2} \\
-\frac{1}{2} & i \frac{\sqrt{3}}{2}
\end{pmatrix}
\begin{pmatrix}
    z_1 \\ z_2
\end{pmatrix},
\end{align}
is equivalent to 
\begin{align}
\begin{pmatrix}
    z_1^{(p)} \\ z_2^{(p)}
\end{pmatrix} \xrightarrow{(ST_1T_2\gamma_P)}
\begin{pmatrix}
    \omega z_2^{(p)} \\ \omega z_1^{(p)}
\end{pmatrix}.
\end{align}
This guarantees that $Z_6$ eigenstates on $T^4/Z_6^{(3)}$ side are $Z_6$ eigenstates on $T^4/Z_6^{(4)}$ side as well. 

In short, magnetized $T^4/Z_6^{(3)}$ and $T^4/Z_6^{(4)}$ are connected by the parity transformation. Positive chirality on one side correspond to negative chirality on the other side. This kind of duality will be useful to analyze the number of negative chirality zero-modes by modular transformation since positive chirality wavefunctions are written by the Riemann theta function whose behaviour under $Sp(4,\mathbb{Z})$ is well known.

\section{Conclusion}
\label{conclusion}
We have studied $T^4/Z_N$ orbifold models with background magnetic fluxes. We saw that $Sp(4,\mathbb{Z})$ modular transformations can be related to  $Z_N$-twists defining the orbifolds. Behaviour of zero-mode wavefunctions under the modular transformation was studied. This enabled systematic analysis of zero-mode numbers. We used results of number theory in mathematics for this.
Our results explicitly show the condition needed to realize three generations of fermions in the effective field theory. Moreover, we saw that the parity transformation on the compact space can elegantly relate positive and negative chirality wavefunctions. Obtained negative chirality modes are consistent with ref.\cite{Antoniadis:2009bg} when real part of the complex structure ${\rm Re}\Omega$ is vanishing. What is more, our result is relevant even when ${\rm Re}\Omega \neq 0$.
This revealed a kind of duality between two different $T^4/Z_6$ orbifolds as we studied in Section \ref{section: duality}. 

This work will activate the study of magnetized $T^{2n}/Z_N$ models. Extension to orbifolds of $T^6$ will be our future work. Also, it will be helpful to develop mathematical formula to evaluate the trace of transformation matrix of zero-modes under modular transformation with arbitrary flux size ${\rm det}N$. 

When we embed our $T^4/Z_N$ orbifold models to superstring theory,
some models may include tachyonic modes in closed string sector, e.g.
$T^4/Z_5$.
Such configurations may be unstable.
Some moduli develop their vacuum expectation values
deforming geometry without changing topology.
If that is a stable manifold, our results on $T^4/Z_5$ could be realized,
because the number of zero-modes is determined by topology.
Such a study on embedding to string theory is beyond our scope.

\vspace{1.5 cm}
\noindent
{\large\bf Acknowledgement}\\
This work was supported by JSPS KAKENHI Grant Numbers
JP20J20388(H.U.) and JP22J10172(S.K.),
and JST SPRING Grant Number JPMJSP2119(K.N.). 


\appendix
\section{Modular transformation of zero-modes}
\label{eq: proof_zero-nodes}

\subsection{$S$ transformation}
\label{appendix: S-zero}
Here we give a proof of eq.(\ref{eq: S_zero-mode}), where it is assumed that $N$ is a symmetric positive definite integer matrix. 
It is known that the Riemann theta function satisfies the following relation \cite{Mumford:1983}, 
\begin{equation}
    \vartheta (- \Omega^{-1}\vec{z}, - \Omega^{-1}) = \sqrt{
    {\rm det}(\Omega / i)}
    \cdot e^{ \pi i  \vec{z}^{\rm T} 
    \Omega^{-1}  \vec{z} } \cdot  \vartheta(\vec{z}, \Omega ). 
\end{equation}
Note that the branch of the  square root is chosen so that it gives positive value when $\Omega$ is purely imaginary.

Firstly, replace the complex coordinates as $\vec{z} \rightarrow \vec{z} + N^{-1} \vec{J}$ and by use of eqs.(\ref{eq: property1}), (\ref{eq: property2}), we obtain
\begin{align}
    \vartheta
     \begin{bmatrix}
     \vec{J}^{\, \rm T}N^{-1} \\ 0
     \end{bmatrix}(- \Omega^{-1}\vec{z}, -\Omega^{-1}) = \sqrt{{\rm det}(\Omega/i)} \cdot
     e^{\pi i \vec{z}^{\rm T} \Omega^{-1}  \vec{z}} \cdot
     \vartheta 
      \begin{bmatrix}
      0 \\ \vec{J}^{\, \rm T} N^{-1}
      \end{bmatrix} (\vec{z}, \Omega).
\end{align}
Secondly, we replace the complex structure moduli as $\Omega \rightarrow N^{-1}\Omega$ to get,
\begin{align}
\label{eq: proof_S}
    \vartheta
     \begin{bmatrix}
     \vec{J}^{\, \rm T}N^{-1} \\ 0
     \end{bmatrix}(- \Omega^{-1} N \vec{z}, -\Omega^{-1} N) = \sqrt{{\rm det}(N^{-1}\Omega/i)} \cdot
     e^{\pi i \vec{z}^{\rm T} (N^{-1} \Omega)^{-1}  \vec{z}} \cdot
     \vartheta 
      \begin{bmatrix}
      0 \\ \vec{J}^{\, \rm T} N^{-1}
      \end{bmatrix} (\vec{z}, N^{-1}\Omega).
\end{align}
The Riemann theta function on the right-hand side of eq.(\ref{eq: proof_S}) can be written as 
\begin{align}
 \begin{aligned}
    \vartheta
    \begin{bmatrix}
    0 \\ \vec{J}^{\, \rm T}N^{-1}
    \end{bmatrix}(\vec{z}, {{N}}^{-1} \Omega) 
    & =
    \sum_{\vec{l} \in \mathbb{Z}^2} e^{\pi i \vec{l}^{\rm T} {{N}}^{-1} \Omega \vec{l}} e^{2 \pi i \vec{l}^{\rm T} \cdot (\vec{z} + N^{-1}\vec{J})} \\
    &= \sum_{\vec{K} \in \Lambda_N} e^{2 \pi i \vec{K}^{\rm T}  N^{-1} \vec{J}} \sum_{\vec{s} \in \mathbb{Z}^2} e^{\pi i (N \vec{s} + \vec{K})^{\rm T} \cdot {{N}}^{-1} \Omega \cdot ({{N}} \vec{s} + \vec{K})} \cdot e^{2 \pi i ({{N}} \vec{s} + \vec{K})^{\rm T} \cdot \vec{z}} \\
    & = \sum_{\vec{K} \in \Lambda_N} 
    e^{2 \pi i \vec{J}^{\rm T}  N^{-1} \vec{K}} \cdot
    \vartheta 
     \begin{bmatrix}
     \vec{K}^{\rm T} {{N}}^{-1} \\ 0
     \end{bmatrix}({{N}} \vec{z}, {{N}} \Omega).
     \end{aligned}
\end{align}
We have changed the summation variable as $\vec{l} = {{N}} \vec{s} + \vec{K}$, where $\vec{K}$'s are integer points inside the lattice $\Lambda_N$ spanned by ${{N}} \vec{e}_n$. 
Thus, we get
\begin{align}
\begin{aligned}
    \label{eq: S_1st}
 & \vartheta
     \begin{bmatrix}
     \vec{J}^{\, \rm T}N^{-1} \\ 0
     \end{bmatrix}(N(- \Omega^{-1} \vec{z}), N(-\Omega^{-1}) )  \\
  &  = 
     \sqrt{{\rm det}(N^{-1}\Omega/i)} \cdot
     e^{\pi i \vec{z}^{\rm T} (N^{-1} \Omega)^{-1}  \vec{z}} 
      \sum_{\vec{K} \in \Lambda_N} 
    e^{2 \pi i \vec{J}^{\rm T}  N^{-1} \vec{K}} \cdot
    \vartheta 
     \begin{bmatrix}
     \vec{K}^{\rm T} {{N}}^{-1} \\ 0
     \end{bmatrix}({{N}} \vec{z}, {{N}} \Omega).
     \end{aligned}
\end{align}
Thirdly, we consider the $S$ transformation of the $\vec{z}$ dependent factor,
\begin{equation}
\label{eq: z_dependent}
    e^{\pi i [N \vec{z}]^{\rm T} ({\rm Im}\Omega)^{-1} {\rm Im}\vec{z}} 
    \xrightarrow{S}
    e^{\pi i [- N \Omega^{-1} \vec{z}]^{\rm T} ({\rm Im}(-\Omega^{-1}))^{-1} {\rm Im}(-\Omega^{-1} \vec{z})}.
\end{equation}
When this factor is multiplied by the factor $ e^{\pi i \vec{z}^{\rm T} (N^{-1} \Omega)^{-1}  \vec{z}} $ which appears in the right-hand side of eq.(\ref{eq: S_1st}), we get
\begin{align}
    \begin{aligned}
    & e^{\pi i [- N \Omega^{-1} \vec{z}]^{\rm T} ({\rm Im}(-\Omega^{-1}))^{-1} {\rm Im}(-\Omega^{-1} \vec{z})} \cdot  e^{\pi i \vec{z}^{\rm T} (N^{-1} \Omega)^{-1}  \vec{z}} \\
    &= 
    e^{- \pi i [N \vec{z}]^{\rm T} \Omega^{-1} ({\rm Im}(-\Omega^{-1}))^{-1} {\rm Im}(-\Omega^{-1} \vec{z}) } e^{\pi i [N \vec{z}]^{\rm T} \Omega^{-1} \vec{z}} \\
    &= 
    e^{\pi i [N \vec{z}]^{\rm T} \Omega^{-1} [- ({\rm Im}(-\Omega^{-1}))^{-1}{\rm Im}(-\Omega^{-1} \vec{z}) + \vec{z}]} \\
    &= e^{\pi i [N \vec{z}]^{\rm T} 
    ({\rm Im}\Omega)^{-1} {\rm Im}\vec{z}}.
    \end{aligned}
\end{align}
Consequently, we obtain the $S$ transformation of the zero-mode wavefunctions,
\begin{equation}
    \psi_N^{\vec{J}}(-\Omega^{-1}\vec{z},-\Omega^{-1}) = \sqrt{{\rm det}(N^{-1}\Omega/i)} \sum_{\vec{K} \in \Lambda_N} e^{2 \pi i \vec{J}^{\, \rm T} \cdot N^{-1}\cdot \vec{K}} \psi_N^{\vec{K}}(\vec{z}, \Omega).
\end{equation}
Note that the normalization constant shown in eq.(\ref{eq: normalization}) is invariant under the $S$ transformation.

\subsection{$T$ transformation}
\label{appendix: T-zero}
Here we give a proof of eq.(\ref{eq: T_zero-mode}). It is known that the Riemann theta function satisfies the following relation provided diagonal matrix elements of $N B_i$ are all even\cite{Mumford:1983},
\begin{equation}
    \vartheta ({N}\vec{z}, { N}(\Omega + B_i)) = 
    \vartheta({ N}\vec{z}, { N}\Omega).
\end{equation}
Firstly, replace the complex coordinates as $\vec{z} \rightarrow \vec{z}+({\Omega} + B_i){N^{-1}}^{\rm T} \vec{J}$ and by use of eq.(\ref{eq: property2}), we obtain
\begin{equation}
    \vartheta 
    \begin{bmatrix}
    \vec{J}^{\, \rm T} {N^{-1}} \\ 0
    \end{bmatrix}({ N}\vec{z}, { N}(\Omega + B_i)) = e^{- \pi i \vec{J}^{\, \rm T} {N}^{-1} B_i \vec{J}} \cdot
    \vartheta 
    \begin{bmatrix}
    \vec{J}^{\, \rm T} N^{-1} \\ 0
    \end{bmatrix}({ N}\vec{z}+B_i\vec{J}, { N}\Omega).
\end{equation}
Secondly, by use of eqs.(\ref{eq: property1}), (\ref{eq: property3}), we get
\begin{align}
\begin{aligned}
    \vartheta 
    \begin{bmatrix}
    \vec{J}^{\, \rm T} {N^{-1}} \\ 0
    \end{bmatrix}({N}\vec{z}, { N}(\Omega + B_i)) 
    &=
    e^{- \pi i \vec{J}^{\rm T} {N}^{-1} B_i \vec{J}} \cdot
    \vartheta 
    \begin{bmatrix}
        \vec{J}^{\, \rm T} {N^{-1}} \\ \vec{J}^{\, \rm T} B_i
    \end{bmatrix}({ N}\vec{z}, { N}\Omega) \\
    &=
    e^{ \pi i \vec{J}^{\rm T} {N}^{-1} B_i \vec{J}} \cdot 
    \vartheta \begin{bmatrix}
     \vec{J}^{\, \rm T} {N^{-1}} \\ 0
    \end{bmatrix}({N}\vec{z}, { N}\Omega).
\end{aligned}
\end{align}
The $\vec{z}$ dependent factor
$e^{\pi i [N\vec{z}]\cdot ({\rm Im} \Omega)^{-1} \cdot {\rm Im}\vec{z} }$ is invariant under the $T_i$ transformation. The normalization constant eq.(\ref{eq: normalization}) is also unchanged. Consequently, we obtain
\begin{equation}
    \psi_{N}^{\vec{J}}(\vec{z}, \Omega + B_i) = e^{\pi i \vec{J}^{\rm T} ({N}^{-1} B_i)\vec{J}} \psi_{ N}^{\vec{J}} (\vec{z}, \Omega).
\end{equation}

\subsection{$A \in GL(2,\mathbb{Z})$ transformation}
\label{appendix: A-zero}
Here we give a proof of eq.(\ref{eq: A_trans_zero}). It is known that the Riemann theta function satisfies the following relation for $A \in { GL(2, \mathbb{Z})}$\cite{Mumford:1983},
\begin{equation}
    \vartheta 
    (A \vec{z}, A \Omega A^{\rm T})=\vartheta 
     (\vec{z}, \Omega).
\end{equation}
Firstly, replace the complex coordinates as $\vec{z} \rightarrow \vec{z} + \Omega A^{\rm T} {N^{-1}}^{\rm T} \vec{J}$ and use eq.(\ref{eq: property2}) to get 
\begin{equation}
        \vartheta 
     \begin{bmatrix}
     \vec{J}^{\, \rm T}N^{-1} \\ 0
     \end{bmatrix}(A \vec{z}, A \Omega A^{\rm T})
     =
     \vartheta 
     \begin{bmatrix}
     \vec{J}^{\, \rm T} {N^{-1}} A \\ 0
     \end{bmatrix}(\vec{z} , \Omega)
     =
     \vartheta 
     \begin{bmatrix}
     \vec{J}^{\, \rm T} A {N^{-1}} \\ 0
     \end{bmatrix}(\vec{z} , \Omega).
\end{equation}
Secondly, replace the complex coordinates and moduli as $\vec{z} \rightarrow {{N}} \vec{z}$,  $ \Omega \rightarrow {{N}} \Omega$ to get 
\begin{equation}
      \vartheta 
     \begin{bmatrix}
     \vec{J}^{\, \rm T}N^{-1} \\ 0 
     \end{bmatrix}(A{{N}} \vec{z}, A {{N}} \Omega A^{\rm T}) = \vartheta 
     \begin{bmatrix}
     \vec{J}^{\, \rm T} A N^{-1}\\ 0
     \end{bmatrix}({{N}} \vec{z}, {{N}} \Omega).
\end{equation}
Note that the $\vec{z}$ dependent factor $e^{\pi i [{{N}}\vec{z}]^{\rm T}\cdot ({\rm Im} \Omega)^{-1} \cdot {\rm Im}\vec{z} }$ is invariant under the $A$ transformation.
The normalization constant eq.(\ref{eq: normalization}) is also unchanged. 

Consequently, we obtain 
\begin{equation}
 \psi_N^{\vec{J}}(A \vec{z}, A \Omega A^{\rm T}) = \psi_N^{A^{\rm T}\vec{J}}(\vec{z}, \Omega),
\end{equation}
under the condition eq.(\ref{eq: A_inv_N}).

\section{Details of zero-mode counting in magnetized $T^4/Z_4$}

\subsection{Existence of $N$}
\label{appendix: existence_of_N}
Here, we prove that there is no symmetric positive definite $2 \times 2$ integer matrix with $\left(\frac{n_i}{7} \right) = -1$
when ${\rm det}N=7$.

\underbar{Proof}

To prove it, we use the Kronecker symbol \cite{Allouche: 2018}.
We denote arbitrary non-zero integer by $n$, with prime factorization,
\begin{equation}
    n = u \cdot 2^{e_0} p_1^{e_1} \cdots p_{k}^{e_k},
\end{equation}
where $u$ is a unit $(\pm 1)$, and $p_i$ are odd primes. Let $a$ be an integer. Then the Kronecker symbol $\left(\frac{a}{n} \right)_K$ is given by
\begin{equation}
    \left(\frac{a}{n} \right)_K =  \left(\frac{a}{u} \right)_K
    \left(\frac{a}{2} \right)_K^{e_0}
    \prod_{i=1}^k  \left(\frac{a}{p_i} \right)_L^{e_i}.
\end{equation}
The factor $\left( \frac{a}{u} \right)_K$ is just 1 when $u=1$. When $u=-1$, we define it by
\begin{align}
         \left(\frac{a}{-1} \right)_K =
         \begin{cases}
            -1: \quad &\text{if $a < 0$},\\
            +1:  \quad &\text{if $a > 0$}.
         \end{cases}
\end{align}
Notice that when $n$ is a positive odd integer, the Kronecker symbol is identical to the Jacobi symbol. For even $n$, we define $\left( \frac{a}{2} \right)_K$ by 
\begin{align}
     \left(\frac{a}{2} \right)_K = 
     \begin{cases}
        0: \ & \text{if $a$ is even}, \\
        +1: \ &  \text{if $a \equiv \pm1 \pmod8$}, \\
        -1: \ &  \text{if $a \equiv \pm3 \pmod8$}.
     \end{cases}
\end{align}

The Kronecker symbol has the following property \cite{Allouche: 2018}. For $a \not\equiv 3 \pmod4$, $a\neq0$, we have 
\begin{equation}
 \label{eq: Kronecker_property1}
         \left(\frac{a}{m} \right)_K =      \left(\frac{a}{n} \right)_K,
\end{equation}
whenever 
\begin{equation}
    m \equiv n \quad {\rm mod}
    \begin{cases}
       4 |a|: \quad  &\text{if}\ a \equiv 2 \pmod4, \\
       |a|: \quad &\text{otherwise}.
    \end{cases}
\end{equation}

Next, we note the following basic fact regarding the quadratic residue.
\begin{itemize}
          \item $a$ is a quadratic residue modulo $n$
          $\leftrightarrow$
          $a$ is a quadratic residue modulo $p^k$ for every prime power dividing $n$.
\end{itemize}
This is easy to check.\footnote{$(\rightarrow):$ $m^2 - a = n r = p_0^{e_0} p_1^{e_1}\cdots p_k^{e_k}r$ for some $r\in \mathbb{Z}$. Then, taking $p_j^{\alpha_j}, (0 \leq j \leq k, 1 \leq \alpha_j \leq e_j)$ gives us $m^2 \equiv a \pmod{p_j^{\alpha_j}}.$\\
$(\leftarrow)$: We have $m^2 \equiv a \pmod{p_j^{e_j}}, (0 \leq j \leq k)$. Then, $m^2 -a = r_j p_j^{e_j},\ r_j \in \mathbb{Z}$ for every $j(=0,...,k)$. If $i\neq j$, ${\rm gcd}(p_i^{e_i},p_j^{e_j})=1$ holds. Thus, $m^2 - a = p_1^{e_1} \cdots p_k^{e_k} r$ for some $r \in \mathbb{Z}$. }

Now, consider the equation
\begin{equation}
{\rm det}
    \begin{pmatrix}
  n_1 & m \\ m & n_2
    \end{pmatrix} = 7,\quad  n_1 = 7s -1,
\end{equation}
where $s \in \mathbb{Z}^+, m \in  \mathbb{Z}$. Notice that $n_1$ is a non-quadratic residue modulo $7$ because $\left(\frac{-1}{7} \right) = -1$. We obtain
\begin{equation}
\label{eq: question}
    m^2 \equiv -7 \pmod{7s-1}.
\end{equation}
We show there is no solution of eq.(\ref{eq: question}) for any positive integer, $s$.
When $s=1$, there is no solution since $-7 \equiv -1 \pmod6$ is not a quadratic residue modulo $6$. Now for arbitrary $s \in  \mathbb{Z}^+$, we have
\begin{align}
    \left(\frac{-7}{7s-1} \right)_K = \left(\frac{-7}{6} \right)_K = 
    -1,
\end{align}
because of the property Eq.(\ref{eq: Kronecker_property1}). 
Suppose $7s-1$ is prime factorized as
\begin{equation}
    7s-1 = 2^{e_0} p_1^{e_1} \cdots p_k^{e_k},
\end{equation}
where $p_i,\ (i=1,...,k)$ denotes odd primes.
 Then it is immediate 
\begin{equation}
    \left(\frac{-7}{p_1^{e_1}\cdots p_k^{e_k}} \right)_J = -1, \quad s\in \mathbb{Z}^+,
\end{equation}
holds because $\left( \frac{-7}{2} \right)_K=1$. When the Jacobi symbol is $-1$, we must have at least one $p_l^{e_l}, (1 \leq l \leq k)$ such that $-7$ is its non-quadratic residue,
\begin{equation}
    \left( \frac{-7}{p_l^{e_l}}
    \right)_J = -1.
\end{equation}
According to the aforementioned basic property of the  quadratic residue, this is sufficient to conclude that eq.(\ref{eq: question}) has no solution.

By the same argument, we conclude that 
\begin{align}
\begin{aligned}
    m^2 \equiv -7 \pmod{7s-2}, \\
    m^2 \equiv -7 \pmod{7s-4},
\end{aligned}
\end{align}
have no integer solution for any $s \in \mathbb{Z}^+$. 

\subsection{Number of zero-modes when ${\rm det}N=12$}
\label{appendix: 12}
Here, we analyze the number of positive chirality zero-modes when ${\rm det}N = 12$. We only need to consider cases when ${\rm gcd}(n_1,m) \geq {\rm gcd}(n_2,m)$ because ${\rm tr}\rho(S)$ is invariant under the interchange of $n_1$ and $n_2$. 

\paragraph{When ${\rm gcd}(n_2,m)=1$:}\ \\
Notice that ${\rm gcd}(n_{2},12)=1$ must hold. Thus, we have
\begin{align}
    \begin{aligned}
      {\rm tr}\rho(S) &=   \frac{1}{\sqrt{12}} \sum_{K=0}^{11} e^{\frac{2 \pi i }{12} n_{2}K^2} \\
      & = \frac{1}{\sqrt{12}}\sum_{t=0}^2 e^{\frac{2 \pi i }{3} n_{2}t^2}
      \sum_{s=0}^3
      e^{- \frac{2 \pi i }{4} n_{2}s^2 } \\
      &= i \left( \frac{n_{2}}{3} \right)_L \cdot \frac{1}{2}  \sum_{s=0}^3
      e^{- \frac{2 \pi i }{4} n_{2}s^2 } ,
    \end{aligned}
\end{align}
where $K = 4t + 3s$. 
Thus, we have
\begin{equation}
    {\rm tr}\rho(S) =
    \begin{cases}
 \left( \frac{n_{2}}{3} \right)_L (1+i) , \quad {\rm if\ } n_{2} \equiv 1 \pmod{4}\\
         -\left( \frac{n_{2}}{3} \right)_L (1-i)
         , \quad {\rm if\ } n_{2} \equiv -1 \pmod{4}
    \end{cases}.
\end{equation}

\paragraph{When ${\rm gcd}(n_{2},m)=3,\  {\rm gcd}(n_{1},m)=4$:}\ \\
Notice that ${\rm gcd}(n_{1},3)={\rm gcd}(n_{2},4)=1$ must hold. Thus, we have
\begin{align}
    \begin{aligned}
        {\rm tr}\rho(S) &=  \frac{1}{\sqrt{12}}
        \sum_{K_1=0}^3 \sum_{K_2=0}^2
        e^{\frac{2\pi i }{12}(n_{1}K_2^2 + n_{2}K_1^2 - 2m K_1 K_2)}\\
        &= \frac{1}{\sqrt{12}} \sum_{K_1=0}^3 \sum_{K_2 = 0}^2 e^{\frac{2 \pi i }{3} (n_{1}/4) K_2^2} e^{\frac{2\pi i}{4} (n_{2}/3)K_1^2} \\
        &= i \left( \frac{(n_{1}/4)}{3} \right)_L \cdot \frac{1}{2} \sum_{K_1=0}^3 e^{\frac{2\pi i}{4} (n_{2}/3)K_1^2}.
    \end{aligned}
\end{align}
Therefore, we find
\begin{align}
    {\rm tr}\rho(S) = \begin{cases}
       i \left( \frac{n_{1}}{3} \right)_L \cdot (1+i),\quad {\rm if\ } {n_{2}} \equiv -1 \pmod{4}, \\
        i \left( \frac{n_{1}}{3} \right)_L \cdot (1-i), \quad {\rm if\ } n_{2} \equiv 1 \pmod{4}.
    \end{cases}
\end{align}

\paragraph{When ${\rm gcd}(n_{2},m)=2,\  {\rm gcd}(n_{1},m)=3$:}\ \\
Notice that ${\rm gcd}(n_{1},2)={\rm gcd}(n_{2},3)=1$ must hold. Thus, we have
\begin{align}
    \begin{aligned}
    {\rm tr}\rho(S) &= \frac{1}{\sqrt{12}} \sum_{K_1=0}^5 \sum_{K_2=0}^1 
    e^{\frac{2 \pi i}{12} (n_{2} K_1^2 + n_{1} K_2^2 - 2 m K_1 K_2)} \\
    &= \frac{1}{\sqrt{6}}\sum_{K_1=0}^5 e^{\frac{2\pi i }{6} (n_{2}/2)K_1^2} \cdot \frac{1}{\sqrt{2}}
    \sum_{K_2=0}^1 e^{\frac{2 \pi i}{4} (n_{1}/3) K_2^2} \\
    &= \frac{1}{\sqrt{2}} \sum_{t=0}^1 e^{\frac{2 \pi i}{2}(n_{2}/2)t^2}
    \cdot 
    \frac{1}{\sqrt{3}} \sum_{s=0}^2 e^{\frac{2 \pi i}{3} (-n_{2}/2)s^2}
    \cdot \frac{1}{\sqrt{2}}
    \sum_{K_2=0}^1 e^{\frac{2 \pi i}{4} (n_{1}/3) K_2^2},
    \end{aligned}
\end{align}
where $K_1 = 3t + 2s$. Notice that $(n_{2}/2) \equiv 0 \pmod{2}$ and $(n_{1}/3) \equiv \pm 1 \pmod{4}$ are only possible. Therefore, we find
\begin{align}
\begin{aligned}
    {\rm tr}\rho(S) &= i \left( \frac{(-n_{2}/2)}{3}\right)_L \cdot (1+e^{\frac{2 \pi i}{4} (n_{1}/3)}) \\
    &= 
    \begin{cases}
       i \left( \frac{n_{2}}{3}\right)_L \cdot (1+i), \quad
       {\rm if\ } n_{1} \equiv -1 \pmod{4},\\
       i \left( \frac{n_{2}}{3}\right)_L \cdot (1-i),\quad
       {\rm if\ } n_{1} \equiv 1 \pmod{4}.
       \end{cases}
\end{aligned}
\end{align}

\paragraph{When
${\rm gcd}(n_{2},m)={\rm gcd}(n_{1},m)=2$:}\ \\
Notice that ${\rm gcd}(n_1,3)={\rm gcd}(n_2,3)=1$ must hold. Thus, we have
\begin{align}
    \begin{aligned}
        {\rm tr}\rho(S)
        = \frac{1}{\sqrt{12}}
        \sum_{K_1=0}^5 \sum_{K_2=0}^1 e^{\frac{2 \pi i}{6}((n_{2}/2)K_1^2 + (n_{1}/2)K_2^2 - 2 (m/2)K_1 K_2)}.
    \end{aligned}
\end{align}
Let us define a new variable by 
\begin{equation}
    K_3 = K_1 - g K_2.
\end{equation}
This leads to 
\begin{align}
    \begin{aligned}
        {\rm tr}\rho(S) &=
        \frac{1}{\sqrt{12}}
        \sum_{K_1=0}^5 \sum_{K_2=0}^1 e^{\frac{2 \pi i}{6}[((n_{1}/2)+g^2(n_{2}/2) - 2 (m/2)g)K_2^2 + (n_{2}/2)K_3^2 + 2 K_2 K_3 (g (n_{2}/2) - (m/2))] } \\
        &=
        \frac{1}{\sqrt{12}}
        \sum_{K_2=0}^1 \sum_{K_3=0}^5 e^{\frac{2 \pi i}{6}[((n_{1}/2)-g^2(n_{2}/2))K_2^2 + (n_{2}/2)K_3^2] },
    \end{aligned}
\end{align}
where we used the fact that we can choose $g \in \mathbb{Z}$ to satisfy $g (n_{2}/2) \equiv (m/2) \pmod{3}$ for any $(m/2) \in \mathbb{Z}$ since $(n_{2}/2) \not\equiv 0 \pmod{3}$. We notice that $ (n_{1}/2)-g^2(n_{2}/2) \equiv 0 \pmod{3}$.\footnote{Consider the determinant,  $n_{1}n_{2}-m^2=12$. Dividing both sides by $4$, we get $(n_{1}/2)(n_{2}/2)-(m/2)^2 = 3$. Then we have
$(n_{2}/2)[(n_{1}/2)-g^2(n_{2}/2)]=3$. Since $(n_{2}/2)$ and 3 are co-prime, our claim is verified.} Here, we define $3 \tilde{n}:= (n_{1}/2)-g^2(n_{2}/2)$. Then we find 
\begin{align}
\begin{aligned}
       {\rm tr}\rho(S) &= \frac{1}{\sqrt{2}} \sum_{K_2=0}^1 e^{ \pi i\tilde{n}K_2^2 } \cdot \frac{1}{\sqrt{6}} \sum_{K_3 = 0}^5 e^{\frac{2\pi i}{6}(n_2/2)K_3^2} \\
    &= \frac{1}{2} (1+(-1)^{\tilde{n}}) \cdot (1+(-1)^{(n_{2}/2)}) \cdot i \left( \frac{-(n_{2}/2)}{3}
    \right)_L \\
    &= \frac{1}{2}(1+(-1)^{(n_{1}/2)}) \cdot (1+(-1)^{(n_{2}/2)}) \cdot i \left( \frac{n_{2}}{3}
    \right)_L,
\end{aligned}
\end{align}
where we used the fact $\tilde{n} \equiv (n_{1}/2) - g^2 (n_{2}/2) \pmod{2}$. 
\paragraph{When ${\rm gcd}(n_{2},m)=2, {\rm gcd}(n_{1},m)=6$:}\ \\
Note that ${\rm gcd}(n_{2},3)=1$ must hold. Thus, we have
\begin{align}
    \begin{aligned}
        {\rm tr}\rho(S) &=
        \frac{1}{\sqrt{12}} \sum_{K_1=0}^5 \sum_{K_2=0}^1 e^{\frac{2 \pi i}{12}(n_{1} K_2^2 + n_{2}K_1^2 - 2m K_1 K_2)} \\
        &= \frac{1}{\sqrt{6}} \sum_{K_1=0}^5 e^{\frac{2 \pi i}{6}(n_{2}/2)K_1^2} \cdot \frac{1}{\sqrt{2}} \sum_{K_2=0}^1 e^{\frac{2 \pi i}{2} (n_{1}/6)K_2^2} \\
        &= \frac{1}{2} (1+(-1)^{(n_{2}/2)})(1+(-1)^{(n_{1}/6)}) \cdot i \left(\frac{n_{2}}{3}\right)_L.
    \end{aligned}
\end{align}
Results are summarized in Table \ref{tb: 12}.

\begin{table}[H]
\begin{center}
\begin{tabular}{|c||c|} \hline
${\rm tr}[\rho(S)]$ & conditions     \\ \hline \hline 
$1+i$ & \begin{tabular}{c}
$n_{k} \stackrel{\rm{mod}4}{\equiv} 1 ,\  \left(\frac{n_{k}}{3} \right)_L = 1, \quad   (  k=1\ {\rm or\ }2)$, \\
{\rm or} \\
$n_{1} \stackrel{\rm{mod}4}{\equiv} 1,\ \left(\frac{n_{2}}{3} \right)_L = 1$, \\
{\rm or} \\
$ n_{2} \stackrel{\rm{mod}4}{\equiv} 1,\ \left(\frac{n_{1}}{3} \right)_L = 1$. 
\end{tabular}   \\ \hline
$1-i$ &  \begin{tabular}{c}
$n_{k} \stackrel{\rm{mod}4}{\equiv} -1 ,\  \left(\frac{n_{k}}{3} \right)_L = -1, \quad   (  k=1\ {\rm or\ }2)$, \\
{\rm or} \\
$n_{1} \stackrel{\rm{mod}4}{\equiv} -1,\ \left(\frac{n_{2}}{3} \right)_L = -1$, \\
{\rm or} \\
$ n_{2} \stackrel{\rm{mod}4}{\equiv} -1,\ \left(\frac{n_{1}}{3} \right)_L = -1$. 
\end{tabular} 
\\ \hline
$-1+i$ & 
 \begin{tabular}{c}
$n_{k} \stackrel{\rm{mod}4}{\equiv} -1 ,\  \left(\frac{n_{k}}{3} \right)_L = 1, \quad   (  k=1\ {\rm or\ }2)$, \\
{\rm or} \\
$n_{1} \stackrel{\rm{mod}4}{\equiv} -1,\ \left(\frac{n_{2}}{3} \right)_L = 1$, \\
{\rm or} \\
$ n_{2} \stackrel{\rm{mod}4}{\equiv} -1,\ \left(\frac{n_{1}}{3} \right)_L = 1$. 
\end{tabular} 
\\ \hline
$-1-i$ & 
 \begin{tabular}{c}
$n_{k} \stackrel{\rm{mod}4}{\equiv} 1 ,\  \left(\frac{n_{k}}{3} \right)_L = -1, \quad   (  k=1\ {\rm or\ }2)$, \\
{\rm or} \\
$n_{1} \stackrel{\rm{mod}4}{\equiv} 1,\ \left(\frac{n_{2}}{3} \right)_L = -1$, \\
{\rm or} \\
$ n_{2} \stackrel{\rm{mod}4}{\equiv} 1,\ \left(\frac{n_{1}}{3} \right)_L = -1$. 
\end{tabular} 
\\ \hline
$2i$ &
 \begin{tabular}{c}
$  n_{1} \stackrel{\rm{mod}4}{\equiv} n_{2} \stackrel{\rm{mod}4}{\equiv} 0,\  
    \left(\frac{n_{k}}{3} \right)_L = 1,\ (k=1,{\rm or}\ 2)$, 
\end{tabular} 
\\
\hline
$-2i$ & $  n_{1} \stackrel{\rm{mod}4}{\equiv} n_{2} \stackrel{\rm{mod}4}{\equiv} 0,\  
    \left(\frac{n_{k}}{3} \right)_L = -1,\ (k=1,{\rm or}\ 2)$,\\
\hline
$0$ &  $n_k \equiv 0 \pmod{4},\quad (k=1, {\rm or}\ 2)$. \\
\hline
\end{tabular}
\end{center}
\caption{${\rm tr}\rho(S)$ when ${\rm det}N = 12$}
\label{tb: 12}
\end{table}

In fact, there are no symmetric positive definite $N$ which correspond to ${\rm tr}\rho(S)=1-i, -1-i, -2i$. This can be proven in a similar way presented in Appendix \ref{appendix: existence_of_N}. Then we can understand the result shown in Table \ref{tb: T4/Z4_1-19}.

\subsection{Number of zero-modes when ${\rm det}N=p^2$ ($p$ is an odd prime)}
\label{appendix: p^2}
Here, we prove eq.(\ref{eq: p^2}). We have ${\rm gcd.}(n_{1},m)= {\rm gcd}(n_{2},m)=p$. The trace is given by
\begin{equation}
\label{eq: trace_p_p}
    {\rm tr}\rho(S) = \frac{1}{p}\sum_{K_1=0}^{p-1} \sum_{K_2=0}^{p-1} e^{\frac{2 \pi i}{p}(n_{1}' K_2^2 + n_{2}'K_1^2 - 2 m' K_1 K_2)},
\end{equation}
where we defined $n_{1}=p n'_{1}$, $n_{2}=p n_{2}'$, and $m=pm'$. 
\paragraph{When $n_{i}' \not\equiv 0 \pmod{p},\ (i=1\ {\rm or}\ 2)$ :}\ \\
It suffices to consider the case $n_1' \not\equiv 0 \pmod{p}$.
We transform the variable as
\begin{equation}
\label{eq: trans}
    K_3 := K_2 - g K_1,
\end{equation}
where $g \in \mathbb{Z}$. Then we can write eq.(\ref{eq: trace_p_p}) as 
\begin{equation}
     {\rm tr}\rho(S) = \frac{1}{p}\sum_{K_1=0}^{p-1} \sum_{K_3=0}^{p-1} e^{\frac{2 \pi i}{p} [ (g^2 n_{1}' + n_{2}' - 2 m' g) K_1^2+ n_{1}' K_3^2 + 2(n_{1}' g - m') K_1 K_3 ]}.
\end{equation}
We can always find $g \in \mathbb{Z}$ such that $n_{1}'g - m' \equiv 0 \pmod{p}$.\footnote{$g$ can take $\{ 0,1,2,...,p-1, \pmod{p} \}$. Then we have $\{ n_{1}'g \} =\{ 0, n_{1}', 2n_{1}',...,(p-1)n_{1}', \pmod{p}\}$. Since ${\rm gcd}(n_{1}',p)=1$, we get an equality of set as $\{ n_{1}'g \}=\{ 0, 1, 2,...,(p-1), \pmod{p}\}$. This shows the existence of $g$ satisfying $n_{1}'g-m'\equiv 0$ for any given $m'$.} Having chosen such a $g$, we obtain
\begin{align}
\begin{aligned}
      {\rm tr}\rho(S) &= \frac{1}{p}\sum_{K_1=0}^{p-1} \sum_{K_3=0}^{p-1} e^{\frac{2 \pi i}{p} [ (g^2 n_{1}' + n_{2}' - 2 m' g) K_1^2+ n_{1}' K_3^2 ]}\\
      &= (-1)^{\frac{p-1}{2}} \left( \frac{g^2 n_{1}' + n_{2}' - 2 m' g}{p}
      \right)_L \left( \frac{n_{1}'}{p} \right)_L \\
      &= (-1)^{\frac{p-1}{2}} \left( \frac{ n_{2}' -  m' g}{p}
      \right)_L \left( \frac{n_{1}'}{p} \right)_L \\
      &=(-1)^{\frac{p-1}{2}} \left( \frac{n_{1}'n_{2}' - { m'}^2}{p}
      \right)_L \\
      &=(-1)^{\frac{p-1}{2}} \left( \frac{1}{p} \right)_L= (-1)^{\frac{p-1}{2}}.
\end{aligned}
\end{align}

\paragraph{When $n_{i}' \equiv 0 \pmod{p},\ (i=1\ {\rm and}\ 2)$ :}\ \\
We consider the case $n_{1}' \equiv n_{2}' \equiv 0 \pmod{p}$.\footnote{This means we have ${m'}^2 + 1 = 0 \pmod{p}$ where $p$ is an odd prime. From eq.(\ref{eq: (-1/p)}), it follows that there is no solution $m' \in \mathbb{Z}$ if $p \equiv 3 \pmod{4}$. The solution exists if and only if $p \equiv 1 \pmod{4}$. Thus, we could have restricted our discussion to odd primes congruent to 1 modulo 4.} In this case, we have
\begin{equation}
    {\rm tr}\rho(S) = \frac{1}{p} \sum_{K_1=0}^{p-1} \sum_{K_2 = 0}^{p-1} e^{\frac{2 \pi i}{p}(-2m'K_1K_2)}.
\end{equation}
  Let us define $L_1$ and $L_2$ as
  \begin{equation}
  \label{eq: magnification}
      \begin{pmatrix}
      K_1 \\ K_2
      \end{pmatrix}
       = L_1 
      \begin{pmatrix}
      p-1 \\ 1 
      \end{pmatrix}
      + L_2
      \begin{pmatrix}
      -1 \\ p-1
      \end{pmatrix}.
  \end{equation}
  Even after this transformation, the new variables $L_1, L_2$ move from $0$ to $p-1$. To understand it, consider 
  \begin{align}
  \label{eq: L_1}
        L_1(p-1) - L_2 &= L_1' (p-1) - L_2' + p r, \\
  \label{eq: L_2}
      L_1 + (p-1)L_2 &= L_1' + (p-1)L_2' + p s,
  \end{align}
 where $r, s \in \mathbb{Z}$. We verify our claim if we could show that all solutions of above equations satisfy $L_1 \equiv L_1' \pmod{p}$ and $L_2 \equiv L_2' \pmod{p}$ simultaneously. 
 Eq.(\ref{eq: L_1})$\times (p-1)$ + Eq.(\ref{eq: L_2}) gives us
 \begin{equation}
     2 (L_1 - L_1') \equiv 0 \pmod{p}.
 \end{equation}
Since we have ${\rm gcd}(p,2)=1$, the relation $L_1 \equiv L_1' \pmod{p}$ must be satisfied. Substituting this result into Eq.(\ref{eq: L_1}) yields $L_2 \equiv L_2' \pmod{p}$. Thus, we verified our claim. 
Consequently, we get
\begin{align}
    \begin{aligned}
        {\rm tr}\rho(S) &= \sum_{L_1=0}^{p-1} \sum_{L_2=0}^{p-1} e^{\frac{2 \pi i}{p}[-2m'(p-1)^2 L_1^2 + 2 m' (p-1) L_2^2]} \\
        &= \left(\frac{-2m'(p-1)}{p} \right)_L \left( \frac{2m'(p-1)}{p}
        \right)_L \\
        &= \left( \frac{-1}{p}
        \right)_L = (-1)^{\frac{p-1}{2}}.
    \end{aligned}
\end{align}

\section{$T^4/Z_2$ permutation orbifold}
\label{appendix: z2_per}
Here, we study the $T^4/Z_2$ permutation orbifold and zero-mode number on it. $(T^2 \times T^2) /Z_2$ permutation orbifold was studied in 
Ref. \cite{Kikuchi:2020nxn}. Our results are consistent with the previous works and more general in the sense that we took into account oblique components of fluxes and complex structure moduli. Moreover, we will see that orbifolding of $T^4$ by the $Z_2$ permutation is related to the modular transformation.

\subsection{$Z_2$ permutation}
We define $Z_2$ permutation as ${z}_1 \leftrightarrow z_2$. This can be realized by the modular transformation shown in eq.(\ref{eq: permutation_trans}).
The two complex coordinates are interchanged 
\begin{equation}
    \vec{z} = \begin{pmatrix}
    z_1 \\ z_2
    \end{pmatrix}
    \xrightarrow{\gamma_P} A_P \vec{z}
    = 
    \begin{pmatrix}
    z_2 \\ z_1
    \end{pmatrix}.
\end{equation}

Now, let us look at the conditions of ${ N}$ and $\Omega$ for being invariant under the $\gamma_P$ transformation.

For ${ N}$,
\begin{equation}
    {{N}} = 
    \begin{pmatrix}
    n_{11} & n_{12} \\
    n_{21} & n_{22}
    \end{pmatrix}
    \xrightarrow{\gamma_P} 
    A_P {N} A_P^{-1} =
    \begin{pmatrix}
    n_{22} & n_{21} \\
    n_{12} & n_{11}
    \end{pmatrix},
\end{equation}
therefore we need 
\begin{equation}
    n_{11} = n_{22},\ n_{12} = n_{21}.
\end{equation}

For $\Omega$,
\begin{equation}
    \Omega = 
    \begin{pmatrix}
    \tau_1 & \tau_3 \\
    \tau_3 & \tau_2
    \end{pmatrix}
     \xrightarrow{\gamma_P}
    A_P \Omega A_P^{\rm T} =
    \begin{pmatrix}
    \tau_2 & \tau_3 \\
    \tau_3 & \tau_1
    \end{pmatrix},
\end{equation}
therefore we need
\begin{equation}
    \tau_1 = \tau_2.
\end{equation}

\subsection{$T^4/Z_{2}$ permutation orbifold}
The algebraic relation, $\gamma_{P}^2 = I$ suggests that we can construct a $Z_2$ permutation orbifold where two complex coordinates $z_1$ and $z_2$ are identified. Then the flux must be of the form 
\begin{equation}
    { N} = 
    \begin{pmatrix}
    n & m \\ m & n
    \end{pmatrix},
\end{equation}
for the $Z_2^{(per)}$ invariance. The complex structure needs to be 
\begin{equation}
    \Omega = 
    \begin{pmatrix}
    \tau_1 & \tau_3 \\
    \tau_3 & \tau_1
    \end{pmatrix},
\end{equation}
for the $Z_2^{(per)}$ invariance. The number of $Z_2^{(per)}$ zero-modes is
\begin{align}
    \begin{aligned}
        \# (Z_2^{(per)}\ {\rm even\ mode}) &=  \frac{{\rm det}N+(n+m)}{2},\\
        \# (Z_2^{(per)}\ {\rm odd\ mode}) &= \frac{{\rm det}N-(n+m)}{2}.
    \end{aligned}
\end{align}
The above results can be understood by referring to eq.(\ref{eq: A_trans_zero}). $Z_2^{(per)}$ invariant modes lie on the line $J_1 = J_2$ on the $\vec{J}$ plane. Its number corresponds to the length of the diagonal line of the diamond spanned by vectors $N^{\rm T} \vec{e}_i, (i=1,2)$.

\section{Negative chirality zero-mode wavefunction}
\label{appendix: negative}
Here we study the negative chirality zero-mode wavefunctions which become non-zero when ${\rm det}N <0$. Our results are consistent with ref.\cite{Antoniadis:2009bg} when the complex structure moduli are purely imaginary. We have also studied when ${\rm Re}\Omega$ is non-vanishing, and obtained solutions satisfying both Dirac equation and proper boundary conditions even in such a case.
\subsection{Dirac equation for negative chirality wavefunction}
We have already seen that two negative chirality components $\psi_-^1$ and $\psi_-^2$ satisfy the Dirac equation, eqs.(\ref{eq: 3}) and (\ref{eq: 4}). 
As it is clarified in ref.\cite{Antoniadis:2009bg}, $\psi_-^1$ and $\psi_-^2$ need to be excited simultaneously when oblique fluxes are present.\footnote{This is traced back to the non-commutativity, $[D_{z^i}, \bar{D}_{\bar{z}^j}] = F_{z^i \bar{z}^j} \neq 0$ as in eq.(\ref{eq: Fzbz}).} Thus, we write
\begin{equation}
\label{eq: negative_wave}
    \psi_-^1 = \alpha \psi,\quad 
    \psi_-^2 = \beta \psi,
\end{equation}
where $\alpha$ and $\beta$ are constants of amplitudes. Substituting eq.(\ref{eq: negative_wave}) into eqs.(\ref{eq: 3}) and (\ref{eq: 4}) gives us 
\begin{align}
    \begin{aligned}
\label{eq: negative_psi}
  (\beta D_{z^1} + \alpha D_{z^2}) \psi &= 0, \\
    ( \beta \bar{D}_{\bar{z}^2} \psi - \alpha  \bar{D}_{\bar{z}^1} )\psi &= 0.
    \end{aligned}
\end{align}
For $\psi$ to be non-zero,
\begin{equation}
\label{eq: negative_condition}
    - \alpha \beta F_{z^1\bar{z}^{1}} 
    - \alpha^2 F_{z^2 \bar{z}^1}
    + \beta^2 F_{z^1 \bar{z}^2}
    + \alpha \beta F_{z^2 \bar{z}^2}
    = 0,
\end{equation}
is necessary. 
When we introduce flux $N, ({\rm det}N < 0)$, $q=\frac{\beta}{\alpha}$ is fixed to satisfy eq.(\ref{eq: negative_condition}).

\subsection{Solution}
Let us solve the Dirac equations under the boundary conditions in eq.(\ref{eq: boundaryT4}).
Our argument is based on the  basic fact that a parity transformation flips the chirality of spinors. 
 Here, we study when $N$ and $\Omega$ are of the form,
\begin{align}
N &= 
  \begin{pmatrix}
  n & m \\
  m & n
  \end{pmatrix},\ {\rm det}N < 0,\\ 
\Omega &= 
\begin{pmatrix}
\tau_1 & \tau_3 \\
\tau_3 & \tau_1
\end{pmatrix} \in \mathcal{H}_2,
\end{align}
for simplicity. Note that our choice of $N$ and $\Omega$ is consistent with the F-flat condition in eq.(\ref{eq: SUSY_condition}). 

Firstly, eq.(\ref{eq: negative_condition}) gives us two solutions for $q$,
\begin{equation}
    q = \frac{\beta}{\alpha} = \pm 1.
\end{equation}

Secondly, consider the parity transformation of the complex coordinates as shown in eq.(\ref{eq: P_2}). 
In the new coordinate system $\vec{z}^{\, (p)}$, the flux is
\begin{equation}
    N_{(p)}=
    \begin{pmatrix}
    m & n \\ n & m
    \end{pmatrix},\ {\rm det}N_{(p)} > 0.
\end{equation}
Complex structure moduli are
\begin{align}
\begin{aligned}
\Omega_{(p)} = 
    \begin{pmatrix}
{\rm Re}\tau_3 + i {\rm Im}\tau_1 & {\rm Re}\tau_1 + i {\rm Im}\tau_3 \\
{\rm Re}\tau_1 + i {\rm Im}\tau_3 & 
{\rm Re}\tau_3 + i {\rm Im}\tau_1
    \end{pmatrix}
&= \begin{pmatrix}
0 & 1 \\ 1 & 0 
\end{pmatrix}
{\rm Re}\Omega + i {\rm Im}\Omega \in \mathcal{H}_2.
\end{aligned} 
\end{align}
This shows that we obtain  positive chirality solutions in the new coordinate system. We should also note that the F-flat condition is maintained even after the parity transformation. 
 
If $N_{(p)} {\rm Im}{\Omega}_{(p)} > 0$, we obtain
\begin{equation}
\label{eq: Parity_+}
    \psi^{\vec{j}_{(p)}}_{N_{(p)}} (\vec{z}^{\, (p)}, \Omega_{(p)}) 
    = 
    \mathcal{N}_{\vec{j}_{(p)}} \cdot e^{\pi i [N_{(p)} \vec{z}^{\, (p)}]^{\rm T} \cdot ({\rm Im}\Omega_{(p)})^{-1} \cdot {\rm Im}\vec{z}^{\, (p)}} \cdot \vartheta
    \begin{bmatrix}
     {\vec{j}_{(p)}}^{\, \rm T} \\ 0
    \end{bmatrix}({N}_{(p)}\vec{z}^{\, (p)}, {N}_{(p)} \Omega_{(p)}),\quad  N_{(p)}^{\rm T} \vec{j}_{(p)} \in \mathbb{Z}^2,
\end{equation}
in the first positive chirality component, $\psi_+^1$ as we have reviewed in Section \ref{sec:Revew}.\footnote{ Here, normalization of the labels $\vec{j}^{(p)}$ for zero-modes is the same as refs.\cite{Cremades:2004wa, Antoniadis:2009bg} for explicit comparisons.
We use the notation $\Vec{J}$ in the main text, and they are related by $\vec{J}=
N^{\rm T} \vec{j}\pmod{N^{\rm T} \vec{e}_n}$. }
Maintaining the profile of wavefunctions, we describe positions on $T^4$ in terms of the initial complex coordinates $\vec{z}$. Then we expect that negative chirality wavefunctions we sought will be obtained. 
The results are as follows,
\begin{align}
\begin{aligned}
\label{eq: N_chiral_sol}
  \psi_{N,M}^{\vec{j}_{(p)}}(\vec{z},\vec{\bar{z}}) 
  & :=
  \psi^{\vec{j}_{(p)}}_{N^{(p)}} (\vec{z}^{\, (p)}, \Omega^{(p)})  
  = 
  \mathcal{N} \cdot f(\vec{z},\vec{\bar{z}}) \cdot  \hat{\Theta}(\vec{z},\vec{\bar{z}}), \\
   f(\vec{z},\vec{\bar{z}}) &=
    e^{{\pi i} ([\hat{N}\vec{z}]^{\rm T}({\rm Im}\Omega)^{-1}{\rm Im}\vec{z} - [\tilde{N}\vec{\bar{z}}]^{\rm T}({\rm Im}\Omega)^{-1}{\rm Im}\vec{\bar{z}})}, \\
    \hat{\Theta}(\vec{z},\vec{\bar{z}})
    &=
    \sum_{\vec{m} \in \mathbb{Z}^2}
    e^{\pi i  [(\vec{m} + \vec{j}_{(p)})^{\rm T}(Mi {\rm Im}\Omega + N{\rm Re}\Omega)(\vec{m} + \vec{j}_{(p)})]}
    e^{2 \pi i (\vec{m} + \vec{j}_{(p)})^{\rm T} [\hat{N}\vec{z}]} 
    e^{2 \pi i (\vec{m} + \vec{j}_{(p)})^{\rm T}[\tilde{N}\vec{\bar{z}}]},
\end{aligned}
\end{align}
where
\begin{align}
\hat{N} = \frac{m+n}{2}
\begin{pmatrix}
1 & 1 \\ 1 & 1
\end{pmatrix},\quad
\tilde{N} =
\frac{n-m}{2} 
\begin{pmatrix}
1 & - 1\\ -1 & 1
\end{pmatrix},\quad
 M = 
\begin{pmatrix}
    m & n \\ n & m
\end{pmatrix}.
\end{align}
When ${\rm Re}\Omega = 0$, our solution
is consistent with the result obtained in ref.\cite{Antoniadis:2009bg}. 

Thirdly, we should confirm that eq.(\ref{eq: N_chiral_sol}) indeed satisfies the Dirac equation in eq.(\ref{eq: negative_psi}).
We can check that the obtained solutions are correct
with the choice $q=\frac{\beta}{\alpha}=-1$ through direct substitutions. On the other hand,  another solution, $q=+1$ is selected when ${N_{(p)}}{\rm Im}\Omega_{(p)}<0$.
In this case, the negative chirality solution is modified from eq.(\ref{eq: N_chiral_sol}). However, it can be easily obtained by noting the fact that eq.(\ref{eq: negative_definite}) is the right choice instead of eq.(\ref{eq: Parity_+}) if ${N_{(p)}}{\rm Im}\Omega_{(p)}$ is negative definite.

Lastly, boundary conditions shown in eq.(\ref{eq: boundaryT4}) need to be checked. We can check this directly, but it is easier to use the parity transformation.
Following translations in the coordinate system $\vec{z}^{\, (p)}$,
\begin{equation}
    \vec{z}^{\, (p)} \rightarrow \vec{z}^{\, (p)} + \vec{e}_{1, (2)},
\end{equation}
correspond to 
\begin{equation}
    \vec{z}\rightarrow \vec{z} + \vec{e}_{2, (1)},
\end{equation}
in the initial coordinate system $\vec{z}$. Thus, we just need to confirm
\begin{align}
\begin{aligned}
\chi_{\vec{e}_1}^{(p)} &:= \pi [N_{(p)} ({\rm Im}\Omega_{(p)})^{-1} {\rm Im}\vec{z}^{\, (p)}]_1 = \pi [N ({\rm Im}\Omega)^{-1} {\rm Im}\vec{z}]_2 
=
\chi_{\vec{e}_2}, \\
\chi_{\vec{e}_2}^{(p)} &:= \pi [N_{(p)} ({\rm Im}\Omega_{(p)})^{-1} {\rm Im}\vec{z}^{\, (p)}]_2 = \pi [N ({\rm Im}\Omega)^{-1} {\rm Im}\vec{z}]_1 
=
\chi_{\vec{e}_1},
\end{aligned}
\end{align}
so that wavefunctions receive correct $U(1)$ phase transformations. Similarly, 
\begin{equation}
    \vec{z}^{\, (p)} \rightarrow \vec{z}^{\, (p)} + \Omega_{(p)} \vec{e}_k,\ (k=1,2)
\end{equation}
is equivalent to 
\begin{equation}
    \vec{z}\rightarrow \vec{z} + \Omega \vec{e}_{k},\ \ (k=1,2).
\end{equation}
Thus, we just need to verify 
\begin{align}
\begin{aligned}
\chi_{\Omega_{(p)} \vec{e}_k}^{(p)} 
&:=
\pi {\rm Im} [N_{(p)} \bar{\Omega}_{(p)} ({\rm Im}\Omega_{(p)})^{-1} \vec{z}^{\, (p)}]_k = 
\pi{\rm Im}[N \bar{\Omega} ({\rm Im}\Omega)^{-1} \vec{z}]_k
=
\chi_{\Omega \vec{e}_k},
\end{aligned}
\end{align}
which is straightforward.

\section{F-flatness supersymmetry condition}
\label{appendix: F-flat}
We review the F-flatness supersymmetry condition in the 10D $\mathcal{N}=1$ super Yang-Mills theory (SYM).
\subsection{10D $\mathcal{N}=1$ SYM}
The action of 10D $\mathcal{N}=1$ SYM with $U(N)$ gauge symmetry is given by 
\begin{equation}
\label{eq: SYM_action}
    S_{10D} = \int d^{10}X \sqrt{-G} \frac{1}{g^2} {\rm Tr} \left[ -\frac{1}{4}F^{MN}F_{MN} + \frac{i}{2}\bar{\lambda} \Gamma^M D_M \lambda
    \right],
\end{equation}
where the trace is taken over the gauge space of the adjoint representation. The gauge coupling constant is denoted by $g$.
We first consider $T^4 \times T^2 \simeq \mathbb{R}^4/\Lambda_4 \times \mathbb{R}^2 / \Lambda_2$ compact space, where $\Lambda_{4}$ and $\Lambda_2$ represent 4D and 2D lattices respectively. Thus, we can take real orthogonal coordinates over the 10D space-time and they are denoted by $X^M, (M=0,1,...,9)$. Metric tensor in this basis is given by $G_{MN}=\eta_{MN}={\rm diag}(-1,1,\cdots,1)$. 10D gauge field is denoted by $A_M$ and 10D Majorana-Weyl spinor is denoted by $\lambda$. Its field strength and the covariant derivatives are given by
\begin{align}
\begin{aligned}
    F_{MN} &= \partial_{M}A_N - \partial_N A_M -i[A_M, A_N], \\
    D_M \lambda &= \partial_M \lambda -i[A_M,\lambda].
\end{aligned}
\end{align}
 Suppose that $\Lambda_4$ is spanned by $e_i \in \mathbb{C}^2, (i=1,2,3,4)$ where $e_1$ and $e_3$ have the same norm and perpendicular to each other. Then, by a $SO(4)$ rotation in $\mathbb{R}^4$, we can write their components as eq.(\ref{eq: lattice_vectors}). Similarly, $\Lambda_2$ is spanned by $e_5 = 2\pi r$ and $e_6 = (2\pi r) \tau$ where $\tau \in \mathbb{C}$ and $r$ is a scale factor. 


For convenience, we define complex coordinates on $T^6 \simeq T^4 \times T^2$ as,
\begin{align}
    \begin{aligned}
          z^1 =\frac{1}{2\pi R} (X^{4}+iX^5),\ z^2 =\frac{1}{2\pi R} (X^6 + iX^7),\ z^3 =\frac{1}{2\pi r} (X^8 + iX^9).
    \end{aligned}
\end{align}
It follows that corresponding gauge fields are given by
\begin{equation}
A_{z^1} = \frac{2\pi R}{2}(A_4 - i A_5),\ 
A_{z^2} = \frac{2\pi R}{2}(A_6 - i A_7),\ 
A_{z^3} = \frac{2\pi r}{2}(A_8 - i A_9).
\end{equation}
Note that $z^1$ and $z^2$ are identified as those defined on the right-hand side of eq.(\ref{eq: complex_corrdinates_T4}). We have shift identifications of the form $\vec{z} \sim \vec{z} + m^a \vec{e}_a + n^b \Omega \vec{e}_b,\ (m^{a(=1,2,3)}, n^{b(=1,2,3)} \in \mathbb{Z})$, where
\begin{equation}
    \Omega = 
    \begin{pmatrix}
    \tau_1 & \tau_3 & 0 \\
    \tau_4 & \tau_2 & 0 \\
    0 & 0 & \tau
    \end{pmatrix}.
\end{equation}
The metric corresponding to the complex coordinates is written as,
\begin{equation}
\label{eq: metric_tori1}
    ds_{6D}^2 = 2 h_{i \bar{j}} dz^i d\bar{z}^{{j}},
\end{equation}
where $h_{i\bar{j}}=\frac{1}{2}(2\pi R_i)^2\  \delta_{i\bar{j}}, (i,j=1,2,3)$ and $R_1=R_2=R$ and $ R_3 = r$. 
Then we find
\begin{equation}
\label{eq: metric_tori2}
    h_{i\bar{j}} = (2\pi R_i)^2 \delta_{i\bar{j}} = \delta_{\rm i\bar{j}}\ e_{i}^{\ \rm i}\  \bar{e}_{\bar{j}}^{\ \rm {\bar{j}}},
\end{equation}
where vielbeins are given by 
\begin{align}
    e_i^{\ \rm i} &= (2\pi R_i)\delta_i^{\ \rm i},\quad  (i = 1,2,3),
\end{align}
therefore Roman indices correspond to the local Lorentz frame. We write the inverse and the complex conjugates of the vielbeins as
\begin{equation}
    e_{\rm i}^{\ i} e_{i}^{\ \rm j} = \delta_{\rm i}^{\ \rm j},\quad 
    e_{i}^{\ \rm i}e_{\rm i}^{\ j} = \delta_{i}^{\ j},\quad \bar{e}_{\bar{i}}^{\ \bar{\rm i}} = ({e}_{{i}}^{\ {\rm i}})^*.
\end{equation}
We raise and lower Italic indices by $h^{\bar{i} j}$ and $h_{i \bar{j}}$ respectively. Similarly, Roman indices are raised and lowered by $\delta^{\rm  \bar{i} j }$ and $\delta_{\rm i \bar{j}}$ respectively.

Orbifolds of tori are obtained by identifying positions related by rotational symmetries. Then the moduli are usually fixed. We still have the same metric as eqs.(\ref{eq: metric_tori1}) and (\ref{eq: metric_tori2}) in the bulk.

\subsection{Superfield description}
We discuss the F-flat condition based on the 4D $\mathcal{N} = 1$ superfield description of the 10D SYM action\cite{Abe:2012ya}. 
The 10D Majorana-Weyl spinor $\lambda$ is decomposed into four 4D Weyl spinors as
\begin{equation}
    (\lambda_0,\lambda_{z^1},\lambda_{z^2},\lambda_{z^3}) = (\lambda_{+++},\lambda_{+--},\lambda_{-+-},\lambda_{--+}),
\end{equation}
where $\pm$ represents the chirality. More specifically, we have
\begin{equation}
\Gamma^{(i)}_c \lambda_0 = +\lambda_0,\quad \Gamma^{(i)}_c \lambda_{z^j} = + \lambda_{z^j}\ (i=j),\quad \Gamma^{(i)}_c \lambda_{z^j} = - \lambda_{z^j}\ (i\neq j) ,
\end{equation}
where $\Gamma^{(i)}_c$ denotes the chirality operators defined on each of the complex coordinates $z^{i=1,2,3}$. Note that other four spinor components $(\lambda_{---}, \lambda_{-++},\lambda_{+-+}, \lambda_{++-})$ are projected out by the Majorana-Weyl condition. The 10D gauge field $A_M$ is decomposed into a 4D vector $A_{\mu}$ and complex scalars $A_{z^1},A_{z^2},A_{z^3}$,
where $\mu(=0,1,2,3)$ denotes the indices of the 4D Minkowski space-time.
These fermionic and bosonic fields become component fields of one vector $V$ and three chiral supermultiplets $\phi_{z^i}$ in the 4D $\mathcal{N}=1$ superfield description,
\begin{align}
\begin{aligned}
V &:= -\theta \sigma^{\mu} \bar{\theta} A_{\mu} + i\bar{\theta}\bar{\theta}\theta \lambda_0 - i \theta \theta \bar{\theta} \bar{\lambda}_0 + \frac{1}{2}\theta \theta \bar{\theta} \bar{\theta} D, \\
\phi_{z^i} &:= \frac{1}{\sqrt{2}} A_{z^i} + \sqrt{2}\theta \lambda_{z^i} + \theta \theta F_{z^i},
\end{aligned}
\end{align}
where $D$ and $F_{z^i}$ are auxiliary fields. Grassmann coordinates of 4D $\mathcal{N}=1$ superspace are denoted by $\theta$ and $\bar{\theta}$.\footnote{We mostly follow the conventions in ref.\cite{Wess:1992cp}.} In terms of these superfields, the action in eq.(\ref{eq: SYM_action}) can be  expressed as \cite{ArkaniHamed:2001tb}
\begin{equation}
    S_{10D} = \int d^{10}X \sqrt{-G} 
   \left[ \int d^4\theta\ \mathcal{K} + 
   \left\{ \int d^2\theta \left( \frac{1}{4g^2} \mathcal{W}^{\alpha} \mathcal{W}_{\alpha} + \mathcal{W} \right) + {\rm h.c.} \right\}    \right],
\end{equation}
where
\begin{align}
\begin{aligned}
    \mathcal{K} &= \frac{2}{g^2}h^{\bar{i}j} {\rm Tr} \left[ \left(\sqrt{2}\bar{\partial}_{\bar{z}^{{i}}}+\bar{\phi}_{\bar{z}^{{i}}}\right)  e^{-V} \left(-\sqrt{2}{\partial}_{z^j}+{\phi}_{z^j} \right) e^V + \bar{\partial}_{\bar{z}^{{i}}} e^{-V} \partial_{z^j} e^V  \right] + \mathcal{K}_{WZW}, \\
\mathcal{W} &= \frac{1}{g^2} \epsilon^{\rm ijk}e_{\rm i}^{\ i} e_{\rm j}^{\ j} e_{\rm k}^{\ k} \ {\rm Tr} \left[ \sqrt{2} \phi_{z^i} \left( \partial_{z^j} \phi_{z^k} - \frac{1}{3\sqrt{2}} [\phi_{z^j}, \phi_{z^k}]
\right)
\right].
\end{aligned}
\end{align}
 The field strength superfield is given by $\mathcal{W}_{\alpha} := -\frac{1}{4} \bar{D} \bar{D} e^{-V} D_{\alpha} e^V$ where $D_{\alpha}$ and $ \bar{D}_{\dot{\alpha}}$ are the supercovariant derivatives with 4D spinor indices denoted by $\alpha$ and its conjugate $\dot{\alpha}$. The Wess-Zumino-Witten term is denoted by $\mathcal{K}_{WZW}$ which vanishes under the Wess-Zumino gauge.
In the superpotential $\mathcal{W}$, we have a totally anti-symmetric tensor $\epsilon^{\rm ijk}$ where $\epsilon^{123}=1$.

From the equations of motion for the auxiliary fields $F_{z^i}$, we find 
\begin{align}
\begin{aligned}
\bar{F}_{\bar{z}^{{i}}} &= - h_{j \bar{i}}\ \epsilon^{\rm{jkl}} e_{\rm j}^{\ j} e_{\rm k}^{\ k} e_{\rm l}^{\ l} \left( \partial_{z^k} A_{z^l} -\frac{1}{4} [A_{z^k},A_{z^l}] \right),
\end{aligned}
\end{align}
as in ref.\cite{Abe:2012ya}.
The F-flat SUSY condition is to require $\langle F_{z^i} \rangle = 0$, where the angle brackets indicate to take the vacuum expectation value. Noting that we have only assumed the Abelian background magnetic flux, the F-flat condition is reduced to 
\begin{equation}
    \langle   \bar{F}_{\bar{z}^{{i}}}  \rangle = -  h_{j \bar{i}}\ \epsilon^{\rm{jkl}} e_{\rm j}^{\ j} e_{\rm k}^{\ k}
    e_{\rm l}^{\ l}  \partial_{z^k} \langle A_{z^l} \rangle = 0.
\end{equation}
This shows that $(2,0)$- and $(0,2)$-components of the background flux are vanishing. For example, if we take $i=3$,
\begin{align}
\begin{aligned}
\langle \bar{F}_{\bar{z}^{{3}}}  \rangle &= -  h_{j \bar{3}}\ \epsilon^{\rm{jkl}} e_{\rm j}^{\ j} e_{\rm k}^{\ k}
    e_{\rm l}^{\ l}  \partial_{z^k} \langle A_{z^l} \rangle  \\
    &= - \frac{1}{(2\pi R)^2} \frac{1}{2\pi r} h_{3 \bar{3}}\ \left( \partial_{z^1} \langle A_{z^2} \rangle - \partial_{z^2} \langle A_{z^1} \rangle \right) \\
    &\propto \langle F_{z^1 z^2} \rangle.
\end{aligned}
\end{align}
Thus, the background flux is constrained to be a $(1,1)$-form.

So far, we have assumed torus compactifications.
However, if there are no tachyonic modes before the orbifold projection,
our orbifold models have no tachyonic modes in the gauge sector.

\section{Properties of Riemann theta function}
Following properties of the Riemann theta functions are useful\cite{Mumford:1983, Alvarez-Gaume:1986}. For $\vec{a'}, \vec{b'} \in \mathbb{Q}^2$, we have
\begin{align}
 \label{eq: property1}
    \vartheta
     \begin{bmatrix}
     \vec{a}^{\, \rm T} \\ \vec{b}^{\, \rm T} + \vec{b^'}^{\rm T}
     \end{bmatrix}(\vec{z}, \Omega)
     &=
     \vartheta 
      \begin{bmatrix}
      \vec{a}^{\, \rm T} \\ \vec{b}^{\, \rm T}
      \end{bmatrix} (\vec{z} + \vec{b}', \Omega), \\
 \label{eq: property2}      
     \vartheta 
      \begin{bmatrix}
      \vec{a}^{\, \rm T} + \vec{a'}^{\rm T} \\ \vec{b}^{\, \rm T}
      \end{bmatrix}(\vec{z}, \Omega) 
     &=
     e^{\pi i \vec{a'}^{\rm T} \cdot \Omega \cdot \vec{a'} + 2 \pi i \vec{a'}^{\rm T} \cdot (\vec{z} + \vec{b})} \cdot 
     \vartheta 
      \begin{bmatrix}
      \vec{a}^{\, \rm T} \\ \vec{b}^{\, \rm T}
      \end{bmatrix}(\vec{z} + \Omega \vec{a'}, \Omega).
\end{align}
For $\vec{k}, \vec{l} \in \mathbb{Z}^2$, we have
\begin{equation}
\label{eq: property3}
    \vartheta 
      \begin{bmatrix}
      {\vec{a}}^{\, \rm T} + \vec{k}^{\, \rm T} \\ \vec{b}^{\, \rm T} + {\vec{l}}^{\, \rm T}
      \end{bmatrix}(\vec{z}, \Omega) 
     =
     e^{2 \pi i \vec{a}^{\rm T} \vec{l}} \cdot 
     \vartheta 
      \begin{bmatrix}
      \vec{a}^{\, \rm T} \\ \vec{b}^{\, \rm T}
      \end{bmatrix}(\vec{z}, \Omega).
\end{equation}


 \end{document}